\documentclass[aps,prd,a4paper,amsfonts,amssymb,nofootinbib,onecolumn,longbibliography,preprintnumbers,10pt]{revtex4-2}

\usepackage[utf8]{inputenc}
\usepackage{blindtext}
\usepackage{anysize}
\marginsize{2cm}{2cm}{1.5cm}{1.5cm}
\usepackage[pdftex]{graphicx}
\usepackage{float}
\usepackage{amsmath}
\usepackage{amssymb}
\usepackage{amsfonts}
\usepackage{accents}
\usepackage{xcolor}
\usepackage[hidelinks]{hyperref}
\usepackage{multirow}
\usepackage{booktabs}

\usepackage{tikz}
\usepackage{wrapfig}
\usetikzlibrary{quantikz2}

\newcommand{\bal}{\begin{align}}
\newcommand{\eal}{\end{align}}
\newcommand{\beq}{\begin{eqnarray}}
\newcommand{\eeq}{\end{eqnarray}}
\newcommand{\nneeq}{\nonumber \end{eqnarray}}

\newcommand{\ketbrac}[2] { {\text{$\mathfrak{C}_{{#1}{#2}}$}} }
\newcommand{\projector}[2]{{\text{$\mathbb{P}^{({#1})}_{{#2}}$}}}
\newcommand{\set}[1]{{\text{$\mathfrak{s}^{\dagger}_{{#1}}$}}}
\newcommand{\scrap}[1]{{\text{$\mathfrak{s}_{{#1}}$}}}

\newcommand{\stepsym}[1] {{\text{$\mathcal{S}_{{#1}\leftarrow{#1}-1} $}}}
\newcommand{\stepasym}[1] {{\text{$\mathcal{A}_{{#1}\leftarrow{#1}-1} $}}}

\newcommand{\innerid} {{\text{$\mathfrak{i}$}}}
\newcommand{\outeridentity} {{\text{$\mathbb{I}$}}}
\newcommand{\bigO}[1] {{\text{$\mathcal{O}\left({#1}\right)$}}}

\newcommand{\fSWAP}[3] {{\text{$\text{\tt fS}_{{#1}}\left({#2},{#3}\right)$}}}
\newcommand{\fSWAPdg}[3] {{\text{$\text{\tt fS}^{\dagger}_{{#1}}\left({#2},{#3}\right)$}}}
\newcommand{\fX}[2] {{\text{$\text{\tt fX}_{{#1}}\left({#2}\right)$}}}
\newcommand{\fXdg}[2] {{\text{$\text{\tt fX}^{\dagger}_{{#1}}\left({#2}\right)$}}}
\newcommand{\fSp}[3] {{\text{$\text{\tt fSp}_{{#1}}\left({#2},{#3}\right)$}}}

\newcommand{\bSWAP}[3] {{\text{$\text{\tt bS}_{{#1}}\left({#2},{#3}\right)$}}}

\newcommand{\bX}[2] {{\text{$\text{\tt bX}_{{#1}}\left({#2}\right)$}}}
\newcommand{\bXdg}[2] {{\text{$\text{\tt bX}^{\dagger}_{{#1}}\left({#2}\right)$}}}

\newcommand{\bmat}{\left[\begin{array}}
\newcommand{\emat}{\end{array}\right]}

\begin{document}

\title{A dynamical implementation of canonical second quantization\\
on a quantum computer 
}
\author{Juan Jos\'e G\'alvez-Viruet and Felipe J. Llanes-Estrada \\
Depto. de F\'{\i}sica Te\'orica \& IPARCOS, Fac. CC. F\'{\i}sicas,\\ Plaza de las Ciencias 1 28040 Universidad Complutense de Madrid, Spain.}
\date{\today}

\begin{abstract}
We develop theoretical methods for the implementation of creation and destruction operators in separate registers of a quantum computer, allowing for a transparent and dynamical creation and destruction of particle modes in second quantization in problems with variable particle number.  We establish theorems for the commutation (anticommutation) relations on a finite memory bank and provide the needed symmetrizing and antisymmetrizing operators. Finally, we provide formulae in terms of these operators for unitary evolution under conventional two- and four-body Hamiltonian terms, as well as terms varying the particle number.
 In this formalism, the number of qubits needed to codify $n$ particles with $N_p$ modes each is of order $n\log_2 N_p$.
Such scaling is more efficient than the Jordan-Wigner transformation which requires $O(N_p)$ qubits, whenever there are a modest number of particles with a large number of states available to each (and less advantageous for a large number of particles with few states available to each). And although less efficient, it is also less cumbersome than compact encoding. 
\end{abstract}

\maketitle

\centerline{{\vbox{\hbox{Preprint issued as IPARCOS-UCM-23-140}}}}

\newpage 

\tableofcontents

\newpage

\section{Introduction}
Quantum computing is widely held as a promising tool to advance many-body physics in all areas where quantum phenomena are of importance, particularly nuclear and strongly coupled particle physics~\cite{Beck:2023xhh}.
Quantum mechanical applications at fixed particle number abound~\cite{Lanyon:2009tha}, importantly for quantum chemistry and typical condensed matter Hamiltonians, and increasingly nuclear physics~\cite{Wang:2024scd,perez-obiol_nuclear_2023,Drissi:2024cnn}. In particular, quantum simulation of fermion systems is one of the most advanced applications of quantum computers. First algorithms were proposed twenty years ago \cite{abrams_009711_1997,ortiz_quantum_2001,bravyi_fermionic_2002} and most of them assume the second quantization formalism, in which the identification between the fermion occupation-number basis and a quantum memory of qubits is straightforward. Remarkably, \cite{abrams_009711_1997} also considered the use of ``first quantization'' and already discussed algorithms to antisymmetrize a quantum memory. Nowadays there are very efficient algorithms for fixed particle-number problems, in particular to solve the electronic structure problem, see for example \cite{Lanyon:2009tha,Whitfield:2011mp,toloui_quantum_2013,di_matteo_improving_2021,kirby_second-quantized_2022,shee_qubit-efficient_2022,huang_quantum_2023} or \cite{mcardle_quantum_2020} for a recent review. Almost all of them use the occupation number basis, but as recent studies have shown, first quantization, in which the quantum memory is antisymmetrized explicitly (as it is in this work), can scale better in terms of qubit counts and gate complexities \cite{babbush_exponentially_2016,bravyi_tapering_2017,berry_improved_2018,babbush_quantum_2019,su_fault-tolerant_2021}.


Many of the techniques developed in quantum chemistry can be used for the simulation of quantum field theories, but have to be generalized to accommodate a varying number of particles, to encode bosons as well as fermions and to consider the exponentiation of general Hermitian operators. Usually these problems are solved by leaving aside particles and encoding fields instead, with lattice gauge theories a salient example~\cite{Kogut:1974ag,jordan_quantum_2012,Zohar:2013zla,macridin_digital_2018,Farrell:2022wyt,Stetina:2022simulatingeffective,macridin_bosonic_2022,farrell_preparations_2023,farrell_scalable_2024}. In this work, we have adopted the few- and many- body point of view in which only a few Fock-sectors are taken into account and the basic object is the particle \cite{weinberg_quantum_2014}. This is convenient, for example, for the study of jet broadening \cite{barata_220806750v1_2022,barata_230701792v1_2023}, heavy quarkonia \cite{glazek_renormalized_2017} and more generally on systems where a mass-gap is present or spontaneously arises, as the number of modes that need to be taken into account to a given precision is small.


We have developed a natural implementation of the formalism of second quantization  in a way that is scalable to large model spaces and computer memories, without relying on the Jordan-Wigner transformation which, at every step of a calculation, puts the translation of the physical system into the quantum computer out of the reach of many possible users, due to its bookkeeping and improvable scalability. 

Our philosophy here is to take one step of abstraction further so that the memory of the quantum computer is organized in a way similar to quantum field theory in canonical second quantization in the particle basis. This is intended to somewhat free the physicist from low-level thinking on the computer qubits but allow for higher level objects, the particle registers, to become the basic stones for building physical implementations. This register-particle codification was first introduced in \cite{Barata:2020jtq} to simulate a boson theory and it is similar to ``compact encoding'' \cite{Kirby:2021PRA}. The main differences with respect to previous studies is that our implementation of creation and annihilation operators exactly  fulfills both commutation (or anticommutation) relations provided the memory (not completely occupied) is filled following certain order. Furthermore, our operators proceed changing only a small amount of registers on each time-evolution step, avoiding the costs of having to advance the entire memory. We expect this particle-register with clear physical interpretation to be useful for formulations of field theory in light-front gauge~\cite{Qian:2021jxp,Yao:2022eqm}, axial gauge~\cite{Farrell:2022wyt} or eventually Coulomb gauge~\cite{Rocha:2010nab} theories.

Readers familiar with object-oriented programming languages will quickly recognize the parallel among the particle creation/annihilation operators and the 
creator/destructor methods of a class, here the ``particle'', whose attributes are its physical quantum numbers. This is a natural analogy. The particular feature of second quantization is that the creator and annihilation operators need to follow specific rules, in the form of (anti)commutation relations that need to be mapped to the computer's methods.

This is certainly not the most efficient way to code field theory problems for a quantum computer, with many intelligent ideas~\cite{Zohar:2013zla} proposed in the last decades, but with fast hardware improvements it becomes attractive to think of user-friendly ways of  coding physical problems that will reduce the overhead time for developers.

Section~\ref{sec:1body} of this article is dedicated to implementing one-body operators. Then, section~\ref{sec:2bosons} discusses the simultaneous occupation of two registers, and how Bose symmetry among those two particles can be described: the addition of an extra register to accommodate two particles must comply with the requirements of quantum statistics. In it we also start addressing the canonical commutation relations among creation and destruction operators.

In section~\ref{sec:nbosons} we generalize the concepts to a computer with an arbitrary number of registers and holding an arbitrary (but small) number of particles. The commutator of the earlier creation and annihilation operators will there be seen to be well implemented over the multiparticle Fock space, except for the boundary term that arises when the number of particles is large enough to saturate the number of qubits available.

We follow with a section, \ref{sec:anticommuting}, dedicated to the implementation of fermion anticommuting operators. The creation operators must now avoid filling two separate registers with the same quantum numbers, which would be forbidden by Pauli's exclusion principle, and properly implementing the anticommutation relations among the creation and destruction operators is a way to guarantee it.

The idea of the effort is to make a way to dynamically handle variable numbers of particles in running time
available, as well as allowing the computer to maintain superpositions of states with different particle numbers during execution.
Section~\ref{sec:evolution} then explains the implementation of evolution operators via exponentiation of the operators in the so-constructed Fock-algebra and their products in the case that the particle number is constant. A discussion of 
the computational cost and scalability of the construction is also specified.
This section is followed by an equivalent discussion for the more general case of particle-number changing operators in section~\ref{sec:evolution2}. 
Further discussion about needed memory size and scalability is given in section~\ref{subsec:costs}, and a final discussion is laid out in section~\ref{sec:conclusions}.
Several demonstrations too extensive for the main text are presented, in the spirit of exhaustiveness, in the  appendices. This includes a glossary of symbols in appendix~\ref{app:glossary} that may prove useful to the reader given the notation-intensive presentation.

\section{Implementation of single-particle operators}
\label{sec:1body}

The abstraction that we pursue is to assign a ``register'' to each particle active in a process. 
Such registers are sets of qubits encoding the quantum numbers $q$ of the represented particle, with one additional qubit that we dub the ``presence/absence'' qubit denoted as $P/A$, which is a handle allowing to decide whether one particle is represented by that register, or whether it should be considered empty and available. 
If an additional particle is needed, an empty register has to be activated and its presence qubit set to 1.

For a particle with momentum, spin, and internal quantum numbers such as a color, the vacuum state (corresponding to the absence of the particle) is conveniently taken as
\begin{equation}
\ket{\Omega} \equiv \ket{0}_{\text{P/A}}\otimes \ket{0}_{\text{spin}}\otimes\ket{00}_{\text{color}} \otimes \ket{0 ... 0}_{\text{momentum}} 
\label{eq:creation-1}
\end{equation}
the number of possible values of a quantum number dictates how many qubits are necessary for its representation: codification of spin requires one qubit, codification of color~\cite{Chawdhry:2023jks} two qubits and codification of momenta, which is a continuous variable, requires discretization on a variable number of qubits to reach the wished precision. From here on we consider $N_p$ different values of momenta codified in $n_p=\lceil\log_2 N_p\rceil$ qubits (or in three dimensions, remark which we will usually omit in the rest of the article, $(\lceil\log_2 N_p\rceil)^3$ qubits).
In the vacuum state, each register is in a reference state, here taken to be the null ket, that does not represent any valid quantum number.

Our first goal is to represent
the action of the basic creation operator in Fock space, a staple of many-body theory, $a_{s,c,p}^{\dagger}$.

To implement it, we write it in terms of \textit{set} operators $\mathfrak{s}$ acting on each of the product Hilbert spaces describing each degree of freedom,  and control operators $\ketbrac{i}{j}$
\begin{equation}
a_{s,c,p}^{\dagger} \equiv \ketbrac{1}{0}\otimes \mathfrak{s}^{\dagger}_{s}\otimes \mathfrak{s}^{\dagger}_{c} \otimes \mathfrak{s}^{\dagger}_{p}, 
\label{eq:creation-2}
\end{equation}
so that 
\begin{align}
a_{s,c,p}^{\dagger}\ket{\Omega} & \equiv \ketbrac{1}{0}\ket{0}_{\text{P/A}}\otimes \mathfrak{s}^{\dagger}_{s}\ket{0}\otimes \mathfrak{s}^{\dagger}_{c}\ket{00} \otimes \mathfrak{s}^{\dagger}_{p} \ket{0...0} \\
&  = \ket{1}_{\text{P/A}}\otimes \ket{s}\otimes \ket{c} \otimes  \ket{p} \equiv \ket{1scp}\ .
\label{eq:creation-3}
\end{align}
The first qubit, that indicated the presence or absence of a particle in the register, has now being set to $|1\rangle_{P/A}$. The control operator $\ketbrac{1}{0}$ also prevents the attempt of creating a quantum if the register is already encoding another one (``occupied register''). Control operators acting over the presence/absence qubits fulfill
\begin{equation}
    \ketbrac{i}{j}\ket{k} = \delta_{jk}\ket{i},\,\,\,\ketbrac{i}{k}\ketbrac{l}{m} = \delta_{kl}\ketbrac{i}{m}.
\label{def:control-operators}
\end{equation}

The Hermitian conjugate of the abstract $a^\dagger$ operators in many-body theory are the annihilation operators; they can be implemented as in
\begin{equation}
a_{s,c,q} = (a_{s,c,q}^{\dagger})^{\dagger}=\ketbrac{0}{1}\otimes \mathfrak{s}_{s}\otimes \mathfrak{s}_{c} \otimes \mathfrak{s}_{q},
\label{eq:annihilation}
\end{equation}
where, in intuitive notation, $\mathfrak{s}$ are \textit{scrap} operators which can clear the quantum numbers: $\mathfrak{s}_{s}$ returns states of spin $\ket{s}$ (spin up or down) to the reference spin state $\ket{s_{0}}$ (spin down, for example), with analogous behaviour for $\mathfrak{s}_{c}$ and $\mathfrak{s}_{q}$.
(Because the quantum computer will encode the unitary operators arising from exponentiation of combinations of $a^\dagger$ and $a$, the fact that, {\it e.g.} $\ketbrac{1}{0}\ket{1}=0$ yields an unrepresentable null vector will pose no problem, as it exponentiates to the identity.)

In summary, the operations that this implementation of the Fock algebra can be used for are:
\begin{itemize}
\item To occupy an unassigned register, encoding a particle in it,
\begin{equation}
a_{s,c,p}^{\dagger}\ket{\Omega}  \ =\  \ketbrac{1}{0}\ket{0}_{\text{P/A}}\otimes \mathfrak{s}^{\dagger}_{s}\ket{0}\otimes \mathfrak{s}^{\dagger}_{c}\ket{00} \otimes \mathfrak{s}^{\dagger}_{p} \ket{0...0} \ = \  \ket{1scp}\ .
\label{def:set-operators}
\end{equation}
\item To vacate an occupied register if its quantum numbers match those of the destruction operator, resetting them to vacuum default,
\begin{equation} 
a_{s',c',q'}\ket{1scp} \ =\  \ket{0}\otimes \mathfrak{s}_{s'}\ket{s}\otimes \mathfrak{s}_{c'}\ket{c} \otimes \mathfrak{s}_{q'}\ket{q} = \delta_{s's}\delta_{c'c}\delta_{q'q}\ket{\Omega}
\label{def:scrap-operators}
\end{equation}
\item To replace the quantum numbers of an existing particle occupying a register, employing combinations that conserve particle number (or, as a specific case, to pass on as with the identity) such as
\begin{align}
(a^{\dagger}_{s_{2},c_{2},q_{2}}a_{s'_{2},c'_{2},q'_{2}})\cdot (a^{\dagger}_{s_{1},c_{1},q_{1}}a_{s'_{1},c'_{1},q'_{1}}) \ket{1s_0c_0q_0} & = \delta_{s_0,s'_1}\delta_{c_0,c'_1}\delta_{q_0,q'_1}a^{\dagger}_{s_{2},c_{2},q_{2}}a_{s'_{2},c'_{2},q'_{2}}\ket{1s_1c_1q_1} \nonumber \\
& \delta_{s_0,s'_1}\delta_{c_0,c'_1}\delta_{q_0,q'_1}\delta_{s_1,s'_2}\delta_{c_1,c'_2}\delta_{q_1,q'_2}\ket{1s_2c_2q_2}
\end{align}
or, acting on an empty register,
\begin{align}
(a_{s'_{2},c'_{2},q'_{2}}a^{\dagger}_{s_{2},c_{2},q_{2}})\cdot(a_{s'_{1},c'_{1},q'_{1}}a^{\dagger}_{s_{1},c_{1},q_{1}})\ket{\Omega} & = \delta_{s_1,s'_1}\delta_{c_1,c'_1}\delta_{q_1,q'_1}a_{s'_{2},c'_{2},q'_{2}}a^{\dagger}_{s_{2},c_{2},q_{2}}\ket{\Omega} \nonumber \\
& = \delta_{s_2,s'_2}\delta_{c_2,c'_2}\delta_{q_2,q'_2}\delta_{s_1,s'_1}\delta_{c_1,c'_1}\delta_{q_1,q'_1}\ket{\Omega} \ .
\end{align}

\end{itemize}

\section{Implementation of two-particle boson operators}
\label{sec:2bosons}
\subsection{Basic necessary properties}
In this section we address the setting and handling of two bosons in the memory. 
The new complication, respect to dealing with only one particle, is the 
need to represent the abstract symmetry under boson exchange independently of the underlying hardware.

Let us start by the vacuum states, omitting quantum numbers other than momentum to lighten the expressions.
\begin{equation}
    \ket{\Omega} = \ket{\Omega}_{2}\otimes \ket{\Omega}_{1} \equiv \ket{0}_{\text{P/A}}\ket{0...0}_{\text{momentum}}\otimes\ket{0}_{\text{P/A}}\ket{0...0}_{\text{momentum}},
\end{equation}
We have highlighted the two product states by the presence/absence and a ``momentum'' subscripts, that we now drop, with the ordering of the two states specifying which is which.

 The action of any creation operators $a^{(2)\dagger}_i$ over this pair of registers is defined to emulate their corresponding action on Fock space. (The superindex $(2)$ indicates that it acts over a two-register memory, which will be useful to keep track of to distinguish the various operators discussed). 
 
 If applied over the vacuum, it creates a first particle, which we assign to the rightmost register labelled (1), 
\begin{equation}
a_{q_{1}}^{(2)\dagger}\ket{\Omega}  =\ket{q_{1}}\equiv \ket{\Omega}_{2}\otimes \left(\ket{1}\ket{q_{1}}\right)_1 = \ket{\Omega}_2 \otimes\ket{1q_{1}}_1\ .
\label{eq:2Reg-creation-1}
\end{equation}

If that register is already occupied by one boson, the creation operator needs to assign the second (note Bose's symmetry):
\begin{equation}
a_{q_{1}}^{(2)\dagger}\ket{\Omega}_2 \otimes \ket{1p_{1}}_1 \equiv \frac{1}{\sqrt{2}}\left(\ket{1q_{1}}_2\otimes \ket{1p_{1}}_1+\ket{1p_{1}}_2\otimes \ket{1q_{1}}_1\right)\ .
\label{eq:2Reg-creation-2}
\end{equation}
Finally, if both registers already contain a boson, no further particle can be encoded in this computer with only two registers. 

The creation operator can then do nothing and we naturally choose to set its action to null,
\begin{equation}
a_{q_{1}}^{(2)\dagger}\ket{1p_{2}}_2\otimes \ket{1p_{1}}_1  = 0.
\label{eq:2Reg-creation-3}
\end{equation}
This is a condition at the boundary of the computer memory: its size must truncate the Fock space, and as usual, introduce a discretization error in the second quantized problem being treated. Adding further memory will alleviate it, so we will return to this below in section~\ref{sec:nbosons}.

In the case in which 
only one object sits in the two--register  memory, we need to resolve the ambiguity of choosing one of them to carry the active quantum numbers.

We select to stack particles from right to left, entailing
the exclusion of states such as $\ket{1p}_2\otimes\ket{\Omega}_1$, i.e., with the first register empty and the second occupied.
Note that no symmetry is here necessary, since the registers are just sections of memory and do not represent physical particles: here there is just one, and nothing to symmetrize.

The action of the corresponding annihilation operators is also defined to mimic their behaviour in Fock space,  
\begin{equation}
\begin{array}{ccc}
a^{(2)}_{q_{1}}\ket{\Omega}  = 0; & \phantom{hazsitio} & a_{q_{1}}^{(2)}\ket{\Omega}_2\otimes \ket{1p_{1}}_1 \  = \ \delta_{q_{1}p_{1}}\ket{\Omega}_2\otimes \ket{\Omega}_1\  = \ \delta_{q_{1}p_{1}}\ket{\Omega},
\end{array}    
\label{eq:2Reg-anni-1}
\end{equation}
and 
\begin{equation}
a^{(2)}_{q_{1}}\left(\ket{1p_{2}}_2 \otimes \ket{1p_{1}}_1 \right)_S= \delta_{q_{1}p_{2}}\ket{\Omega}_2 \otimes \ket{1p_{1}}_1 + \delta_{q_{1}p_{1}}\ket{\Omega}_2 \otimes \ket{1p_{2}}_1,
\label{eq:2Reg-anni-2}
\end{equation}
where $\left(\ket{1p_{2}}_2 \otimes \ket{1p_{1}}_1 \right)_S$ is a two-particle symmetric state with momenta $p_1$ and $p_2$.

With these requirements we can now try to write down the higher level creation/annihilation operators in terms of the 
lower level set/scrap operators from section~\ref{sec:1body}.

\subsection{Implementation}
We find convenient to split the Fock operator acting on the two-register memory into two auxiliary suboperators, one acting on each of the two registers when that register is the last occupied one (which regrettably requires an additional subindex),
\begin{align}
a_{q_{1},1}^{(2)\dagger} & = \left( \ketbrac{0}{0}\otimes\innerid\right)_{2}\otimes \left(\ketbrac{1}{0}\otimes \mathfrak{s}_{q_{1}}^{\dagger} \right)_{1}, \nonumber \\ 
a_{q_{1},2}^{(2)\dagger} & = \left(\ketbrac{1}{0}\otimes \mathfrak{s}_{q_{1}}^{\dagger} \right)_{2}\otimes \left(\ketbrac{1}{1}\otimes \innerid\right)_{1}.
\label{2bodyauxiliary}
\end{align}
(Parentheses isolate operators acting on the first or second register; the control terms $\ketbrac{i}{j}$ act on presence qubits and the identity $\innerid$ and set $\mathfrak{s}$ operators act on momentum qubits.)
The construction of these auxiliary operators is straightforward: the first one sets the quantum numbers of the first, empty, register to the particle being added and flips its presence qubit from absence to presence, while leaving the second register untouched. 

The second operator leaves the first register, assumed occupied, untouched, and adds the particle to the second. Before summing it with the first to form the total creation operator, we need to note that its action produces a two-particle state that does need to be symmetric under exchange. Hence, we multiply it by a symmetrizer
\begin{equation}
    a_{q_{1}}^{(2)\dagger} = a^{(2)\dagger}_{q_{1},1}+ \frac{1}{\sqrt{2}}\left(\outeridentity\otimes \outeridentity+\mathcal{P}_{21}\right)a^{(2)\dagger}_{q_{1},2}\ .
    \label{2bodycreator}
\end{equation}
(Note that symmetrization cannot be trivially implemented in runtime after the initial state preparation, because it projects out the antisymmetric parts of the Hilbert space, which is not a unitary operation germane to a quantum computer. We can circumvent the problem as shown in Eq.~(\ref{eq:antisymmetric-action}) below.)

 The $\mathcal{P}_{21}$ permutation operation swaps the quantum numbers encoded in each of the two registers:
\begin{equation}
\mathcal{P}_{21}\ket{1p_{1}}\ket{1p_{2}} \equiv \ket{1p_{2}}\ket{1p_{1}}\ ,
\label{def:12-permutator}
\end{equation}
whereas $\outeridentity$ is the ``register-level identity'' which leaves the quantum numbers in the corresponding register unchanged.

If we substitute the auxiliary operators of Eq.~(\ref{2bodyauxiliary}) into Eq.~(\ref{2bodycreator}) we can give a direct expression for the creation operator over the two-register memory
\begin{align}
    a_{q_{1}}^{(2)\dagger} & = \left(\ketbrac{0}{0}\otimes \innerid \right)_{2}\otimes \left(\ketbrac{1}{0}\otimes \mathfrak{s}_{q_{1}}^{\dagger}\right)_{1} \nonumber \\
    & + \frac{1}{\sqrt{2}}\left(\outeridentity\otimes\outeridentity + \mathcal{P}_{21} \right)\left(\ketbrac{1}{0}\otimes \mathfrak{s}_{q_{1}}^{\dagger} \right)_{2}\otimes \left(\ketbrac{1}{1}\otimes \innerid\right)_{1}.
    \label{eq:2Reg-creation-4}
\end{align}
This equation fulfills the quantum field theory requirements of Eq.~(\ref{eq:2Reg-creation-1}) 
 and (\ref{eq:2Reg-creation-2}) in the two-body sector. 
 
The annihilation operators are then found by Hermitian conjugation
\begin{align}
    a_{q_{1}}^{(2)} & = \left(\ketbrac{0}{0}\otimes \innerid \right)_{2}\otimes \left(\ketbrac{0}{1}\otimes \mathfrak{s}_{q_{1}}\right)_{1} \nonumber \\
    & + \left(\ketbrac{0}{1}\otimes \mathfrak{s}_{q_{1}} \right)_{2}\otimes \left(\ketbrac{1}{1}\otimes \innerid\right)_{1}\frac{1}{\sqrt{2}}\left(\outeridentity\otimes\outeridentity + \mathcal{P}_{21} \right).
    \label{2bodyannihilation}
\end{align}
The first piece acts when only the first register is occupied, annihilating that one particle; but if both are filled, its action is null, and it is the second piece that is active and erases the particle in the second register.
We now turn to showing that this construction satisfies the canonical commutation relations in the Fock space restricted to two memory registers. 

\subsection{Canonical commutation relations among bosons}

boson operators must fulfill, at the operator level, \begin{equation}
\left[a_{q_{1}}a^{\dagger}_{q_{2}}\right] = a_{q_{1}}a^{\dagger}_{q_{2}}-a^{\dagger}_{q_{2}}a_{q_{1}} = \delta_{q_{1}q_{2}}\ .
\end{equation}
In this subsection we show that, in a two--register memory, this is satisfied for occupation numbers zero and one, but that the  two-particle case takes a boundary term (which can only be avoided by adding further registers).

The first of the two terms of the commutator $\left[a_{q_{1}},a^{\dagger}_{q_{2}}\right]$ is, from Eq.~(\ref{eq:2Reg-creation-4}) and Eq.~(\ref{2bodyannihilation}),
\begin{align}
    a_{q_{1}}^{(2)}a^{(2)\dagger}_{q_{2}} & = \left[\left(\ketbrac{0}{0}\otimes \innerid \right)_{2}\otimes \left(\ketbrac{0}{1}\otimes \mathfrak{s}_{q_{1}}^{\phantom{\dagger}\! }\right)_{1}\right]\left[\left(\ketbrac{0}{0}\otimes \innerid \right)_{2}\otimes \left(\ketbrac{1}{0}\otimes \mathfrak{s}_{q_{2}}^{\dagger}\right)_{1}\right]\nonumber \\
    & + \left[\left(\ketbrac{0}{1}\otimes \mathfrak{s}_{q_{1}}^{\phantom{\dagger}\! } \right)_{2}\otimes \left(\ketbrac{1}{1}\otimes \innerid\right)_1\right]\left(\outeridentity\otimes\outeridentity + \mathcal{P}_{21} \right)\left[\left(\ketbrac{1}{0}\otimes \mathfrak{s}_{q_{2}}^{\dagger} \right)_{2}\otimes \left(\ketbrac{1}{1}\otimes \innerid\right)_{1}\right]
\end{align}
(the crossed terms cancel because of Eq.~(\ref{def:control-operators})).
Associating together the operators over each subspace allows this to be written as
\begin{equation}
a_{q_{1}}^{(2)}a^{(2)\dagger}_{q_{2}} = \left(\ketbrac{0}{0}\otimes \innerid\right)_2\otimes \left(\ketbrac{0}{0}\otimes \mathfrak{s}_{q_{1}}\mathfrak{s}^{\dagger}_{q_{2}}\right)_1 + \left(\ketbrac{0}{0}\otimes \mathfrak{s}_{q_{1}}\mathfrak{s}^{\dagger}_{q_{2}}\right)\otimes \left(\ketbrac{1}{1}\otimes\innerid\right)+ \mathcal{P}_{21}\left(\ketbrac{1}{0}\otimes \mathfrak{s}^{\dagger}_{q_{2}}\right)\otimes \left(\ketbrac{0}{1}\otimes \mathfrak{s}_{q_{1}}\right)
    \label{eq:commutator-1}
\end{equation}
The reversed-order product is
\begin{align}
    a^{(2)\dagger}_{q_{2}}a_{q_{1}}^{(2)} & = \left[\left(\ketbrac{0}{0}\otimes \innerid \right)_{2}\otimes \left(\ketbrac{1}{0}\otimes \mathfrak{s}_{q_{2}}^{\dagger}\right)_{1}\right]\left[\left(\ketbrac{0}{0}\otimes \innerid \right)_{2}\otimes \left(\ketbrac{0}{1}\otimes \mathfrak{s}_{q_{1}}^{\phantom{\dagger}\! }\right)_{1}\right]\\
    & + \left(\outeridentity\otimes\outeridentity + \mathcal{P}_{21} \right)\left[\left(\ketbrac{1}{0}\otimes \mathfrak{s}_{q_{2}}^{\dagger} \right)_{2}\otimes \left(\ketbrac{1}{1}\otimes \innerid\right)_{1}\right]\left[\left(\ketbrac{0}{1}\otimes \mathfrak{s}_{q_{1}}^{\phantom{\dagger}\! } \right)_{2}\otimes \left(\ketbrac{1}{1}\otimes \innerid\right)_{1}\right],
\end{align}
or, upon expanding the product and employing again the simplification of the factor $\mathcal{P}_{12}$  because of the preprogrammed symmetrization of two-particle states, 
\begin{align}
a^{(2)\dagger}_{q_{2}}a^{(2)}_{q_{1}} & = \left(\ketbrac{0}{0}\otimes \innerid\right)_2\otimes \left(\ketbrac{1}{1}\otimes \mathfrak{s}^{\dagger}_{q_{2}}\mathfrak{s}_{q_{1}}\right)_1+ \left(\ketbrac{1}{1}\otimes \mathfrak{s}^{\dagger}_{q_{2}}\mathfrak{s}_{q_{1}}\right)_2\otimes \left(\ketbrac{1}{1}\otimes\innerid\right)_1+ \left(\ketbrac{1}{1}\otimes \innerid\right)_2\otimes \left(\ketbrac{1}{1}\otimes \mathfrak{s}^{\dagger}_{q_{2}}\mathfrak{s}_{q_{1}}\right)_1,
    \label{eq:commutator-2}
\end{align}
Note the action of the diagonal operator with $q_1=q_2$ as a number operator: 
if only the rightmost register is occupied, it returns the same state multiplied by 1 due to the leftmost term. If both are occupied with bosons of equal $q_1$, it will return twice the state due to the middle and right pieces. 
With these expressions in Eq.~(\ref{eq:commutator-2}) we can now act the commutator over each of the three sectors of the Fock space abstracted from two particle registers. 

\paragraph{Zero-particle state.}
The zero-particle state is the vacuum, and the action thereon of the commutator reads
\begin{align}
    \left[a^{(2)}_{q_{1}},a^{(2)\dagger}_{q_{2}}\right]\ket{\Omega} & = a^{(2)}_{q_{1}}a^{(2)\dagger}_{q_{2}}\ket{\Omega}_2\ket{\Omega}_1 - a^{(2)\dagger}_{q_{2}}a^{(2)}_{q_{1}}\ket{\Omega}_2\ket{\Omega}_1\nonumber \\
    & = \ket{\Omega}_2\otimes \left(\ket{0}\mathfrak{s}_{q_{1}}\mathfrak{s}^{\dagger}_{q_{2}}\ket{0...0}\right)_1\nonumber \\
    & = \delta_{q_{1}q_{2}}\ket{\Omega},
\end{align}
where in the last line we used Eqs.~(\ref{eq:creation-3}) and~(\ref{eq:annihilation}).
\paragraph{One-particle states.}
In this case we also have a faithful representation of what is expected in second quantization, 
\begin{align}
    \left[a^{(2)}_{q_{1}},a^{(2)\dagger}_{q_{2}}\right]\ket{p} & = a^{(2)}_{q_{1}}a^{(2)\dagger}_{q_{2}}\ket{\Omega}_2\ket{1p}_1 - a^{(2)\dagger}_{q_{2}}a^{(2)}_{q_{1}}\ket{\Omega}_2\ket{1p}_1 \nonumber \\
    & = \left(\ket{0}\mathfrak{s}_{q_{1}}\mathfrak{s}^{\dagger}_{q_{2}}\ket{0...0}\right)_2\otimes\ket{1p}_1+\left(\ket{0}\mathfrak{s}_{q_{1}}\ket{p}\right)_2\otimes\left(\ket{1}\mathfrak{s}^{\dagger}_{q_{2}}\ket{0...0}\right)_1-\ket{\Omega}_2\left(\ket{1}\mathfrak{s}^{\dagger}_{q_{2}}\mathfrak{s}_{q_{1}}\ket{p}\right)_1 \nonumber \\
    & = \delta_{q_{1}q_{2}}\ket{p}+\delta_{q_{1}p}\ket{q_{2}}-\delta_{q_{1}p}\ket{q_{2}}\nonumber \\
    & =  \delta_{q_{1}q_{2}}\ket{p}\ ;
\end{align}
so far, these operators act as the standard commutation relations do.

\paragraph{Two-particle states.}
Finally we show the action of the commutator on the doubly occupied and symmetrized (as denoted by the subindex $S$) two-particle register,
\begin{align}
    \left[a^{(2)}_{q_{1}},a^{(2)\dagger}_{q_{2}}\right]\left(\ket{1q}\ket{1p}\right)_S 
    & = \left(a^{(2)}_{q_{1}}a^{(2)\dagger}_{q_{2}}-a^{(2)\dagger}_{q_{2}}a^{(2)}_{q_{1}}\right)\frac{1}{\sqrt{2}}\left(\ket{1q}_2\ket{1p}_1+\ket{1p}_2\ket{1q}_1\right) \nonumber\\& = -a^{(2)\dagger}_{q_{2}}a^{(2)}_{q_{1}}\frac{1}{\sqrt{2}}\left(\ket{1q}_2\ket{1p}_1+\ket{1p}_2\ket{1q}_1\right) \nonumber\\
    & = - \frac{1}{\sqrt{2}}\left(\left(\ket{1}\mathfrak{s}^{\dagger}_{q_2}\mathfrak{s}_{q_1}\ket{q}\right)_2\ket{1p}_1+\left(\ket{1}\mathfrak{s}^{\dagger}_{q_2}\mathfrak{s}_{q_1}\ket{p}\right)_2\ket{1q}_1\right.\nonumber \\
    & \left.+\ket{1q}_2\left(\ket{1}\mathfrak{s}^{\dagger}_{q_2}\mathfrak{s}_{q_1}\ket{p}\right)_1+\ket{1p}_2\left(\ket{1}\mathfrak{s}^{\dagger}_{q_2}\mathfrak{s}_{q_1}\ket{q}\right)_1\right)\nonumber \\
    & = -\frac{1}{\sqrt{2}}\left(\delta_{q_1 q} \ket{q_2 p} + \delta_{q_1 p}\ket{q_2 q} + \delta_{q_1 p} \ket{q q_2} + \delta_{q_1 q}\ket{p q_2}\right) \nonumber \\
    & = -\delta_{q_1 q}\ket{q_2 p}_S - \delta_{q_1 p}\ket{q_2 q}_S.
    \label{2bodycomm}
\end{align}
As expected due to the boundary term in Eq.~(\ref{eq:2Reg-creation-1}), in which the creation operator acting over two occupied registers is by necessity defined to be 0, 
a boundary mismatch  appears in this
calculation, that in textbook field-theory should yield $\delta_{q_1q_2}\left( \ket{1q}\ket{1p}\right)_S$
instead of the outcome of Eq.~(\ref{2bodycomm}).

If a third, unoccupied register is added to the memory,  the commutator acting over two-particles states can then be faithfully reproduced.  This leads us to generalizing to $n$ particles in the next section~\ref{sec:nbosons}.

\section{Implementation of \textit{n}-particle creation \& annihilation operators}

\subsection{Commuting operators}
\label{sec:nbosons}
We are now ready to explain the general case for a large memory bank containing an unspecified, variable number of particles. In this section we present the generalization of the preceding codifications in section~\ref{sec:1body} and~\ref{sec:2bosons} to the $n$-register case. 

Consider scalar particles with only their momentum as quantum number, with such momenta codified in $n_{P}$ qubits.
The null ket is taken as a reference and does not represent any valid momentum
\begin{equation}
\ket{0...0} \equiv \underbrace{\ket{0}_{P}...\ket{0}_{P}}_{n_{p}} .
    \label{def:nReg-p0}
\end{equation}
Consider also presence/absence qubits initially in the state $\ket{0}_{P/A}$: 
the \textit{n-register vacuum} is then written as
\begin{equation}
    \ket{\Omega}\equiv\ket{\Omega}_{n}\otimes ... \otimes \ket{\Omega}_{1} = \underbrace{\left(\ket{0}_{P/A}\ket{0...0}\right)_n\otimes ... \otimes\left(\ket{0}_{P/A}\ket{0...0}\right)_1}_{n},
    \label{def:nReg-vacuum}
\end{equation}
Generally, operators acting over $n$-registers consist of projection and control operators $\ketbrac{i}{j} = \ket{i}\bra{j}$ defined in Eq.(\ref{def:control-operators}) over the presence/absence qubits and set/scrap operators $\mathfrak{s}^{\dagger}$, $\mathfrak{s}$ over qubits representing quantum numbers as
 defined in Eq.~(\ref{def:set-operators}) and Eq.~(\ref{def:scrap-operators}). For example, an operator that creates a particle in the first register in an $n$-register quantum computer is
\begin{equation}
a^{(n)\dagger}_{p,1} \equiv \left(\ketbrac{0}{0}\otimes\innerid\right)_n...\left(\ketbrac{0}{0}\otimes\innerid\right)_2\left(\ketbrac{1}{0}\otimes\mathfrak{s}^{\dagger}_{p}\right)_1,
\end{equation}
indeed
\begin{equation}
a^{(n)\dagger}_{p,1}\ket{\Omega}  = \left(\ketbrac{0}{0}\ket{0}_{P/A}\innerid\ket{0...0}\right)_n...\left(\ketbrac{1}{0}\ket{0}_{P/A} \mathfrak{s}^{\dagger}_{p}\ket{0...0}\right)_1 = \left(\ket{0}_{P/A}\ket{0...0}\right)_n...\left(\ket{1}_{P/A}\ket{p}\right)_1\ .
\end{equation}

We next define the generalization of creation and annihilation operators for arbitrary number of particles. Note that in Eq. (\ref{eq:2Reg-creation-1}) the two-register operator
\begin{equation} \label{2partsymmetrizer}
    \mathcal{S}_{2} = \frac{1}{\sqrt{2}}\left(\outeridentity\otimes\outeridentity+\mathcal{P}_{21}\right)
\end{equation}
is a two-particle \textit{symmetrizer}. Thus it is tempting (but inefficient) to generalize the expression in Eq.(\ref{eq:2Reg-creation-3}) adding operators that create particles in each of the registers and higher-order symmetrizers, for example, in the case of three registers, changing $\mathcal{S}_{2}\rightarrow \outeridentity\otimes \mathcal{S}_{2}$ and introducing a three-particles symmetrizer $\mathcal{S}_{3}$: 
\begin{align}
a^{\dagger}_{q}|_3 \equiv & \, \outeridentity^{\otimes 3} \cdot \left(\ketbrac{0}{0}\otimes \innerid\right)_{3}\otimes \left(\ketbrac{0}{0}\otimes \innerid\right)_{2}\otimes \left(\ketbrac{1}{0}\otimes \mathfrak{s}^{\dagger}_{q}\right)_{1} \nonumber \\
+ & \,\outeridentity\otimes S_{2} \cdot  \left(\ketbrac{0}{0}\otimes \innerid\right)_{3}\otimes\left(\ketbrac{1}{0}\otimes \mathfrak{s}^{\dagger}_{q}\right)_{2}\otimes \left(\ketbrac{1}{1}\otimes \innerid\right)_{1} \nonumber \\
+ & \,S_{3} \cdot  \left(\ketbrac{1}{0}\otimes \mathfrak{s}^{\dagger}_{q}\right)_{3}\otimes\left(\ketbrac{1}{1}\otimes \innerid\right)_{2} \otimes \left( \ketbrac{1}{1}\otimes \innerid\right)_{1},
\end{align}
this generalization would be correct if applied to unsymmetrized two-particle states, but in the case of symmetric states it produces extra contributions: 
\begin{align}
    a^{\dagger}_{q}\ket{00}_3\left(\ket{1p_{1}}_2\ket{1p_{2}}_1\right)_{\mathcal{S}} = &\, \mathcal{S}_3 \ket{1q}_3\left(\ket{1p_{1}}_2\ket{1p_{2}}_1\right)_{\mathcal{S}} \nonumber \\
     = & \, \frac{1}{\sqrt{6}}\left(\outeridentity^{\otimes 3}+\mathcal{P}_{21}+\mathcal{P}_{31}+\mathcal{P}_{32}+\mathcal{P}_{231}+\mathcal{P}_{312}\right)\ket{1q}\left(\ket{1p_{1}}\ket{1p_{2}}\right)_{\mathcal{S}}\\
    = & \, 2\left(\ket{1q}\ket{1p_{1}}\ket{1p_{2}}\right)_{\mathcal{S}},
\end{align}
since we are symmetrizing again the already symmetrized part. Additionally, even if designing adequate normalizations to overcome this, numerous redundant operations would be wasted resymmetrizing already symmetric parts of the wavefunction.

Thus, in the case of three or more registers, symmetrization or antisymmetrization of states will be implemented in a step-wise fashion, assuming that earlier creation operations already symmetrized a state with $n-1$ particles, and only the newly added one requires symmetrization with the others. Imposing symmetric states, the three-particle symmetrizer should be changed to a ``step-symmetrizer''
\begin{equation}
    \mathcal{S}_{3}\rightarrow \mathcal{S}_{3\leftarrow 2} \equiv \frac{1}{\sqrt{3}}\left(\outeridentity^{\otimes 3}+\mathcal{P}_{32}\otimes \outeridentity_ 1 +\mathcal{P}_{31}\otimes \outeridentity_2 \right)
\end{equation}
for the general case, and omitting the identities for conciseness, we have $\mathcal{S}_{n\leftarrow n-1}$, which promotes an $(n-1)$-particle symmetric state to an $n$-particle symmetric one:
\begin{equation}
    \mathcal{S}_{n\leftarrow n-1} \equiv \frac{1}{\sqrt{n}}\left(\outeridentity^{\otimes n}+\mathcal{P}_{n(n-1))}+...+\mathcal{P}_{n2}+\mathcal{P}_{n1}\right).
\label{def:step-symmetrizer}
\end{equation}
Note that with this choice, the symmetrizer is not idempotent, but instead $S^2_{n\leftarrow n-1} =
\sqrt{n}\ S_{n\leftarrow n-1}$.

\paragraph{Implementation of creation operators.}
The auxiliary operator that attempts to create a particle on the $j$th register in an $n$-register memory segment now reads
\begin{align}
a^{(n)\dagger }_{q,j} & = \outeridentity^{\otimes n-j} \otimes \mathcal{S}_{j\leftarrow(j-1)} \cdot \left(\ketbrac{0}{0}\otimes \innerid\right)_{n}\otimes ...\otimes \underbrace{\left(\ketbrac{1}{0}\otimes \mathfrak{s}^{\dagger}_{q}\right)_{j} } \otimes \left(\ketbrac{1}{1}\otimes \innerid\right)_{j-1}\otimes ... \otimes\left(\ketbrac{1}{1}\otimes \innerid\right)_{1}\nonumber \\
& = \mathcal{S}_{j\leftarrow(j-1)} \cdot \projector{n-j}{0}\otimes \underbrace{\left(\ketbrac{1}{0}\otimes \mathfrak{s}^{\dagger}_{q}\right)_{j} } \otimes \,\projector{j-1}{j-1},
\label{def:nReg-bosoncreator-j}
\end{align}
where the identity multiplying the $j$-register step symmetrizer is omitted to alleviate notation and  $\projector{n}{j}$ is the projector onto states with exactly $j$ particles on an $n$-register quantum computer: 
\begin{equation}
\projector{n}{j}= \projector{n-j}{0}\otimes \projector{j}{j}=
\left[ \left(\ketbrac{0}{0}\otimes \innerid\right)_{n}\otimes...\otimes\left(\ketbrac{0}{0}\otimes \innerid\right)_{j+1} \right]\otimes \left[ \left(\ketbrac{1}{1}\otimes \innerid\right)_{j}\otimes ... \otimes\left(\ketbrac{1}{1}\otimes \innerid\right)_{1}\right]\ ,
\label{def:projectors}
\end{equation}
which fullfil $\projector{n}{j}\projector{n}{i}=\delta_{ij}$ (because of the unequal number of particle presence/absence control operator $\ketbrac{0}{0}$ and $\ketbrac{1}{1}$ for different values of $i$ and $j$).
The action of this operator over a register that already contains $i-1$ particles is automatically restricted to filling $j=i$ (last particle on the stack) by its very definition, so that  
\begin{align}
&a^{(n)\dagger }_{q,j}\ket{\Omega}^{\otimes n-i+1}\otimes\left(\ket{1p_{i-1}}_{i-1}...\ket{1p_1}_1\right)_{\mathcal{S}}  \nonumber \\ 
& = a^{(n)\dagger }_{q,j}\cdot \projector{n-i+1}{0}\otimes\projector{i-1}{i-1}\cdot\ket{\Omega}^{\otimes n-i+1}\otimes\left(\ket{1p_{i-1}}_{i-1}...\ket{1p_1}_1\right)_{\mathcal{S}}   \nonumber \\  
& =  \mathcal{S}_{j\leftarrow(j-1)} \cdot \projector{n-j}{0}\otimes \underbrace{\left(\ketbrac{1}{0}\otimes \mathfrak{s}^{\dagger}_{q}\right)_{j} } \otimes \,\projector{j-1}{j-1}\cdot \projector{n-i+1}{0}\otimes\projector{i-1}{i-1}\cdot\ket{\Omega}^{\otimes n-i+1}\otimes\left(\ket{1p_{i-1}}_{i-1}...\ket{1p_1}_1\right)_{\mathcal{S}} \nonumber \\
& = \delta_{i,j}\ket{\Omega}^{\otimes n-i+1}\otimes \left[\mathcal{S}_{i\leftarrow i-1}\ket{1q}\left(\ket{1p_{i-1}}_{i-1}...\ket{1p_1}_1\right)_{\mathcal{S}}\right]\nonumber \\
& =\delta_{i,j}\ket{\Omega}_n...\left(\ket{1q}_{i}\ket{1p_{i-1}}_{i-1}...\ket{1p_1}_1\right)_{\mathcal{S}},
\end{align}
where in the third line we used the orthogonality of projectors to produce a $\delta_{i,j}$. 

The total creation operator that implements the field-theory requirements is the addition of auxiliary creation operators over each register: 
\begin{equation}
a^{(n)\dagger }_{q} = \sum^{n}_{i = 1}a^{(n)\dagger }_{q,i}  =  \sum^{n}_{i = 1}  \mathcal{S}_{i\leftarrow(i-1)} \cdot \projector{n-i}{0}\otimes \left(\ketbrac{1}{0}\otimes \mathfrak{s}^{\dagger}_{q}\right)_{i} \otimes \,\projector{i-1}{i-1}
    \label{def:nReg-bosoncreation}
\end{equation}
note that $S_{2\leftarrow 1} = S_{2}$, so for the case $n=2$ we recover the two-particle case studied earlier in section~\ref{sec:2bosons}. The action of  the full $a_q^{(n)\dagger}$ over an \textit{i-particle} state $\ket{\Omega}_n...\left(\ket{1p_{i}}_i...\ket{1p_1}_1\right)_{\mathcal{S}}$ is then
\begin{align}
& a^{(n)\dagger }_{q}  \ket{\Omega}_n...\left(\ket{1p_{i}}_i...\ket{1p_1}_1\right)_{\mathcal{S}}  = \sum^{n}_{j = 1}a^{\dagger (n)}_{q,j}\ket{\Omega}_n...\left(\ket{1p_{i}}_i...\ket{1p_1}_1\right)_{\mathcal{S}} \nonumber \\
& = \sum^{n}_{j = 1} \delta_{j,i+1}\ket{\Omega}_n...\left(\ket{q}_{i+1}\ket{1p_{i}}_{i}...\ket{1p_1}_1\right)_{\mathcal{S}} \nonumber \\
& = \ket{\Omega}_n...\left(\ket{q}_{i+1}\ket{1p_{i}}_{i}...\ket{1p_1}_1\right)_{\mathcal{S}}
\end{align}

At last, its action over an $n$-particle state the result is by the definition zero:
\begin{equation}
    a^{(n)\dagger }_{p} \left(\ket{1p_{n}}\ket{1p_{n-1}}...\ket{1p_{1}}\right)_{\mathcal{S}} = 0,
\end{equation}
which is no more than a boundary term due to the quantum computer's finite memory.

\paragraph{Implementation of annihilation operators.} Annihilation operators are found by complex conjugation of creation operators. For the auxiliary ones acting over the $j$-register on an $n$-register quantum computer we (quite directly) find
\begin{align}
a^{(n)}_{p,j} & =  \left(\ketbrac{0}{0}\otimes \innerid\right)_{n}\otimes ...\otimes \underbrace{\left(\ketbrac{0}{1}\otimes \mathfrak{s}_{p}\right)_{j}}\otimes \left(\ketbrac{1}{1}\otimes \innerid\right)_{j-1}\otimes ... \otimes\left(\ketbrac{1}{1}\otimes \innerid\right)_{1} \cdot \mathcal{S}_{j\leftarrow(j-1)} \nonumber \\
& = \projector{n-j}{0}\otimes \underbrace{\left(\ketbrac{0}{1}\otimes \scrap{p}\right)_{j} } \otimes \,\projector{j-1}{j-1} \cdot\mathcal{S}_{j\leftarrow(j-1)}  ,
\label{def:nReg-bosonannihilation-j}
\end{align}
its action over an $i$-particle state, with $j > i$, reads
\begin{align} 
a^{(n)}_{p,j} & \ket{\Omega}^{\otimes n-i}\left(\ket{1p_i}_i...\ket{1p_1}_1\right)_S =\nonumber \\
& \projector{n-j}{0}\otimes\left(\ketbrac{0}{1}\otimes \mathfrak{s}_{p}\right)_{j}\otimes\projector{j-1}{j-1} \ket{\Omega}^{\otimes n-j} \left[ \stepsym{j}\ket{\Omega}^{\otimes j-i}\left(\ket{1p_i}_i...\ket{1p_1}_1\right)_S\right] = 0;
\label{bosonannihil}
\end{align}
the step-symmetrizer in the last line mixes the vacuum registers with the occupied ones without changing their quantum numbers, which means that $j-i$ of the $j$ registers are always empty registers. This being so, the term between square brackets, which acts upon the first $j$ registers, will always produce a 0 because all its presence/absence operators are of the form $\ketbrac{i}{1}$, which always find at least one empty register and thus yield null.
When the index $j\leq i$ the annihilation reads,
\begin{align}
a^{(n)}_{p,j} & \ket{\Omega}^{\otimes n-i}\left(\ket{1p_i}_i...\ket{1p_1}_1\right)_S  \nonumber \\
& =  \projector{n-j}{0}\otimes \left(\ketbrac{0}{1}\otimes \scrap{p}\right)_j\otimes\projector{j-1}{j-1}\cdot \stepsym{j}\ket{\Omega}^{\otimes n-i}\left(\ket{1p_i}_i...\ket{1p_1}_1\right)_S\nonumber \\
&  = \sqrt{j} \, \projector{n-j}{0}\otimes \left(\ketbrac{0}{1}\otimes \scrap{p}\right)_j\otimes\projector{j-1}{j-1}\ket{\Omega}^{\otimes n-i}\left(\ket{1p_i}_i...\ket{1p_1}_1\right)_S\nonumber \\
& = \delta_{ij}\, \sqrt{i}\, \frac{\sqrt{(i-1)!}}{\sqrt{i!}}\sum^{i}_{l=1} \ket{\Omega}^{\otimes n-i}\left(\ketbrac{0}{1}\ket{1}\otimes \scrap{p}\ket{p_l}\right)_i\left(\ket{1p_i}_{i-1}...\ket{1p_{l+1}}_{l}\ket{1p_{l-1}}_{l-1}...\ket{1p_1}_1\right)_S\nonumber \\
& = \delta_{ij}\sum^{i}_{l=1} \delta_{p p_l}\ket{\Omega}^{\otimes n-i+1}\left(\ket{1p_i}_{i-1}...\ket{1p_{l+1}}_{l}\ket{1p_{l-1}}_{l-1}...\ket{1p_1}_1\right)_S
\label{def:nReg-bosonannihilation}
\end{align}

Summing over the particle number $j$ we arrive at the annihilation operator
\begin{equation}
a^{(n)}_{q} = \sum^{n}_{j=1}a^{(n)}_{q,j} = \sum^{n}_{j=1} \projector{n-j}{0}\otimes \left(\ketbrac{0}{1}\otimes \scrap{p}\right)_{j} \otimes \,\projector{j-1}{j-1} \cdot\mathcal{S}_{j\leftarrow(j-1)}.
\end{equation}
The action of the operator over an $i$-particle symmetric state is
\begin{align}
a^{(n)}_{q} & \ket{\Omega}_n...\ket{\Omega}_{i+1}\left(\ket{1p_i}_i...\ket{1p_1}_1\right)_S  =  \sum^{n}_{j = 1}a^{(n)}_{q,j}\ket{\Omega}_n...\ket{\Omega}_{i+1}\left(\ket{1p_i}_i...\ket{1p_1}_1\right)_S\nonumber \\
& = \sum^{i}_{l=1}\delta_{p p_l}\ket{\Omega}^{\otimes n-i+1}\left(\ket{1p_i}_{i-1}...\ket{1p_{l+1}}_{l}\ket{1p_{l-1}}_{l-1}...\ket{1p_1}_1\right)_S
\end{align}
which is exactly what we expect from an annihilation operator acting over an $i$-particle symmetric state.
Note that this leaves behind the initial memory piece, with $i$ registers, occupied by only $i-1$ particles (properly symmetrized) and precisely the last $i$th register vacant.

\paragraph{Number operator.} 
We next check the adequacy of the resulting number operator\footnote{$\sum_p\set{p}\scrap{p} = \mathfrak{i}-\ket{0...0}\bra{0...0}$, but $\ket{0...0}\bra{0...0}$ can be ignored because occupied registers with $\ket{0...0}$ never occur} 
\begin{align}
    \hat{N}^{(n)} & := \sum_{p} a^{(n)\dagger}_{p}a^{(n)}_{p} = \sum_{p}\sum_{j,j'} a^{(n)\dagger}_{p,j}a^{(n)}_{p,j'}\nonumber  \\
    &  = \sum_{p,j,j'} \mathcal{S}_{j\leftarrow j-1} \cdot \projector{n-j}{0}\otimes \left(\ketbrac{1}{0}\otimes \set{p}\right)_{j} \otimes \,\projector{j-1}{j-1} \cdot \projector{n-j'}{0}\otimes \left(\ketbrac{0}{1}\otimes \scrap{p}\right)_{j'} \otimes \,\projector{j'-1}{j'-1} \cdot \mathcal{S}_{j'\leftarrow j'-1} \nonumber  \\
    & = \sum_{j} \mathcal{S}_{j\leftarrow j-1} \cdot \projector{n-j}{0}\otimes \left(\ketbrac{1}{1}\otimes \sum_{p} \set{p}\scrap{p}\right)_{j} \otimes \,\projector{j-1}{j-1} \cdot \mathcal{S}_{j\leftarrow j-1} \nonumber \\
    & = \sum_{j}\mathcal{S}_{j\leftarrow j-1} \cdot \projector{n-j}{0}\otimes \left(\ketbrac{1}{1}\otimes \innerid \right)_{j} \otimes \,\projector{j-1}{j-1} \cdot \mathcal{S}_{j\leftarrow j-1} \nonumber \\
    & = \sum_{j} \mathcal{S}_{j\leftarrow j-1} \cdot \projector{n}{j} \cdot \mathcal{S}_{j\leftarrow j-1},
    \label{Number-operator}
\end{align}
where the sum over $j'$ is simplified using a $\delta_{j.j'}$ emerging from multiplication of projectors in the second line. The operator between symmetrizers is just a projector over $j$-occupied registers, so a permutation between terms acting over registers $j$ and $k\in \left\{1,..,j-1\right\}$ gives the same operator. Thus $\projector{n-j}{j}$ and $ \outeridentity^{\otimes n-j} \otimes \mathcal{S}_{j\leftarrow j-1}$ commute, and 
\begin{align}
    \hat{N}^{(n)} & = \sum_{j} \mathcal{S}_{j\leftarrow j-1} \cdot \projector{n}{j} \cdot \mathcal{S}_{j\leftarrow j-1} = \sum_{j}\mathcal{S}^2_{j\leftarrow j-1} \cdot \projector{n}{j}   = \sum_{j} \sqrt{j}\ \mathcal{S}_{j\leftarrow j-1} \cdot \projector{n}{j},
\end{align}
when acting over symmetric states with $j'$ particles. We see that $N^{(n)}$ is diagonal with the expected eigenvalues $j'$,
\begin{align}
\hat{N}^{(n)} \ket{\Omega}_{n}...\left(\ket{1p_{j'}}_{j'}...\ket{1p_1}_{1}\right)_{S} & =   \sum_{j'}\sqrt{j'}\ket{\Omega}_{n}...\mathcal{S}_{j'\leftarrow j'-1}\left(\ket{1p_{j'}}_{j'}...\ket{1p_1}_{1}\right)_S \nonumber \\
& = j'\ \ket{\Omega}_{n}...\left(\ket{1p_{j'}}_{j'}...\ket{1p_1}_{1}\right)_{S}.
\end{align}

Having established the validity of the number operator, we can next check that $a^{(n)\dagger}$ and $a^{(n)}$ for $n$ particles form a canonically conjugate pair.

\paragraph{Commutation relations.} 
We now adopt Eq.(\ref{def:nReg-bosoncreation}) and Eq.(\ref{def:nReg-bosonannihilation-j}) as the definition of the creation and annihilation operators for $n$ registers.

These operators satisfy the following commutation relations in the Hilbert subspace of symmetric states.
\begin{equation}
    \left[a^{(n)}_{q_{1}},a^{(n)\dagger}_{q_{2}}\right]  = \delta_{q_{1},q_{2}}\left(\ketbrac{0}{0}\otimes\innerid\right)_{n}\otimes \sum^{n-1}_{j=0}\projector{n-1}{j}\  -\ \,S_{n\leftarrow n-1}\cdot \left(\ketbrac{1}{1}\otimes \mathfrak{s}^{\dagger}_{q_{1}}\mathfrak{s}_{q_{2}}\right)_n \otimes\projector{n-1}{n-1}\cdot S_{n\leftarrow n-1} \ ,
    \label{eq:nReg-commutation}
\end{equation}
the operator $\sum^{n-1}_{j=0}\projector{n-1}{j}$ projects over properly filled memories, in which registers are filled from right to left. Thus $\mathbb{I}^{(n-1)}-\sum^{n-1}_{j=0}\projector{n-1}{j}$, with $\mathbb{I}^{(n-1)}$ the identity over $(n-1)$ registers, is zero over properly filled memories, and we can write
\begin{equation}
\left[a^{(n)}_{q_{1}},a^{(n)\dagger}_{q_{2}}\right]_{\text {\rm properly\  filled}} = \delta_{q_{1},q_{2}}\left(\ketbrac{0}{0}\otimes\innerid\right)_{n}\otimes \mathbb{I}^{(n-1)}\  -\ \,S_{n\leftarrow n-1}\cdot \left(\ketbrac{1}{1}\otimes \mathfrak{s}^{\dagger}_{q_{1}}\mathfrak{s}_{q_{2}}\right)_n \otimes\projector{n-1}{n-1}\cdot S_{n\leftarrow n-1} \ ,
    \label{eq:nReg-commutation-good}
\end{equation}

The first term on the right-hand side of Eq.~(\ref{eq:nReg-commutation}) is, in this notation, what one would expect of canonical commutation relations, the identity in a Fock space of up to $n-1$ particles (the $\ketbrac{0}{0}$ control guarantees that the $n$th register is empty).

The second deviates from second quantization and emerges due to the quantum computer's finite memory: it vanishes if less than $n$ particles occupy the memory. As a corollary thereof, the canonical commutation relations are exactly satisfied for symmetric states of up to $n-1$ particles (in an $n$-register quantum computer) when only the first term contributes.

The demonstration of Eq.~(\ref{eq:nReg-commutation}), that requires a few bookkeeping steps, is presented in appendix~\ref{proof:commutation}.

\subsection{Anticommuting operators} \label{sec:anticommuting}

With the experience of the boson operators in the proceeding section~\ref{sec:nbosons}, it is now easy to implement anticommuting ones and direct (if tedious) to derive their properties. The most notable difference is that the recursive step-symmetrizers of Eq.~(\ref{def:step-symmetrizer}) should be changed to step-antisymetrizers, which  (unsurprisingly) read
\begin{equation}
    \stepasym{n} = \frac{1}{\sqrt{n}}\left(\outeridentity^{\otimes n}-\mathcal{P}_{n(n-1)}-\mathcal{P}_{n(n-2)}-... -\mathcal{P}_{n2}-\mathcal{P}_{n1}\right)
    \label{def:step-asym}
\end{equation}
that antisymmetrize any dynamically added particle with all the previously filled ones.

Totally antisymmetric states can be written as
\begin{equation}
\left(\ket{1p_i}_i...\ket{1p_1}_1\right)_{\mathcal{A}} =\sum_{\sigma}(-1)^{\eta_\sigma}\sigma\left(\ket{1p_i}_i...\ket{1p_1}_1\right), 
\end{equation}
where $\sigma$ runs over the permutations of $i$-elements and $\eta_{\sigma}$ is the parity of the permutation $\sigma$: $(-1)^{\eta_\sigma}=1$ if the permutations is even, and $(-1)$ if  odd. The action of the step-antisymmetrizer over this state is
\begin{align}
\stepasym{i} &\left(\ket{1p_i}_i...\ket{1p_1}_1\right)_{\mathcal{A}}  =  \frac{1}{\sqrt{i}}\left(\outeridentity^{\otimes n}-\mathcal{P}_{i(i-1)}-...-\mathcal{P}_{i1}\right) \sum_{\sigma}(-1)^{\eta_\sigma}\sigma\left(\ket{1p_i}_i...\ket{1p_1}_1\right)  \nonumber \\
& =  \frac{1}{\sqrt{i}}\left[\sum_{\sigma}(-1)^{\eta_\sigma}\sigma -\sum_{\sigma}(-1)^{\eta_\sigma}\mathcal{P}_{i(i-1)}\sigma -...-\sum_{\sigma}(-1)^{\eta_\sigma}\mathcal{P}_{i1}\sigma \right]\ket{1p_i}_i...\ket{1p_1}_1\ .
\label{permutations}
\end{align}
By the rearrangement lemma, no two  $\mathcal{P}_{ij}\sigma$ permutations are equal, so the sums $\sum_{\sigma}\mathcal{P}_{ij}\sigma$ cover all possible permutations. Also, since $\mathcal{P}_{ij}$ is a simple permutation, the parity of $\mathcal{P}_{ij}\sigma$ is the negative of the parity of $\sigma$. Using the same index, $\sigma$, for all the new permutation we have that all the terms in Eq.~(\ref{permutations}) are identical, so
\begin{align}
\stepasym{i} \left(\ket{1p_i}_i...\ket{1p_1}_1\right)_{\mathcal{A}} & = \frac{1}{\sqrt{i}}\left[\sum_{\sigma}(-1)^{\eta_\sigma}\sigma +\sum_{\sigma}(-1)^{\eta_\sigma}\sigma +...+\sum_{\sigma}(-1)^{\eta_\sigma}\sigma \right]\ket{1p_i}_i...\ket{1p_1}_1 \nonumber \\
& = \sqrt{i} \left(\ket{1p_i}_i...\ket{1p_1}_1\right)_{A}.
\label{antisymmetricstates}
\end{align}

\paragraph{Implementation of the Creation Operators.} 
We start again by an auxiliary operator that creates a particle precisely at the $j$th register in an $n$-register quantum computer, which now reads
\begin{align}
b^{(n)\dagger}_{q,j} & = \outeridentity^{\otimes n-j} \otimes \mathcal{A}_{j\leftarrow j-1}\cdot  \left(\ketbrac{0}{0}\otimes \innerid\right)_{n}\otimes ...\otimes \underbrace{\left(\ketbrac{1}{0}\otimes \set{p}\right)_{j}}\otimes \left(\ketbrac{1}{1}\otimes \innerid\right)_{j-1}\otimes ... \otimes\left(\ketbrac{1}{1}\otimes \innerid\right)_{1}  \nonumber \\ 
& = \stepasym{j}\cdot \projector{n-j}{0}\otimes \underbrace{\left(\ketbrac{1}{0}\otimes \mathfrak{s}^{\dagger}_{q}\right)_{j}}
\otimes \,\projector{j-1}{j-1}\ .
\label{def:nReg-fermioncreation-j}
\end{align}
Its action over a register that contains $i$-particles is similar to the commuting case, but symmetric states should be changed to antysimmetric ones:
\begin{align}
& b^{(n)\dagger }_{q,j}\ket{\Omega}^{\otimes n-i+1}\left(\ket{1p_{i-1}}_{i-1}...\ket{1p_1}_1\right)_{\mathcal{A}} = \delta_{i,j}\ket{\Omega}^{\otimes n-i+1}\otimes \left[\mathcal{A}_{i\leftarrow i-1}\ket{1q}\left(\ket{1p_{i-1}}_{i-1}...\ket{1p_1}_1\right)_{\mathcal{A}}\right]\nonumber \\
& =\delta_{i,j}\ket{\Omega}^{\otimes n-i}\left(\ket{1q}_{i}\ket{1p_{i-1}}_{i-1}...\ket{1p_1}_1\right)_{\mathcal{A}},
\end{align}
where the Kronecker delta appears due to the same reasons it was generated among commuting operators. The field-theoretic creation operator is implemented by the addition of the creation operators over each register 
\begin{equation}
b^{(n)\dagger }_{q} = \sum^{n}_{j = 1}b^{(n)\dagger }_{q,j} =  \sum^{n}_{j = 1}\stepasym{j}\cdot \projector{n-j}{0}\otimes \underbrace{\left(\ketbrac{1}{0}\otimes \mathfrak{s}^{\dagger}_{q}\right)_{j}}
\otimes \,\projector{j-1}{j-1}
    \label{def:nReg-fermioncreation}
\end{equation}
its action over an $i$-particle antisymmetric state $\ket{\Omega}_n...\left(\ket{1p_{i}}_i...\ket{1p_1}_1\right)_{\mathcal{A}}$ is
\begin{align}
b^{(n)\dagger }_{q}&\ket{\Omega}^{\otimes n-i+1}\left(\ket{1p_{i}}_i...\ket{1p_1}_1\right)_{\mathcal{A}}  = \sum^{n}_{j = 1}b^{\dagger (n)}_{q,j}\ket{\Omega}^{\otimes n-i+1}\left(\ket{1p_{i}}_i...\ket{1p_1}_1\right)_{\mathcal{A}}   \nonumber \\
&=\sum^{n}_{j = 1} \delta_{j,i+1}\ket{\Omega}^{\otimes n-i}\left(\ket{q}_{i+1}\ket{1p_{i}}_{i}...\ket{1p_1}_1\right)_{\mathcal{A}} = \ket{\Omega}^{\otimes n-i}\left(\ket{q}_{i+1}\ket{1p_{i}}_{i}...\ket{1p_1}_1\right)_{\mathcal{A}}\ .
\label{eq:nReg-asym-creation-2}
\end{align}
Finally, we need to adopt a boundary condition upon reaching a  full memory, and we again choose
\begin{equation}
b^{(n)\dagger }_{q}\left(\ket{1p_{n}}_n...\ket{1p_1}_1\right)_{\mathcal{A}} = 0.
\end{equation}

\paragraph{Implementation of the Annihilation Operators.} 
By Hermitian conjugation of the creation operators in Eq.~(\ref{def:nReg-fermioncreation-j}), we directly obtain
\begin{align}
b^{(n)}_{p,j} &  = \left(\ketbrac{0}{0}\otimes \innerid\right)_{n}\otimes ...\otimes \underbrace{\left(\ketbrac{0}{1}\otimes \mathfrak{s}_{p}\right)_{j}}\otimes \left(\ketbrac{1}{1}\otimes \innerid\right)_{j-1}\otimes ... \otimes\left(\ketbrac{1}{1}\otimes \innerid\right)_{1} \cdot  \mathcal{A}_{j\leftarrow(j-1)} \nonumber \\
& = \projector{n-j}{0}\otimes \underbrace{\left(\ketbrac{0}{1}\otimes \scrap{p}\right)_{j}}
\otimes \,\projector{j-1}{j-1} \cdot \mathcal{A}_{j\leftarrow(j-1)},
\label{def:nreg-fermionannihilation-j}
\end{align}
whose action over an $i$-particle state with $j>i$ is null by the same reasons argued in the commuting case, c.f. Eq.~(\ref{bosonannihil}). The case with $j\leq$ is similar, but some care is due for minus signs coming from antisymmetry
\begin{align}
b^{(n)}_{p,j} & \ket{\Omega}^{\otimes n-i}\left(\ket{1p_i}_i...\ket{1p_1}_1\right)_{\mathcal{A}}  \nonumber \\
&  = \sqrt{j} \, \projector{n-j}{0}\otimes \left(\ketbrac{0}{1}\otimes \scrap{p}\right)_j\otimes\projector{j-1}{j-1}\ket{\Omega}^{\otimes n-i}\left(\ket{1p_i}_i...\ket{1p_1}_1\right)_{\mathcal{A}} \nonumber \\
&  = \sqrt{j} \, \projector{n-j}{0}\otimes \left(\ketbrac{0}{1}\otimes \scrap{p}\right)_j\otimes\projector{j-1}{j-1}\ket{\Omega}^{\otimes n-i}\frac{\sqrt{(i-1)!}}{\sqrt{i!}}\left[\ket{1p_i}_i\left(\ket{1p_{i-1}}_{i-1}...\ket{1p_1}_1\right)_{\mathcal{A}}\right.\nonumber \\
&  \left.-\sum^{i-1}_{l = 1}\ket{1p_l}_i\left(\ket{1p_{i-1}}_{i-1}...\ket{1p_i}_l...\ket{1p_1}_1\right)_{\mathcal{A}}\right]\nonumber \\
& = \delta_{ij}\, \sum^{i}_{l=1}\left(-1\right)^{i-l} \ket{\Omega}^{\otimes n-i}\left(\ketbrac{0}{1}\ket{1}\otimes \scrap{p}\ket{p_l}\right)_i\left(\ket{1p_i}_{i-1}...\ket{1p_{l+1}}_{l}\ket{1p_{l-1}}_{l-1}...\ket{1p_1}_1\right)_{\mathcal{A}}\nonumber \\
& = \delta_{ij}\sum^{i}_{l=1} \left(-1\right)^{i-l}\delta_{p p_l}\ket{\Omega}^{\otimes n-i+1}\left(\ket{1p_i}_{i-1}...\ket{1p_{l+1}}_{l}\ket{1p_{l-1}}_{l-1}...\ket{1p_1}_1\right)_{\mathcal{A}}
\label{eq:nreg-acom-annihilation}
\end{align}
As in the commuting case, the action of $\left(\ketbrac{0}{1}\otimes \mathfrak{s}_{p}\right)_{i}$ ``extracts'' all the different momenta from $\left(\ket{1p_i}_i...\ket{1p_1}_1\right)_{\mathcal{A}}$. To complete the discussion we can give the annihilation operator in register-operator notation as 
\begin{equation}
b^{(n)}_{q} = \sum^{n}_{j=1}b^{(n)}_{q,j} = \sum^{n}_{j=1} \projector{n-j}{0}\otimes \left(\ketbrac{0}{1}\otimes \scrap{p}\right)_{j}
\otimes \,\projector{j-1}{j-1} \cdot \mathcal{A}_{j\leftarrow j-1} 
\label{def:nReg-fermionannihilation}
\end{equation}
which is the same expression as for a boson operator upon change of step-symmetrizers for step-antisymmetrizers. We note again that $a^{(n)}_{q}$ is defined to act over symmetric states, whereas $b^{(n)}_{q}$ is the appropriate one on antisymmetric states. The action of the annihilation operator over an $i$-particle antisymmetric state is
\begin{align}
b^{(n)}_{q} & \ket{\Omega}_n...\ket{\Omega}_{i+1}\left(\ket{1p_i}_i...\ket{1p_1}_1\right)_{\mathcal{A}} = \sum^{n}_{j=1}b^{(n)}_{q,j}\ket{\Omega}_n...\ket{\Omega}_{i+1}\left(\ket{1p_i}_i...\ket{1p_1}_1\right)_{\mathcal{A}} \nonumber \\
& =\sum_j\delta_{ij}\sum^{i}_{l=1} \left(-1\right)^{i-l}\delta_{p p_l}\ket{\Omega}^{\otimes n-i+1}\left(\ket{1p_i}_{i-1}...\ket{1p_{l+1}}_{l}\ket{1p_{l-1}}_{l-1}...\ket{1p_1}_1\right)_{\mathcal{A}}\nonumber \\
& =\delta_{p p_l} \sum^{i}_{l=1} \left(-1\right)^{i-l}\ket{\Omega}^{\otimes n-i+1}\left(\ket{1p_i}_{i-1}...\ket{1p_{l+1}}_{l}\ket{1p_{l-1}}_{l-1}...\ket{1p_1}_1\right)_{\mathcal{A}}\ .
\end{align}
\paragraph{Anticommutation relations.} Creation and annihilation operators then fulfill the anticommutation relations
\begin{equation}
\left\{b^{(n)}_{q_{1}},b^{(n)\dagger}_{q_{2}}\right\}\  = \ \ \delta_{q_{1},q_{2}}\left(\ketbrac{0}{0}\otimes\innerid\right)_{n}\otimes \sum^{n-1}_{j=0}\projector{n-1}{j}\ \  +\  \ \mathcal{A}_{n\leftarrow n-1}\cdot \left(\ketbrac{1}{1}\otimes \mathfrak{s}^{\dagger}_{q_{1}}\mathfrak{s}_{q_{2}}\right)_n\otimes\projector{n-1}{n-1}\cdot \mathcal{A}_{n\leftarrow n-1} ,
    \label{eq:nReg-anticommutation}
\end{equation}    
where $\projector{n}{j}$ is the projector over states with $j$-particles on an $n$-register section of memory and $\mathbb{I}^{(n-1)}$ is the identity over $(n-1)$-registers.
The demonstration is somewhat cumbersome, so as in the case of bosons it is relegated to appendix~\ref{proof:commutation}.

Again, we see that the first term reproduces the canonical second quantization relation, and the second one is a boundary term appearing only when the memory is full.

\section{Exponentiation of fixed particle-number operators} \label{sec:evolution}
In this section we are going to exemplify the encoding developed with simple few-body Hamiltonian evolution.
The dynamics is implemented in terms of unitary operators, which are Lie exponentials of Hermitian operators written in terms of creation and annihilation operators in momentum space. The two basic cases that we examine are the free-evolution and the momentum-exchange kernel
without particle--number variation.

We shall concentrate on fermion operators $b$, $b^\dagger$ and only on occasion quote analogous results for bosons. 

\subsection{Free-evolution operator}
\label{subsec:free}
The free-evolution operator for fermionic particles on a $N$-register quantum computer is
\begin{eqnarray}
        \mathcal{U}^{f}_{11}\left(\Delta t\right) &=& \exp\left(-i\Delta t \sum_{q}E_q b^{(n)\dagger}_q b^{(n)}_q\right) \nonumber \\ &=& \exp\left(-i\Delta t \sum_{q}E_{q}\sum_{j'j}b^{(n)\dagger}_{q,j'} b^{(n)}_{q,j}\right),
\label{def:free-evolution-fermions}
\end{eqnarray}
while the corresponding operator for bosons, $\mathcal{U}^b_{11}\left(\Delta t\right)$, is obtained by replacing fermion for boson operators. 
(The $11$ subindex indicates that both ket and brac contain exactly one particle.)
To implement the exponential we should first write it in terms of set/scrap operators, by Eq.~(\ref{def:nReg-fermionannihilation})
\begin{align}
b^{(n)\dagger}_{q,j'}b^{(n)}_{q,j}  
= & \, \mathcal{A}_{j'\leftarrow j'-1}\cdot \projector{n-j'}{0} \otimes\left(\ketbrac{1}{0}\otimes \mathfrak{s}^{\dagger}_{q}\right)_{j'}\otimes \projector{j'-1}{j'-1}\cdot \projector{n-j}{0}\otimes\left(\ketbrac{0}{1}\otimes \mathfrak{s}_{q}\right)_j\otimes \projector{j-1}{j-1}\cdot \mathcal{A}_{j\leftarrow j-1} \nonumber \\
= & \, \delta_{j'j}\mathcal{A}_{j\leftarrow j-1}\cdot \projector{n-j}{0} \otimes\left(\ketbrac{0}{1}\otimes \mathfrak{s}^{\dagger}_{q}\mathfrak{s}_{q}\right)_j\otimes \projector{j-1}{j-1}\cdot \mathcal{A}_{j\leftarrow j-1}
\end{align}
where the tensor product of identities multiplying the step-antisymmetrizers is omitted. For boson operators
\begin{equation}
a^{(n)\dagger}_{q,j'}a^{(n)}_{q,j} = \delta_{j',j}\mathcal{S}_{j\leftarrow j-1} \cdot \projector{n-j}{0} \otimes \left(\ketbrac{1}{1}\otimes \mathfrak{s}^{\dagger}_{q}\mathfrak{s}_{q}\right)_j\otimes\projector{j-1}{j-1}\cdot \mathcal{S}_{j\leftarrow j-1}.
\end{equation}
The operator preserves the number of particles (it has neither $\ketbrac{0}{1}$ nor $\ketbrac{1}{0}$ terms), and  the set-scrap pair selects configurations with a momentum $q$ in the $j$th register and discards the rest.  The  subsequent antisymmetrizer (symmetrizer) to its left then ensures a superposition over all registers from $1$ to $j$ so that each may have that momentum $q$. The same effect can be achieved  by applying the operator $\left(\ketbrac{1}{1}\otimes \mathfrak{s}^{\dagger}_{q}\mathfrak{s}_{q}\right)$ to each register, assuming the memory is already antisymmetrized (symmetrized):
\begin{equation}
 \mathcal{A}_{j\leftarrow j-1}\cdot \projector{n-j}{0} \otimes \left(\ketbrac{1}{1}\otimes \mathfrak{s}^{\dagger}_{q}\mathfrak{s}_{q}\right)_j\otimes\projector{j-1}{j-1}\cdot \mathcal{A}_{j\leftarrow j-1}=
\sum_{k=0}^{j-1}\projector{n-j+k}{k}\otimes \left(\ketbrac{1}{1}\otimes \mathfrak{s}^{\dagger}_{q}\mathfrak{s}_{q}\right)_{j-k}\otimes \projector{j-1-k}{j-1-k}\cdot \frac{\mathcal{A}_{j\leftarrow j-1}}{\sqrt{j}} \ ,
\label{eq:antisymmetric-action}
\end{equation}
and correspondingly
\begin{equation}
 \mathcal{S}_{j\leftarrow j-1} \cdot \projector{n-j}{0} \otimes \left(\ketbrac{1}{1}\otimes \mathfrak{s}^{\dagger}_{q}\mathfrak{s}_{q}\right)_j\otimes\projector{j-1}{j-1}\cdot \mathcal{S}_{j\leftarrow j-1}=
\sum_{k=0}^{j-1}\projector{n-j+k}{k}\otimes \left(\ketbrac{1}{1}\otimes \mathfrak{s}^{\dagger}_{q}\mathfrak{s}_{q}\right)_{j-k}\otimes \projector{j-1-k}{j-1-k}\cdot \frac{\mathcal{S}_{j\leftarrow j-1}}{\sqrt{j}} \ ,
\label{eq:symmetric-action}
\end{equation}
note that both expressions are valid because $\sum_{k=0}^{i-1}\projector{n-i+k}{k}\otimes \left(\ketbrac{1}{1}\otimes \mathfrak{s}^{\dagger}_{q}\mathfrak{s}_{q}\right)_{i-k}\otimes \projector{i-1-k}{i-1-k}$ is a symmetry-preserving operator.

Because number-preserving operators over different registers commute, the Baker-Campbell-Hausdorff formula becomes trivial, $e^{A+B}=e^Ae^B$ and the free-evolution operator from Eq.~(\ref{def:free-evolution-fermions}) can then be factorized
\begin{align}
     \mathcal{U}^{f}_{11}(\Delta t) & = \exp\left(-i\Delta t \sum_q E_q \sum_{j'j} \delta_{j'j}\mathcal{A}_{j\leftarrow j-1}\cdot \projector{n-j}{0} \otimes \left(\ketbrac{1}{1}\otimes \mathfrak{s}^{\dagger}_{q}\mathfrak{s}_{q}\right)_j\otimes\projector{j-1}{j-1}\cdot \mathcal{A}_{j\leftarrow j-1}\right) \nonumber \\
    & = \prod_{j} \exp \left(\sum_{k=0}^{j-1}\projector{n-j+k}{k}\otimes \left(\ketbrac{1}{1}\otimes \mathfrak{s}^{\dagger}_{q}\mathfrak{s}_{q}\right)_{j-k}\otimes \projector{j-1-k}{j-1-k}\cdot \frac{\mathcal{A}_{j\leftarrow j-1}}{\sqrt{j}}\right)\ .
\end{align}

Using the symmetry-preserving property of the exponent
\begin{equation}
     \mathcal{U}^{f}_{11}(\Delta t) =\prod_j \left\{
    \mathbb{I}^{(n)}+\sum^{\infty}_{m=1}\left[\sum_k \projector{n-j+k}{k}\otimes \left(\ketbrac{1}{1}\otimes (-i\Delta t) \sum_q E_q \set{q}\scrap{q}\right)_{j-k}\otimes \projector{j-1-k}{j-1-k}\right]^m \cdot \frac{\mathcal{A}_{j\leftarrow j-1}}{\sqrt{j}}\right\},
\end{equation}
where we have separated the $m=0$ term of the sum. Because of the step antisymmetrizer $\mathcal{A}$, this operator reduces to the identity when applied to symmetric states. But this is never the case, since by construction the fermion states are antisymmetrized upon creation of each particle. Thus, 
with all input states being antisymmetric, $\mathcal{A}$ becomes redundant, and the operator may be written down as a double product,
\begin{align}
 \mathcal{U}^{f}_{11}(\Delta t) & = \prod_j \left\{\mathbb{I}^{(n)}+\sum^{\infty}_{m=1}\left[\sum_k \projector{n-j+k}{k}\otimes \left(\ketbrac{1}{1}\otimes (-i\Delta t)- \sum_q E_q \set{q}\scrap{q}\right)_{j-k}\otimes \projector{j-1-k}{j-1-k}\right]^m\right\} \nonumber \\
& = \prod_{jk} \exp\left[\projector{n-j+k}{k}\otimes \left(\ketbrac{1}{1}\otimes \mathfrak{s}^{\dagger}_{q}\mathfrak{s}_{q}\right)_{j-k}\otimes \projector{j-1-k}{j-1-k}\right],
\end{align}
where in the last step we used commutativity of terms with different index $k$.

The exponential can be further simplified using the idempotence of projector operators,  $\left(\projector{m}{l}\right)^n  = \projector{m}{l}$ and of diagonal control operators $\left(\ketbrac{1}{1}\right)^n = \ketbrac{1}{1}$: 
\begin{equation}
 \mathcal{U}^{f}_{11}(\Delta t) =\prod_{ik} \left\{\sum_{\substack{l = 0\\ l\neq i}}\projector{l}{n}+ \projector{n-i+k}{k}\otimes \left(\ketbrac{1}{1}\otimes \mathfrak{U}\left(\Delta t\right)\right)_{i-k}\otimes \projector{i-1-k}{i-1-k}\right\},
\end{equation}
where we have defined an auxiliary register-level exponentiation (we can think of it as a composite gate) leading to an operator in terms of set and scrap operators which formally looks  like an evolution operator itself,
\begin{equation}
\mathfrak{U}_{11}(\Delta t)\equiv \exp\left[-i\Delta t \sum_{q}E_q \set{q}\scrap{q}\right].
\label{def:free-evolution-onregister}
\end{equation}
The products over $j$ and $k$ can be transformed in a sum over $j$ using the orthogonality properties of projectors, finally obtaining
\begin{equation}
\mathcal{U}^{f}_{11}\left(\Delta t\right) = \projector{n}{0} + \sum^{n}_{i=1}\projector{n-i}{0}\prod^{1}_{k=i}\otimes\left(\ketbrac{1}{1}\otimes  \mathfrak{U}_{11}\left(\Delta t\right)\right)_{k},
\label{eq:free-evolution}
\end{equation}
for the case of bosons the memory is instead symmetric, but the operator has the same formal structure. In cases in which both fermions and bosons need to be taken into account, each type of particle would have a separately assigned memory space. We will return to this issue in  subsection~\ref{subsec:splitting} below.

The sums over $i$ in expressions such as Eq.~(\ref{eq:free-evolution}) are complete decompositions and guarantee that no state will be projected out to 0 (which is not possible in unitary evolution) in spite of all the control and projector operators present in the expression which could apparently make it vanish.

\begin{figure}[ht]
\centering
\begin{minipage}{0.3\textwidth}
\centering
\includegraphics[scale=1]{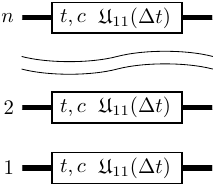}
\end{minipage}
\begin{minipage}{0.6\textwidth}
\centering
\includegraphics[scale=1]{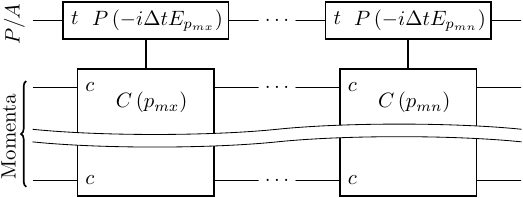}
\end{minipage}

\caption{Left. Schematic circuit implementing the free-evolution term of Eq.~(\ref{def:free-evolution-fermions}), thick lines represent registers. Right. Direct implementation of $\mathfrak{U}_{11}$ in terms of controlled phase gates: $C(p_i)$ indicates that the corresponding phase gate applies only when $p_i$ is stored on memory; $p_{mx}$ is the maximum momentum on the grid, $l=\Lambda_{max}$ on Eq.~(\ref{Momentum-discretization}), while $p_{mn}$ is the minimum, $l=\Lambda_{min}$. On a gate, $t$ denotes target registers whereas $c/ac$ denotes control/anticontrol registers; often, only presence qubits are used as controls.}
    \label{circuit:free-evolution}
\end{figure}

\textbf{Gate-level implementation and costs.} 
Eq.~(\ref{eq:free-evolution}) has a clear interpretations in terms of quantum circuits, see Fig.~(\ref{circuit:free-evolution}). Therein, each square represents an implementation of Eq.~(\ref{def:free-evolution-onregister}) controlled by the relevant momentum qubits. 

We now consider a direct implementation to give upper bounds in terms of {\tt CNOT}s and single-qubit gates. From  Eq.~(\ref{eq:free-evolution}) it is clear that there are $\bigO{n}$ $ \mathfrak{U}_{11}$ gate applications, with $n$ the number of registers, each being an exponential of $\bigO{N_p}$ diagonal terms, one for each codified momenta. As diagonal transformations commute and therefore can be applied one at a time, one possible implementation consists on  looping over the different momenta and controlling on each, applying the corresponding phase transformation. Since momenta is codified in $\bigO{\log_2 N_p}$ qubits, applying a gate controlled on an specific momentum costs $\bigO{\log_2 N_p}$  {\tt CNOT}s and single-qubit gates. Finally, as there are $\bigO{N_p}$ momenta, the total cost scales as $\bigO{n N_p\log_2 N_p}$  {\tt CNOT}s and single-qubit gates. See chapter 4 of \cite{Nielsen:2012yss} for a discussion of how to implement general controlled transformations.

\subsection{Momentum-exchange operator}
\label{subsec:pexchange}

We now consider one of the  simplest unitary operators whose action changes the values stored in registers. 
4-body terms are a staple in various fields with instantaneous potentials or contact terms exchanging momentum among a conserved number of particles, and allow for contacting with nonrelativistic, potential quantum mechanics at fixed particle number. In this subsection we consider this number-conserving part of quadrilinear operators, largely to illustrate the control of momentum exchanges. 

We adopt a discretization of $p$ with a set of momenta equispaced by $\Delta$ (not to be confused with $\Delta t$, see glossary) which sets the scale of the computation and is an infrared cutoff akin to the lattice side length in a classical gauge theory computation. The coded momenta can be spelled out as 
\begin{equation}
    \left\{p_{min},...,p_{-1},p_0 = 0,p_1,...,p_{max}\right\}
    \label{Momentum-discretization}
\end{equation}
with  $p_{l} = p_0+l\Delta = l\Delta$, $l\in \left\{\Lambda_{min},...,\Lambda_{max}\right\}$ and $|\left\{\Lambda_{min},...,\Lambda_{max}\right\}|=N_p$, for simplicity we keep $-\Lambda_{min} = \Lambda_{max} = \Lambda $. Negative values of $p$ are tagged in the storage by assigning a negative sign to the integer subindex.

The four-body ($2\to 2$) momentum-exchange term between fermion particles exponentiates to the unitary operator
\begin{equation} \label{4bodypotential}
\mathcal{U}^{f}_{22}(\Delta t) = \exp \left[-i \Delta t\sum^{\Lambda}_{\xi=-\Lambda}\left(\lambda_{\xi} \sum_{q,p} b_{q+\xi\Delta}^{\dagger} b^{\dagger}_{p} b_{p+\xi \Delta} b_{q}+h.c.\right)\right] \equiv \exp \left(-i\Delta t h^{f}_{22}\right) \ ,
\end{equation}
where $\lambda_\xi$ controls the probability/intensity of the momentum exchange and the limits of sums over $q$ and $p$ should be chosen so that $s = q+\xi\Delta = q + p_\xi$ and $r = p+\xi \Delta = p + p_\xi$ are codified (both positive and negative values of $\xi$ are included):
\begin{equation}
h^{f}_{22} = \sum^{\Lambda}_{\xi = - \Lambda}\sum_{\frac{p}{\Delta} = \text{max}(-\Lambda,-\Lambda-\xi)}^{\text{min}(\Lambda,\Lambda-\xi)}\sum_{\frac{q}{\Delta} = \text{max}(-\Lambda,-\Lambda+\xi)}^{\text{min}(\Lambda,\Lambda+\xi)}\lambda_{\xi} \left(b^{\dagger}_{q + p_\xi}b^{\dagger}_{p}b_{p + p_\xi}b_{q}+h.c.\right)=  \sum_{\xi qp} \left(b^{\dagger}_{s}b^{\dagger}_{p}b_{r}b_{q}+h.c.\right)\left|_{\substack{s= q+p_\xi \\ r = p + p_\xi}}\right. ,
\end{equation}
each term in the sum, using Eqs.~(\ref{def:nReg-fermioncreation}), (\ref{def:nReg-fermionannihilation}) and the projectors $\projector{j}{i}$ introduced under Eq.~(\ref{def:nReg-bosoncreator-j}) is
\begin{align}
    b^{(n)\dagger}_{p} & = \sum^{n}_{j = 1} \stepasym{j}\cdot\projector{n-j}{0}\otimes \left(\ketbrac{1}{0}\otimes \set{p}\right)_j\otimes\projector{j-1}{j-1},\nonumber \\
    b^{(n)}_{p} & = \sum^{n}_{j = 1}\projector{n-j}{0}\otimes\left(\ketbrac{0}{1}\otimes\scrap{p}\right)_j\otimes \projector{j-1}{j-1}\cdot \stepasym{j},
\end{align}
we have
\begin{equation}
b^{\dagger}_{s}b^{\dagger}_{p}b_{r}b_{q} = \sum^{n-1}_{j}\mathcal{A}_{j+1\leftarrow j-1} \cdot \projector{n-j+1}{0}\otimes \left(\ketbrac{1}{1}\otimes \set{s}\scrap{q}\right)_{j+1}\otimes\left(\ketbrac{1}{1}\otimes \set{p}\scrap{r}\right)_{j}\otimes\projector{j-1}{j-1}\cdot \mathcal{A}_{j+1\leftarrow j-1},
\end{equation}
where $\mathcal{A}_{j+1\leftarrow j-1} = \mathcal{A}_{j+1\leftarrow j} \mathcal{A}_{j\leftarrow j-1}$. By Eq.~(\ref{eq:antisymmetric-action}) applied twice, one for each of these two step-antisymmetrizer operators to the left, two different contributions are generated, one with $\left(\ketbrac{1}{1}\otimes \set{s}\scrap{q}\right)$ to the left of $\left(\ketbrac{1}{1}\otimes \set{p}\scrap{r}\right)$ and another with the opposite ordering,
\begin{align}
b^{\dagger}_{s}b^{\dagger}_{p}b_{r}b_{q} & =\sum^{n-1}_{j=1}\sum^{j-1}_{k=0}\sum^{k}_{l = 0}\projector{n-(j+1)+l}{l}\otimes\left(\ketbrac{1}{1}\otimes\set{s}\scrap{q}\right)_{j+1-l}\otimes\projector{k-l}{k-l}\otimes \left(\ketbrac{1}{1}\otimes\set{p}\scrap{r}\right)_{j-k}\otimes \projector{j-1-k}{j-1-k}\cdot \frac{\mathcal{A}_{j+1\leftarrow j-1}}{\sqrt{j(j+1)}} \nonumber \\
& + (pr) \longleftrightarrow (sq)
\end{align}
The bookkeeping becomes less cumbersome if we use other summation indices:
$l\rightarrow 1-i$, $k\rightarrow j-k$ and $i\rightarrow i-k$: 
\begin{align}
b^{\dagger}_{s}b^{\dagger}_{p}b_{r}b_{q} & =\sum^{n-1}_{j=1}\sum^{j}_{k = 1}\sum^{j+1}_{i=1+k}\projector{n-i}{j+1-i}\otimes\left(\ketbrac{1}{1}\otimes\set{s}\scrap{q}\right)_{i}\otimes\projector{i-k-1}{i-k-1}\otimes \left(\ketbrac{1}{1}\otimes\set{p}\scrap{r}\right)_{k}\otimes \projector{k-1}{k-1}\cdot\frac{\mathcal{A}_{j+1\leftarrow j-1}}{\sqrt{j(j+1)}}\nonumber \\
& +
(pr)\longleftrightarrow (sq)
\label{eq:four-body-sums}
\end{align}
before exponentiation it is convenient to separate projectors acting on presence qubits and operators acting on momentum qubits:
\begin{align}
h^{f}_{22} & =\sum^{n-1}_{j=1}\sum^{j}_{k = 1}\sum^{j+1}_{l=1+k}\projector{n-l}{j+1-i}\otimes\left\{\left(\ketbrac{1}{1}\otimes\right)^{\otimes l-k+1}\right\}\nonumber \\
& \left\{ \sum_{\xi} \lambda_\xi \left(\sum_{pq}\left(\set{s}\scrap{q}\right)_{l}\otimes \innerid\otimes ...\otimes \left(\set{p}\scrap{r}\right)_{k} + \sum_{pq}\left(\set{p}\scrap{r}\right)_{l}\otimes \innerid\otimes ...\otimes \left(\set{s}\scrap{q}\right)_{k}+h.c.\right)\right\}\otimes \projector{k-1}{k-1}\cdot\frac{\mathcal{A}_{j+1\leftarrow j-1}}{\sqrt{j(j+1)}}\nonumber \\
& = \sum^{n-1}_{j=1}\sum^{j}_{k = 1}\sum^{j+1}_{l=1+k}\projector{n-i}{j+1-l}\otimes\left\{\left(\ketbrac{1}{1}\otimes\right)^{\otimes l-k+1}\right\}\left\{\mathfrak{h}_{22,l.k}\right\}\otimes \projector{k-1}{k-1}\cdot\frac{\mathcal{A}_{j+1\leftarrow j-1}}{\sqrt{j(j+1)}},
\label{separatecontrolandp}
\end{align}
the notation $\left\{.\right\}\left\{.\right\}$ reminds us that we have separated the part of the operator that acts on momentum and the part that acts on presence/absence to different terms. In the third line we defined a multiregister-level implementation of the potential that exclusively acts on the momentum part of the registers between the $l$th and the $k$th (identities over momentum qubits have being omitted),
\begin{align}
\mathfrak{h}_{22,l.k} = & \sum_{\xi}\lambda_{\xi}\left[\left(\sum_q\set{s}\scrap{q}\right)_{l}\otimes \left(\sum_p\set{p}\scrap{r}\right)_{k}+\left(\sum_p\set{p}\scrap{r}\right)_{l}\otimes \left(\sum_q\set{s}\scrap{q}\right)_{k}+h.c.\right] \nonumber \\
= & 2\sum_{\xi}\lambda_{\xi}\left[\left(\sum_p\set{r}\scrap{p}\right)_{l}\otimes \left(\sum_p\set{p}\scrap{r}\right)_{k}+\left(\sum_p\set{p}\scrap{r}\right)_{l}\otimes \left(\sum_p\set{r}\scrap{p}\right)_{k}\right]\ ,
\label{h22lk}
\end{align}
which allowed to separate the control operators in Eq.~(\ref{separatecontrolandp}). Returning to the exponential of Eq.~(\ref{4bodypotential})
\begin{equation}
\mathcal{U}^{f}_{22}(\Delta t) = \exp\left[-i\Delta t \sum^{n-1}_{j=1}\sum^{j}_{k = 1}\sum^{j+1}_{l=1+k}\projector{n-i}{j+1-l}\otimes\left\{\left(\ketbrac{1}{1}\otimes\right)^{\otimes l-k+1}\right\}\left\{\mathfrak{h}_{22,l.k}\right\}\otimes \projector{k-1}{k-1}\cdot\frac{\mathcal{A}_{j+1\leftarrow j-1}}{\sqrt{j(j+1)}}\right],
\label{eq:exponential-product}
\end{equation}
here comes a delicate step: exponents with different $j$ commute because of projectors, but terms with different $l$ and $k$ do not in general, although it can be shown they do commute provided that for each term with $\xi$, there exists a term with $-\xi$ and that $\lambda_\xi = \lambda_{-\xi}$. For a central (or otherwise parity-even) potential, it is natural to take $\lambda_\xi = \lambda_{-\xi}$; and considering this to be the case, the following decomposition is exact
\begin{equation}
\mathcal{U}^{f}_{22}(\Delta t) = \prod^{n-1}_{j=1}\prod^{j}_{k= 1}\prod^{j+1}_{l= 1+k}\exp\left[-i\Delta t \, \projector{n-l}{j+1-l}\otimes\left\{\left(\ketbrac{1}{1}\otimes\right)^{\otimes l-k+1}\right\}\left\{\mathfrak{h}_{22,l.k}\right\}\otimes \projector{k-1}{k-1}\cdot\frac{\mathcal{A}_{j+1\leftarrow j-1}}{\sqrt{j(j+1)}}\right].
\end{equation}
From here on, the calculation's steps are similar to the ones used in subsec.~\ref{subsec:free}. First, $\mathcal{A}$ and the operator before the $\cdot$ comute because of Eq.~(\ref{eq:antisymmetric-action}) so the step-antisymmetrizer can be factorized from the exponential: 
\begin{equation}
\mathcal{U}^{f}_{22}(\Delta t) = \prod^{n-1}_{j=1}\prod^{j}_{k= 1}\prod^{j+1}_{l= 1+k}\left\{\mathbb{I}^{(n)}+\sum^{\infty}_{m=1}\left[-i\Delta t \, \projector{n-l}{j+1-l}\otimes\left\{\left(\ketbrac{1}{1}\otimes\right)^{\otimes l-k+1}\right\}\left\{\mathfrak{h}_{22,l.k}\right\}\otimes \projector{k-1}{k-1}\right]^{m}\cdot\frac{\mathcal{A}_{j+1\leftarrow j-1}}{\sqrt{j(j+1)}}\right\},
\end{equation}
note also that the step-antisymmetrizer does nothing on fermion memories and that projectors and control operators are idempotent to arrive to the following:
\begin{align}
\mathcal{U}^{f}_{22}(\Delta t) & = \prod^{n-1}_{j= 1}\prod^{j}_{k= 1}\prod^{j+1}_{l= 1+k}\left\{\mathbb{I}^{(n)} + \projector{n-l}{j+1-l}\otimes\left\{\left(\ketbrac{1}{1}\otimes\right)^{\otimes l-k+1}\right\}\left\{	\sum_{m = 1}\left(-i\Delta t\mathfrak{h}_{22,l.k}\right)^{m}\right\}\otimes \projector{k-1}{k-1}\right\} \\
& = \prod^{n-1}_{j= 1}\prod^{j}_{k= 1}\prod^{j+1}_{l= 1+k}\left\{ \sum^{n}_{\substack{l = 0\\ l\neq j+1}}\projector{n}{l}+ \projector{n-l}{j+1-l}\otimes\left\{\left(\ketbrac{1}{1}\otimes\right)^{\otimes l-k+1}\right\}\left\{\mathfrak{U}_{22,l.k}(\Delta t)\right\}\otimes\projector{k-1}{k-1} \right\},
\end{align}
where in the last step we subtracted from the identity the term corresponding to $m= 0$ and defined
\begin{equation}
\mathfrak{U}_{22,l.k}(\Delta t) = \exp\left(-i\Delta t\, \mathfrak{h}_{22,l.k}\right).
\label{U22lk}
\end{equation}
Finally, products are decomposed as in the free evolution case to arrive at
\begin{equation}
    \mathcal{U}^{f}_{22}(\Delta t) = \projector{n}{0}+\projector{n}{1} + \sum^{n-1}_{j=2}\projector{n-1-j}{0}\otimes\left\{\left(\ketbrac{1}{1}\right)^{\otimes j}\right\}\left\{\prod^{j}_{k=1}\prod^{j+1}_{l=1+k}\mathfrak{U}_{22,(l,k)}(\Delta t)\right\},
\label{eq:four-body-I-fermions}
\end{equation}
with 
\begin{equation}
   \mathfrak{U}_{22,l.k}(\Delta t)= \exp\left(-i2\Delta t\sum_{\xi}\lambda_{\xi}\left\{ \left(\sum_p\set{p +\xi\Delta} \scrap{p}\right)_{l} \otimes \left(\sum_q\set{q}\scrap{q+\xi\Delta}\right)_{k} + p\leftrightarrow q\right\}\right),
    \label{eq:four-body-I-fermions-exp}
\end{equation}
which is an auxiliary ``momentum exchanger'' between register in which all identities have been omitted to alleviate notation. The corresponding boson expression can be derived changing step-antisymmetrizers for step-symmetrizers, and the result is equivalent.

\begin{figure}[ht]
    \centering
    \includegraphics[scale=0.9]{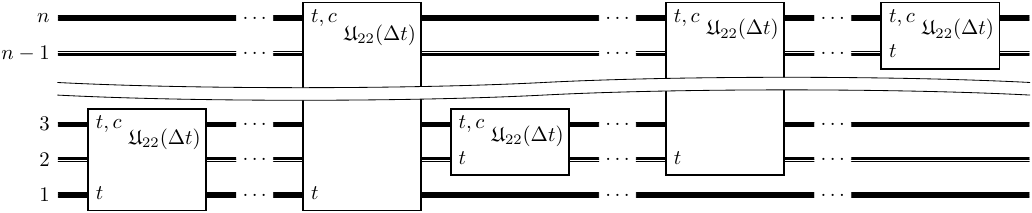}
    \caption{Schematic circuit implementing the four-body term Eq.~(\ref{eq:four-body-I-fermions}). See Fig.(\ref{circuit:free-evolution}) for an explanation of the lettering.}
    \label{circuit:momentum-exchange}
\end{figure}

\textbf{Gate-level implementation and costs.}
The circuit implementation of Eq.~(\ref{eq:four-body-I-fermions}) in terms of the sub-circuits corresponding to Eq.~(\ref{eq:four-body-I-fermions-exp}) is represented in Fig.~(\ref{circuit:momentum-exchange}). The total number of sub-circuits is now the square of the number of registers: for each value of $j$ there are $j-1$ of these subcircuits ($\mathfrak{U}_{22,j.1}$ to $\mathfrak{U}_{22,j.(j-1)}$); in total this gives $\sum^{n}_{j=2}(j-1)=\frac{n(n-1)}{2} = \bigO{n^2}$ (the usual number of exchange terms for pairwise particle interactions).

To bound the costs of each of these sub-circuits we consider a direct but very inefficient procedure. First the Trotter formula is used to decompose the exponential Eq.~(\ref{eq:four-body-I-fermions-exp}) in terms of fixed momenta
\begin{align}
    \mathfrak{U}_{22,l.k}(\Delta t)& =\prod_{\xi}\prod_{\frac{p(q)}{\Delta}=\text{max}(-\Lambda,-\Lambda-(+)\xi)}^{\text{min}(\Lambda,\Lambda-(+)\xi)}\exp\left(-i2\Delta t\lambda_{\xi}\left\{ \left(\set{p+p_\xi}\scrap{p}\right)_{l}\otimes\left(\set{q-p_\xi}\scrap{q}\right)_{k}+h.c.\right\} \right) + \bigO{\Delta t^2}\nonumber \\
    & \equiv\prod_{\xi p q}\mathfrak{U}_{22,l.k}\left(p,q,\xi;\Delta t\right)
\end{align}
each $\mathfrak{U}_{22,l.k}\left(p,q,p_\xi;\Delta t\right)$ is a rotation between states $\ket{p}_{l}\ket{q}_{k}$ and $\ket{p+p_\xi}_{l}\ket{q-p_\xi}_{k}$, so it can be easily diagonalized by a change of basis (generalization of the Hadamard gate) with matrix $T_{(p,q,\xi)}$ given by
\begin{align}
|+\rangle_\xi = T_{(p,q,p_\xi)}^{-1}\ket{p}_{l}\ket{q}_{k}&=\frac{1}{\sqrt{2}}\left(\ket{p}_{l}\ket{q}_{k}+\ket{p+p_\xi}_{l}\ket{q-p_\xi}_{k}\right)\\
|-\rangle_\xi = T_{(p,q,p_\xi)}^{-1}\ket{p+p_\xi}_{l}\ket{q-p_\xi}_{k}&=\frac{1}{\sqrt{2}}\left(\ket{p}_{l}\ket{q}_{k}-\ket{p+p_\xi}_{l}\ket{q-p_\xi}_{k}\right),
\end{align}
which can be implemented by a Gray code and a controlled Hadamard, resulting in an asymptotic cost of $\bigO{\log_2^{2}N_{p}}$ {\tt CNOTs} and single-qubit gates. The transformed rotation is diagonal, e.g.
\begin{equation}
\mathfrak{U}_{22(l,k)}\left(p,q,p_\xi;\Delta t\right)\,|+\rangle_{\xi}=\exp\left(-i2\Delta t\lambda_{\xi}\right)\ket{+}_{\xi}
\end{equation}
and similarly for 
$|-\rangle_\xi$
changing the phase sign. 
Each of these phase transformations controlled with $\bigO{2\log_2 N_p}$ qubits is implementable with $\bigO{\log_2 N_p}$ {\tt CNOTs} and single-qubit gates. The costs are thus dominated by the change of basis, so the full implementation of the four body term is upper bounded by $\bigO{n^2 N_p^3 \log_2^2 N_p}$ {\tt CNOTs} and single-qubit gates. See chapter 4 of \cite{Nielsen:2012yss} for a discussion about Gray codes and the implementation of general unitary transformation.

\section{Exponentiation of particle-number changing operators \label{sec:evolution2}}
The formalism that we have here developed would not be particularly advantageous with evolution operators such as those of subsections~\ref{subsec:free} and~\ref{subsec:pexchange}, that do not really require a dynamical memory but can instead be statically implemented  in other ways at fixed particle number; but it comes into its own when confronted with particle-number changing operators. The difficulty in encoding these comes from the lack of commutativity in the exponentials and the implementation of symmetrizers and antisymmetrizers. We here propose a way of encoding them which is consistent with all the earlier discussion and definitions, without prejudice that more efficient organizations can be thought of in future work.  We have chosen to exemplify with one-body operators (``tadpoles'') and three-body operators (``splitting'') that occupy the rest of this section. 

\subsection{One-body operators (tadpoles)\label{subsubsec:tadpoles}}

One-body operators do not typically appear in field-theoretical Hamiltonian densities formulated on a stable vacuum, as they are linear in the fields, but they can be useful in dynamical situations such as a system rolling down a classical potential. 
In any case, our purpose here is to use the simplest possible operator which can expose the changing number of particles, and these are obviously one-body terms. 
We first address fermion operators which, though not appearing in physical processes (the superselection rule associated with $(2\pi)$ rotations forbids the creation of an odd number of fermions) provide a useful stepping stone for other algorithms. Then we will address the boson tadpole in Subsec.~\ref{subsubsec:boson-tadpole}.

The resulting unitary operator is then
\begin{equation}
\label{def:tadpole}
    \mathcal{U}^{f}_{10}(\Delta t) = \exp\left[-i \Delta t \sum_p \lambda_p \left(b^{(n)\dagger}_{p}+b^{(n)}_{p}\right)\right] = \exp\left[-i \Delta t  \sum_p \lambda_p \sum_{j}\left(b^{(n)\dagger}_{p,j}+b^{(n)}_{p,j}\right)\right]  .
\end{equation}
The first step is to decompose the operator in products of terms separately acting over each register, which amounts to transforming the exponential of the sum over $j$ into a product of exponentials. We iterate to factor each of the $j$-indexed pieces in turn. 
Because of the noncommutativity and the Baker-Campbell-Hausdorff rule for noncommuting exponents, $ e^A e^B = e^{A+B +\frac{1}{2}[A,B]+\dots}$, we expect errors of order $(\Delta t)^2$.
To factor out the first,  $\sum_p \lambda_p(b^{(n)\dagger}_{p,1}+b^{(n)}_{p,1})$ we check its commutation with the rest:
\begin{equation}
   \left[\left(b^{(n)\dagger}_{p,1}+b^{(n)}_{p,1}\right),\sum^{n}_{j=2}\left(b^{(n)\dagger}_{q,j}+b^{(n)}_{q,j}\right)\right]  = b^{(n)\dagger}_{q,2}b^{(n)\dagger}_{p,1}-b^{(n)}_{q,2}b^{(n)}_{p,1},
\end{equation}
as expected (for any operator changing the number of particles), the factorization is not exact, and some error depending on $\lambda_l$ and $\Delta t$ is introduced if particles are created/annihilated one at a time. Such uncertainty is controlled by the Trotter step with small parameter $\Delta t$, whose asymptotically small values would allow the use of the (approximate at finite $\Delta t$) product formula
\begin{equation}
    \mathcal{U}^{f}_{10} =  \exp\left[-i \Delta t \sum_{j}\sum_p \lambda_p (b^{(n)\dagger}_{p,j}+b^{(n)}_{p,j})\right] = \prod_{j}\exp\left[-i \Delta t \sum_p \lambda_p (b^{(n)\dagger}_{p,j}+b^{(n)}_{p,j})\right] + O((\Delta t)^2) 
    \label{eq:tadpole-decomp}
\end{equation}
up to $O((\Delta t)^2)$ in the exponents (thus, the calculation is systematically improvable for $\Delta t\to 0$ independently of $\lambda$, even for strong coupling). A more rigorous error bound should also consider the norm of the first non-vanishing commutators; here there are $n$ of those, each with a norm $\bigO{N_p \bar{\lambda^2} = \sqrt{\sum_p\sum_q\lambda^2_p\lambda^2_q }}$, so controlling the error suggests operating with $\Delta t< <\hbar /\sqrt{n N_p \bar{\lambda^2}}$. In the rest of the manuscript the errors are not written explicitly and in particular its dependence on first commutators is not taken into account.

To continue, creation and  annihilation operators are written in terms of their set and scrap counterparts,
\begin{equation}
b^{(n)\dagger}_{p,j}+b^{(n)}_{p,j} = \mathcal{A}_{j\leftarrow j-1}\cdot \projector{n-j}{0}\otimes \left(\ketbrac{1}{0}\otimes \set{p}\right)_{j}\otimes\projector{j-1}{j-1} +\projector{n-j}{0}\otimes \left(\ketbrac{0}{1}\otimes\scrap{p}\right)_{j}\otimes\projector{j-1}{j-1} \cdot\mathcal{A}_{j\leftarrow j-1},
\label{eq:tadpole-exponent}
\end{equation}
but it is not clear how such a term can be exponentiated, as Eq.~(\ref{eq:antisymmetric-action}) is not applicable because of the change in particle number. The issue can be circumvented if we manage to rewrite that expression in terms of some new unitary operators (note the circumflex hat), the ordered step antisymmetrizer and its conjugate, $\hat{\mathcal{A}}^{\dagger}_{j\leftarrow j-1}$ and $\hat{\mathcal{A}}_{j\leftarrow j-1}$ satisfying 
\begin{align}
    b^{(n)\dagger}_{p,j} & = \mathcal{A}_{j\leftarrow j-1}\cdot \projector{n-j}{0}\otimes \left(\ketbrac{1}{0}\otimes \set{p}\right)_{j}\otimes\projector{j-1}{j-1}\nonumber \\
    & \equiv  \hat{\mathcal{A}}^{\dagger}_{j\leftarrow j-1}\cdot \projector{n-j}{0}\otimes \left(\left(\ketbrac{1}{0}\otimes\set{p}\right)_{j}\otimes\projector{j-1}{j-1}\right)_{f,ord}\cdot \hat{\mathcal{A}}_{j\leftarrow j-1},
    \label{def:ordered-creation-fermions}
\end{align}
where the subscripts $f,ord$ are explained in detail in the next paragraphs. 
It is obvious that the na\"ive antisymmetrizers $\mathcal{A}$ are not invertible (and thus, not possibly unitary) because their application projects out part of the Hilbert space. To construct a unitary operator which is implementable on a quantum computer, we need an invertible rule. This is possible if we force an initial ordering in the registers, so that the antisymmetrizer takes the ordered set producing an antisymmetrized state, whereas its inverse takes the antisymmetrized state and returns it to the unique, properly ordered state in a bijective manner. Such a procedure was already outlined by Abrams and Lloyd 20 years ago \cite{abrams_009711_1997}, see \cite{berry_improved_2018} for a more up to date discussion. Thus we need to distinguish between the $\mathcal{A}$-antisymmetrizer, closer to that of field theory, and the $\hat{\mathcal{A}}$ antisymmetrizer, specific for computer registers which have a well defined order.

To  achieve unitarity, the new operator $\hat{\mathcal{A}}^{\dagger}_{j\leftarrow j-1}$ is defined to act as an antisymmetrizer  over the ordered subset of the $j$-particle states, and as the identity if the $j$th register is empty,\begin{equation}
    \hat{\mathcal{A}}^{\dagger}_{j\leftarrow j-1} \begin{cases}\left[\ket{1p}_{j}\left(\ket{1p_{j-1}}_{j-1}...\ket{1p_{1}}_{1}\right)_{\mathcal{A}}\right]_{ord} & = \left(\ket{1p}_{j}\ket{1p_{j-1}}_{j-1}...\ket{1p_{1}}_{1}\right)_{\mathcal{A}} \\\ \ 
    \ket{\Omega}_{j}\left(\ket{1p_{j-1}}_{j-1}...\ket{1p_{1}}_{1}\right)_{\mathcal{A}} & = \ket{\Omega}_{j}\left(\ket{1p_{j-1}}_{j-1}...\ket{1p_{1}}_{1}\right)_{\mathcal{A}}\ .
    \end{cases}
    \label{def:stepantisymmetrizer-dagger}
\end{equation}
Here, a state $\ket{1p}_{j}\left(\ket{1p_{j-1}}_{j-1}...\ket{1p_{1}}_{1}\right)_{\mathcal{A}}$ is said to be ordered and subscripted so, if the momentum mode on the $j$th-register is the largest, in the sense defined by some appropriate ordering criteria. (Note that, as the antisymmetrization is performed one register at a time, the procedure is different to the one outlined in \cite{abrams_009711_1997}.) We here make the following obvious and natural choices, first on the discretized momenta, subsequently on the ordering of states ensured {\it ad hoc} by the corresponding creation criterion.

\paragraph{Order criterion on momenta. \label{criteria:order}} A mode $\ket{p_i}$ is larger or equal than $\ket{p_j}$ if and only if $i \geq j$. Checked on runtime by an order oracle
\begin{eqnarray}
    O(\ket{x}\ket{p_i}\ket{p_j}) = \begin{cases}
        \ket{x\oplus 1}\ket{p_i}\ket{p_j}&\text{if } i\geq j\\
        \ket{x}\ket{p_i}\ket{p_j}&\text{otherwise,}
    \end{cases}
\label{def:order-oracle}
\end{eqnarray}

\paragraph{Order criterion on creation operators.\label{criteria:creation}}  If a new particle is to be created, the resulting state should have its largest mode, in the sense defined by the order criterion, on the recently occupied register.\\
(This is to facilitate the invertible antisymmetrization). 

From here on, these criteria are enforced and remarked by the subscript $ord$,  as in Eq.~(\ref{def:ordered-creation-fermions}). The other subscript there, $f$, is a reminder that no mode is to be created if it is already on memory, and that $(-)$ signs should occur each time two particle registers are permuted. We now discuss the detailed implementation of each piece, starting from the unitary operator $\hat{\mathcal{A}}$.

\subsubsection{Implementation of unitary step-antisymmetrizers and fermion-{\tt SWAP} gates.}
As for the definition in Eq.~(\ref{def:stepantisymmetrizer-dagger}), the ordered step operator $\hat{\mathcal{A}}^{\dagger}_{j\leftarrow j-1} $ antisymmetrizes states provided the $j$th register is occupied by the most sizeable momentum and acts as the identity if the $j$th register is empty. It would be ideal if the two situations could be distinguished by using controls over the presence qubit of register $j$. 
\paragraph{Implementation of $\hat{\mathcal{A}}^{\dagger}$. \label{par:antisymmetrizers}} The passage from/to an ordered state to/from an antisymmetrized one is realized invoking a subroutine named ``Locate the Largest'' (LL) described in Appendix \ref{miscellanea:LL algorithm} that returns, in an auxiliary register, a true qubit pointing to the largest mode. To illustrate the antisymmetrization step, consider the probe state $\ket{1p_{j}}_j\left(\ket{1p_{j-1}}_{j-1}...\ket{1p_{1}}_1\right)_{\mathcal{A}}$ with the largest mode on the last occupied slot $\ket{p_j}>\ket{p_k},\,\,k\in\left[1,j-1\right]$. The algorithm proceeds as follows
(we show the particle content of the register at each step)
\begin{enumerate}
    \item Add an auxiliary register B with $j$ qubits with the last of them in state $\ket{1}$, so pointing to $\ket{1p_j}$:
    \begin{equation}
        \ket{1,0,...,0}_{B}\ket{1p_{j}}_j\left(\ket{1p_{j-1}}_{j-1}...\ket{1p_{1}}_1\right)_{\mathcal{A}}
    \end{equation}
    \item Create an equally-weighted superposition of states in the auxiliary register with the qubit in state $1$ in each of the positions in turn,
    \begin{equation}
        \frac{1}{\sqrt{j}}\left(\ket{1,0,...,0}_{B}+\ket{0,1,...,0}_{B}+...+\ket{0,...,1}_{B}\right)\ket{1p_{j}}_j\left(\ket{1p_{j-1}}_{j-1}...\ket{1p_{1}}_1\right)_{\mathcal{A}}
    \end{equation}
    This costs $\bigO{\log_2 j}$ single-qubit gates.
    \item Exchange the first momentum register with whichever is pointed at by the single 1 in the auxiliary $B$, and then change the sign to account for fermion antisymmetry. Note that an antisymmetric state of $j-1$ registers can be written as
    \begin{equation}
        \ket{p_j}_j\left(\ket{p_{j-1}}_{j-1}...\ket{p_l}_l...\ket{p_1}_1\right)_{\mathcal{A}} = \sum_{k=1}^{j-1}(-1)^{|l-k|}\ket{p_j}_{j}\left(\ket{p_{j-1}}_{j-1}...\overline{\ket{p_k}}_l...\ket{p_{k+1}}_{k}\ket{p_{k-1}}_{k-1}...\ket{p_1}_1\right)_{\mathcal{A}},
    \label{def:momentum-fixing}
    \end{equation}
    where momentum in register $l$ is fixed to $p_k$. The {\tt SWAP} thus produces an interchange between momenta $p_k$ and $p_j$:
    \begin{align} 
        \frac{1}{\sqrt{j}}&\left\{\ket{1,0,...,0}_{B}\ket{1p_{j}}_j\left(\ket{1p_{j-1}}_{j-1}...\ket{1p_{1}}_1\right)_{\mathcal{A}}\right.\nonumber\\
        &\left.-\sum_{l=1}^{j-1}|0...,\underset{l}{\underbrace{1}},...,0\rangle\sum_{k=1}^{j-1}(-1)^{|l-k|}\ket{p_k}_{j}\left(\ket{p_{j-1}}_{j-1}...\overline{\ket{p_j}}_l...\ket{p_{k+1}}_{k}\ket{p_{k-1}}_{k-1}...\ket{p_1}_1\right)_{\mathcal{A}}\right\}
    \end{align}
    This costs $\bigO{j}$ {\tt SWAP}s and phase gates, each controlled by an auxiliary qubit and the presence qubit of register $j$. For a total of $\bigO{N_p}$ momenta codified in $\bigO{\log_2 N_p}$ qubits, this can be decomposed in $\bigO{j\log_2 N_p}$ {\tt CNOT}s and $\bigO{j\log_2 N_p}$ single-qubit gates.
    \item We now want to disentangle register $B$. For this, we will take (in an orderly manner) all its qubits to $|0\rangle$ and factor it off. 
    By hypothesis, the one qubit in the state $\ket{1}$ was marking the position of the largest mode, and the LL algorithm from appendix~\ref{miscellanea:LL algorithm}  can be inverted to return it to zero:
    \begin{align} 
        \frac{1}{\sqrt{j}}&\left\{\ket{1p_{j}}_j\left(\ket{1p_{j-1}}_{j-1}...\ket{1p_{1}}_1\right)_{\mathcal{A}}\right.\nonumber\\
        &\left.-\sum_{l=1}^{j-1}\sum_{k=1}^{j-1}(-1)^{|l-k|}\ket{p_k}_{j}\left(\ket{p_{j-1}}_{j-1}...\overline{\ket{p_j}}_l...\ket{p_{k+1}}_{k}\ket{p_{k-1}}_{k-1}...\ket{p_1}_1\right)_{\mathcal{A}}\right\},
    \end{align}
    returning the register with momentum $p_j$ to its initial position (summing over $l$) we have 
     \begin{align} 
        \frac{1}{\sqrt{j}}&\left\{\ket{1p_{j}}_j\left(\ket{1p_{j-1}}_{j-1}...\ket{1p_{1}}_1\right)_{\mathcal{A}}-\sum_{k=1}^{j-1}\ket{p_k}_{j}\left(\ket{p_{j-1}}_{j-1}...\ket{p_j}_k...\ket{p_1}_1\right)_{\mathcal{A}}\right\} \nonumber \\
        & = \left(\ket{p_j}_j\ket{p_{j-1}}_{j-1}...\ket{p_1}_1\right)_{\mathcal{A}},
    \end{align}
    Costs: The costs of the LL algorithm over $j$ registers ($j$ particles in memory) involves $\bigO{j^2}$ {\tt CNOT}s, single qubit gates and order-oracle calls.
\end{enumerate}

Thus, the total costs are: $\bigO{j^2 + j\log_2 N_p}$ {\tt CNOT}s and single qubit gates,  $\bigO{j^2}$ order-oracle calls and $\bigO{\log_2 j}$ rotations.

We now face the implementation of the order criterion on creation operators. The ``fermion'' and ``order'' subscripts impose two conditions over the creation/annihilation term of Eq.~(\ref{def:ordered-creation-fermions}): a) mode duplication is forbidden, and b) upon completion, the largest mode must sit on the last occupied register. To fulfill these requirements we proceed one momentum at a time (which implies the use of the Trotter expansion). We will also need a piece of code which allows the agile exchange of entire particle registers, to which we now devote our attention.

\paragraph{Fermion-{\tt SWAP} gate, $\fSWAP{j}{p}{q}$ and fermion-exchange gate, $\fX{j}{p}$.\label{par:fermion-SWAP}}

We need to be able to exchange fermions for two reasons. One is the ability to order the largest mode at the last register so that the ordered step antisymmetrizer can work.  And importantly, to satisfy Pauli exclusion, avoid creating a mode for an existing momentum $p$: when it is to be created, we attempt to bring it to the front to nullify the action of the creation operator in case it is already present in the memory (obviously, by $p$ we mean here the complete fermion orbital including color and other degrees of freedom).

At a lower level of programming, we need to introduce a fermion-{\tt SWAP} gate depending on two modes  $\fSWAP{j}{p}{k}$, the first one being on register $j$.   This function will be overloaded: its implementation is different depending on whether the vacuum mode $0$ or the largest mode $L$ sit at the $j$th register. We then write either $\fSWAP{j}{L}{p}$ or $\fSWAP{j}{0}{p}$ to distinguish these two cases. 

Once both implemented, we can
 define a fermion-{\tt eXchange} operator for momentum $p$ as
\begin{equation}
     \fX{j}{p} = (\ketbrac{0}{0})_j\otimes \fSWAP{j}{0}{p}+(\ketbrac{1}{1})_j\otimes\fSWAP{j}{L}{p}.
\label{def:fTd}
\end{equation}

The gate is implemented by the following sequence of operations:
\begin{enumerate}
    \item By means of an auxiliary register B with $j-1$ qubits, ``mark'' the particle register with modes $k$ for $\fSWAP{j}{p}{k}$; $0$ for $\fSWAP{j}{p}{0}$; and finally the largest mode for $\fSWAP{j}{p}{L}$. This is achieved by controlled-{\tt X} gates or the LL algorithm correspondingly. The costs of this procedure amount to $\bigO{\log_2 N_p}$ Toffoli-gates for an {\tt X}-gate controlled on $\bigO{\log_2 N_p}$ qubits, this can be decomposed in  $\bigO{\log_2 N_p}$ {\tt CNOT}s and  $\bigO{\log_2 N_p}$ single-qubit gates. The LL algorithm adds $\bigO{j^2}$ {\tt CNOT}s, single qubit gates and oracle calls.
    \item Permute the marked register with the $j$th, using the auxiliary qubits in state $\ket{1}$. Account a ($-$) sign for fermion antisymmetry. The associated costs are now as follows: 1-qubit control of {\tt SWAP}s only adds a constant overhead in {\tt CNOT}s and single-qubit gates, so this requires $\bigO{j\log_2 N_p}$ {\tt CNOT}s and $\bigO{j\log_2 N_p}$ single-qubit gates. The (-) signs can be added by applying $\bigO{j}$ phase gates controlled on the auxiliary qubit pointing to the interchanged register.
    \item Apply again the first step but now controlled on momentum $p$, so the auxiliary register returns to its initial state. The cost here is the same as in step 1.
\end{enumerate}
The total costs amounts to  $\bigO{j\log_2 N_p}$ {\tt CNOT} and single-qubit gates for $\fSWAP{j}{p}{k}$ and $\fSWAP{j}{p}{0}$, $\bigO{j^2}$ order-oracle calls and $\bigO{j^2+j\log_2 N_p}$ {\tt CNOT} and single-qubit gates for $\fSWAP{j}{p}{L}$.

\subsubsection{Implementation of Eq.(\ref{def:ordered-creation-fermions}), the qubit-level exponentiation.\label{subsec:7.1.2}}
Introduction of these gates in Eq.~(\ref{def:ordered-creation-fermions}) gives
\begin{align}
b^{(n)\dagger}_{p,j} +b^{(n)}_{p,j} = & \,\hat{\mathcal{A}}^{\dagger}_{j\leftarrow j-1}\cdot \projector{n-j}{0}\otimes \left(\left(\ketbrac{1}{0}\otimes\set{p}\right)_{j}\otimes\projector{j-1}{j-1}+h.c.\right)_{f,ord}\cdot\hat{\mathcal{A}}_{j\leftarrow j-1}\nonumber \\
=  & \,\hat{\mathcal{A}}^{\dagger}_{j\leftarrow j-1}\cdot \left[(\ketbrac{0}{0})_j\otimes \fSWAP{j}{p}{0}+(\ketbrac{1}{1})_j\otimes\fSWAP{j}{p}{L}\right]\cdot \projector{n-j}{0}\otimes \left(\left(\ketbrac{1}{0}\otimes\set{p}\right)_{j}\otimes\projector{j-1}{j-1}+h.c.\right)\nonumber\\
& \,\cdot \left[(\ketbrac{0}{0})_j\otimes \fSWAP{j}{0}{p}+(\ketbrac{1}{1})_j\otimes\fSWAP{j}{L}{p}\right] \cdot \hat{\mathcal{A}}_{j\leftarrow j-1}\nonumber \\
=& \,\hat{\mathcal{A}}^{\dagger}_{j\leftarrow j-1}\cdot\fXdg{j}{p} \cdot \projector{n-j}{0}\otimes \left(\left(\ketbrac{1}{0}\otimes\set{p}\right)_{j}\otimes\projector{j-1}{j-1}+h.c.\right)\cdot \fX{j}{p} \cdot \hat{\mathcal{A}}_{j\leftarrow j-1},
\end{align}

where $\fSWAPdg{j}{p}{k} = \fSWAP{j}{k}{p}$. To arrive to the second equality, note that both the order and no mode-duplication requirements are satisfied: $\hat{\mathcal{A}}_{j\leftarrow j-1}$ leaves the largest mode on register $j$ when it is occupied, so controlled on the presence of $j$ being $\ket{1}$, applying the $\fSWAP{j}{L}{p}$ gate leaves momentum $p$ on register $j$, and it can be annihilated. On the other hand, if register $j$ is empty, $\hat{\mathcal{A}}_{j\leftarrow j-1}$ does nothing and $\fSWAP{j}{0}{p}$ tries to {\tt SWAP} the $p$ and vacuum modes; if this is achieved the $j$ register ends up in the state $\ket{0p}$, which goes to null upon action of $\left(\ketbrac{1}{0}\otimes\set{p}\right)_j$ (the corresponding exponential then goes to the identity) and avoids mode duplication; if otherwise there was no mode $p$ on memory, $\fSWAP{j}{0}{p}$ does nothing and the creation procedure carries on. Subsequent application of $\fXdg{j}{p}$ reorders the memory if register $j$ ends up being occupied, or returns $\ket{0p}$ to $\ket{\Omega}$ if it is empty.

With these operators we may proceed to exponentiation. Since antisymmetrizers do not depend on the created/annihilated momentum and $\hat{\mathcal{A}}^{\dagger}_{j\leftarrow j-1}\hat{\mathcal{A}}_{j\leftarrow j-1}=\mathbb{I}^{(j)}$
\begin{align}
& \exp\left(-i \Delta t\sum_p \lambda_p (b^{(n)\dagger}_{p,j}+b^{(n)}_{p,j})\right) \nonumber\\
&= \hat{\mathcal{A}}^{\dagger}_{j\leftarrow j-1}\cdot \left[\exp\left(-i\Delta t \sum_p \lambda_p\fXdg{j}{p} \cdot \projector{n-j}{0}\otimes \left(\left(\ketbrac{1}{0}\otimes\set{p}\right)_{j}+h.c.\right)\otimes\projector{j-1}{j-1}\cdot \fX{j}{p}\right)\right]\cdot \hat{\mathcal{A}}_{j\leftarrow j-1}
\end{align} 
where operators $\outeridentity^{\otimes n-j}$ multiplying antisymmetrizers have been omitted. To continue with exponentiation, we use the Trotter formula for the sum over $p$ and use $\fXdg{j}{p}\fX{j}{p} = \mathbb{I}^{(j)}$
\begin{align}
\mathcal{U}^{f}_{10}  & = \prod_j\exp\left[-i \Delta t\sum_p \lambda_p (b^{(n)\dagger}_{p,j}+b^{(n)}_{p,j})\right] \nonumber\\
& =  \prod_j\left\{ \hat{\mathcal{A}}^{\dagger}_{j\leftarrow j-1}\cdot \prod_p\left[\fXdg{j}{p}\cdot\exp\left(-i\Delta t \lambda_p \projector{n-j}{0}\otimes \left(\left(\ketbrac{1}{0}\otimes\set{p}\right)_{j}+h.c.\right)\otimes\projector{j-1}{j-1}\right)\cdot \fX{j}{p}\right]\right.\nonumber \\
& \left.\cdot \hat{\mathcal{A}}_{j\leftarrow j-1}\right\}+\bigO{\Delta t^2},
\end{align}

finally, using the idempotency of operators $\ketbrac{1}{1}$, $\ketbrac{0}{0}$ and $\innerid$ and similar transformations to those of the preceeding sections, we can write down the exponential as
\begin{align}
\mathcal{U}^{f}_{10} & = \prod_{j}\left(  \sum_{\substack{l = 0\\ l\neq j,j-1}}\projector{n}{l} +\hat{\mathcal{A}}^{\dagger}_{j\leftarrow j-1}\cdot\prod_{p}\left[ \fXdg{j}{p}\cdot\projector{n-j}{0}\otimes\mathfrak{U}_{10,j}\left(\Delta t,\lambda_p\right)\otimes\projector{j-1}{j-1}\cdot\fX{j}{p}\right]\cdot\hat{\mathcal{A}}_{j\leftarrow j-1}\right) \nonumber\\
& +\bigO{\Delta t^2}.
\label{eq:one-body-fermions-SWAP}
\end{align}
with 
\begin{equation}
\mathfrak{U}_{10,j}\left(\Delta t,\lambda_p\right) = \exp\left(-i\Delta t  \lambda_p \left(\ketbrac{1}{0}\otimes \set{p}+\ketbrac{0}{1}\otimes\scrap{p}\right)_{j}\right).
\label{eq:one-body-fermions-exp}
\end{equation}

\begin{figure}
\centering
\includegraphics[scale=1]{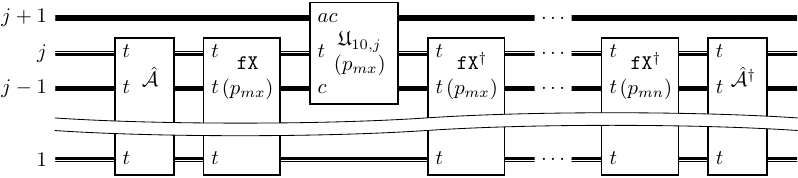}

\caption{Circuit implementation of Eq.~(\ref{eq:one-body-fermions-SWAP}),  each thick line representing a particle register. Lettering as in Fig.(\ref{circuit:free-evolution}) and the glossary.
 With time advancing from left to right (the equation backwards, from right to left), we start with the antisymmetric state which is unpacked by $\hat{\mathcal{A}}$ to one with the largest $p$ in the last occupied register. Then if a fermion in the memory has $p_{i}$ it is brought to the front by a fermion exchange, to be able to block the creation of a second state with equal quantum numbers which would violate Pauli's exclusion. Should that not be the case, $\mathfrak{U}_{10,j}$ introduces the new fermion as indicated by the Hamiltonian. The fermion exchange is then inverted to bring the largest $p$ to the front, and the memory is again antisymmetrized. This row of operators is repeated as needed to complete the wanted number of Trotter steps.
\label{fig:circuitU10}}
\end{figure}

\textbf{Gate-level implementation and costs.}
Each term in the $j$-product on Eq.~(\ref{eq:one-body-fermions-SWAP}) can be represented as depicted in Fig.~\ref{fig:circuitU10} 
(the controlled gates therein use presence qubits).

The operators inside the momentum product in Eq.~(\ref{eq:one-body-fermions-SWAP}) are repeated $\bigO{N_p}$ times, so we have (for the costs of each operation, we refer to the preceding paragraphs or to table \ref{table:implementation-costs-others}):
\begin{itemize}
    \item $\bigO{N_p}$ $\fX{j}{p}$, with a total cost of $\bigO{jN_p\log_2 N_p}$ {\tt CNOTs} and single-qubit gates and $\bigO{j^2}$ oracle-calls (these include {\tt SWAP}s with largest momenta).
    \item $\bigO{1}$ $\hat{\mathcal{A}}^{\dagger}_{j\leftarrow j-1}$, with a total costs of $\bigO{j^2+j\log_2 N_p}$ {\tt CNOTs}, single-qubit gates and $\bigO{j^2}$ order-oracle calls.
    \item $\bigO{N_p}$ $\mathfrak{U}_{10,j}\left(\Delta t,\lambda_p\right)$. Each of these implemented as $\mathfrak{U}_{22}$ back on section \ref{subsec:pexchange}, with a Gray code and a controlled single-qubit rotation that results in a total costs of $\bigO{N_p \log_2^2N_p}$ {\tt CNOT}s and single-qubit gates.
\end{itemize}
Addition from $j=1$ to $j=n$ gives a total cost of $\bigO{n^2N_p\log_2 N_p+N_p\log_2^2 N_p}$ {\tt CNOT}s and single-qubit gates + $\bigO{ n^3 N_p}$ order-oracle calls.
\subsubsection{Boson tadpole \label{subsubsec:boson-tadpole}}
Many aspects of implementing boson operators can be read off from the fermion ones by changing the antisymmetrizing operators for symmetrizing ones, which  directly leads to operators $\hat{\mathcal{S}}$ and $\hat{\mathcal{S}}^{\dagger}$ which symmetrize ``semi''-ordered states, as the largest mode can now be repeated (no Pauli exclusion applies). Thus the preceding algorithms have to be slightly extended. As usual, details are considered in appendix \ref{general-boson-operators}. Here we give examples that illustrate the main steps for the case of three registers. 

\paragraph{Implementation of  $\hat{\mathcal{S}}^{\dagger}$. \label{par:symmetrizer-3reg}} As an example of the implementation of the symmetrizer, consider the symmetrization of the semiordered state $\ket{\phi_0}=\ket{P}_3\left(\ket{P}_2\ket{q}_1\right)_S$ with $\ket{P}>\ket{q}$ and two equal momenta. 
We may proceed as follows.
\begin{enumerate}
\item Add two auxiliary registers of $n=3$ qubits, $C$ (with one qubit on state 1) and $B$ (to null). 
\begin{equation}
    \ket{\phi_1}= \ket{100}_C\ket{000}_B\frac{1}{\sqrt{2}}\left(\ket{PPq}+\ket{PqP}\right)
\end{equation}
\item Apply the {\tt LL} algorithm to $B$ (now there can be more than one qubit marking the largest $P$):
\begin{equation}
      \ket{\phi_2}=\ket{100}_C\frac{1}{\sqrt{2}}\left(\ket{110}_B\ket{PPq}+\ket{101}_B\ket{PqP}\right)
\end{equation}
\item Symmetrize register C:
\begin{equation}
      \ket{\phi_3}=\frac{1}{\sqrt{3}}\left(\ket{100}_C+\ket{010}_C+\ket{001}_C\right)\frac{1}{\sqrt{2}}\left(\ket{110}_B\ket{PPq}+\ket{101}_B\ket{PqP}\right)
\end{equation}
\item Using register $C$, apply Eq.~(\ref{def:step-symmetrizer}) with $j=3$ on both the $B$ and the momentum registers, {\it i.e.}, permute the third of the mode registers and the third of $B$'s qubits with that mode-register and that qubit of $B$'s which are pointed at by register $C$:
\begin{align}
\ket{\phi_4} = & \frac{1}{\sqrt{3}}\ket{100}_C\frac{1}{\sqrt{2}}\left(\ket{110}_B\ket{PPq}+\ket{101}_B\ket{PqP}\right) + \frac{1}{\sqrt{3}}\ket{010}_C\frac{1}{\sqrt{2}}\left(\ket{110}_B\ket{PPq}+\ket{011}_B\ket{qPP}\right)\nonumber \\
+ & \frac{1}{\sqrt{3}}\ket{001}_C\frac{1}{\sqrt{2}}\left(\ket{011}_B\ket{qPP}+\ket{101}_B\ket{PqP}\right),
\end{align}
rearranging
\begin{align}
\ket{\phi_4} = & \frac{1}{\sqrt{3}} \left[\frac{1}{\sqrt{2}}\left(\ket{100}_C+\ket{010}_C\right)\ket{110}_B\ket{PPq} + \frac{1}{\sqrt{2}}\left(\ket{100}_C+\ket{001}_C\right)\ket{101}_B\ket{PqP}\right.\nonumber\\
    + & \left.\frac{1}{\sqrt{2}}\left(\ket{010}_C+\ket{001}_C\right)\ket{011}_C\ket{qPP}\right]
\end{align}
\item Register $B$ can now be used to return $C$ to its initial state: for each state of $B$, we have equal superpositions of states of $C$ with the qubit in state $1$ in the positions pointed by $B$:
\begin{itemize}
    \item If $\ket{B}= \ket{110}$ interchange $  \frac{1}{\sqrt{2}}\left(\ket{100}_C+\ket{010}_C\right) \leftrightarrow \ket{000}_C $
    \item If $\ket{B}= \ket{101}$ interchange $  \frac{1}{\sqrt{2}}\left(\ket{100}_C+\ket{001}_C\right)\leftrightarrow \ket{000}_C$
    \item If $\ket{B}= \ket{011}$ interchange $ \frac{1}{\sqrt{2}}\left(\ket{010}_C+\ket{001}_C\right)\leftrightarrow \ket{000}_C  $
    \item If $\ket{B}= \ket{111}$ interchange $\frac{1}{\sqrt{3}}\left(\ket{100}_C+\ket{010}_C+\ket{001}_C\right)\leftrightarrow \ket{000}_C$
\end{itemize}
so the state now showcases an unentangled $C$,
\begin{equation}
    \ket{\phi_5} = \frac{1}{\sqrt{3}}\ket{000}_C\left(\ket{110}_B\ket{PPq}+\ket{101}_B\ket{PqP}+\ket{011}_B\ket{qPP}\right)\ .
\end{equation}
It is of note  that the cost of this step rises exponentially with the number of registers $n$, since the possible number of states of $B$ grows as $2^n$. We introduce another algorithm named ``Symmetric-Digit-Decomposition (SDD)'' discussed in appendix \ref{general-boson-operators} to circumvent the issue.
\item Finally, apply {\tt LL} in reverse on register $B$:
\begin{equation}
    \ket{\phi_6} =\frac{1}{\sqrt{3}}\ket{000}_C\ket{000}_B\left(\ket{PPq}+\ket{PqP}+\ket{qPP}\right)   \ .
\end{equation}
\end{enumerate}
The main difference is thus the addition of a new register $B$ to store the positions of the (several) largest modes. Note this produces the desired states for the other cases with three registers: if there is no repeated momenta, register $B$ will be initialized to $\ket{100}_B$ on step \textit{2.} and it will be reversed to $\ket{000}_B$ on step \textit{6.}, so no changes are produced in step \textit{5}. If the three modes were equal, $B$ will be initialized to $\ket{111}_B$ and in step \textit{5.} the symmetrization of step \textit{3.} will be applied in reverse, so register $C$ returns to $\ket{000}_C$. Gate costs are discussed on appendix \ref{general-boson-operators} and compiled on table \ref{table:implementation-costs-others}.
\vspace{0.5cm}
\paragraph{Boson {\tt SWAP} 
$\bSWAP{j}{p}{q}$ 
and controlled boson {\tt eXchange} gates.
{\label{par:bSWAP-3reg}}} 
Similarly to the fermion case, we introduce a boson {\tt SWAP} gate depending on two modes ${\tt bS}_j(p,q)$, the first one being on register $j$. This allows to move the largest momentum to and away from the last occupied register, a manoeuvre needed for the ordered step symmetrizer operations to work.

As an example of its implementation we use the probe state $\ket{\phi_0}=\ket{p}_4(\ket{p}_3\ket{q}_2\ket{q}_1)_S \equiv \ket{p}\ket{pqq}_S$:
\begin{enumerate}
    \item Add three auxiliary registers $B$, $C$ and $D$ with three qubits each. Initialize $B$ with the positions of mode $q$ and $C$ with the positions of mode $p$ in the first two registers (count from right to left):
    \begin{equation}
            \ket{\phi_1}=\ket{000}_D\frac{1}{\sqrt{3}}\left(\ket{100}_C\ket{011}_B\ket{ppqq}+\ket{010}_C\ket{101}_B\ket{pqpq}+\ket{001}_C\ket{110}_B\ket{pqqp}\right)
    \end{equation}
    \item Register $B$ marks the positions in which interchanges should occur, so it is used to initialize register $D$ using the transformations of step \textit{5.} of the symmetrizer's algorithm
    \begin{align}
         \ket{\phi_2} = \frac{1}{\sqrt{3}} & \left[\frac{1}{\sqrt{2}}\left(\ket{010}_D+\ket{001}_D\right)\ket{100}_C\ket{011}_B\ket{ppqq}\right. \nonumber \\
        & + \frac{1}{\sqrt{2}}\left(\ket{100}_D+\ket{001}_D\right)\ket{010}_C\ket{101}_B\ket{pqpq}\nonumber \\
        & \left. + \frac{1}{\sqrt{2}}\left(\ket{100}_D+\ket{010}_D\right)\ket{001}_C\ket{110}_B\ket{pqqp}\right]       
    \end{align}
    \item Controlled on each qubit of $D$, apply a {\tt SWAP} gate between the corresponding qubits of $B$ and $C$ and permute the fourth momentum register with the one pointed by each state in the superpositions of $D$:
    \begin{align}
         \ket{\phi_3} =  \frac{1}{\sqrt{3}} & \left[\frac{1}{\sqrt{2}}\ket{010}_D\ket{110}_C\ket{001}_B\ket{qppq}+\frac{1}{\sqrt{2}}\ket{001}_D\ket{101}_C\ket{010}_B\ket{qpqp}\right. \nonumber \\
        & + \frac{1}{\sqrt{2}}\ket{100}_D\ket{110}_C\ket{001}_B\ket{qppq}+\frac{1}{\sqrt{2}}\ket{001}_D\ket{011}_C\ket{100}_B\ket{qqpp}\nonumber \\
        & \left. + \frac{1}{\sqrt{2}}\ket{100}_D\ket{101}_C\ket{010}_B\ket{qpqp}+\frac{1}{\sqrt{2}}\ket{010}_D\ket{011}_C\ket{100}_B\ket{qqpp}\right],       
    \end{align}
    which can be conveniently rearranged:
    \begin{align}
         \ket{\phi_3} =  \frac{1}{\sqrt{3}} & \left[\frac{1}{\sqrt{2}}\left(\ket{010}_D+\ket{100}_D\right)\ket{110}_C\ket{001}_B\ket{qppq}\right. \nonumber \\
        & + \frac{1}{\sqrt{2}}\left(\ket{001}_D+\ket{100}_D\right)\ket{101}_C\ket{010}_B\ket{qpqp}\nonumber \\
        & \left. + \frac{1}{\sqrt{2}}\left(\ket{001}_D+\ket{010}_D\right)\ket{011}_C\ket{100}_B\ket{qqpp}\right],       
    \end{align}    
    \item The states of $C$ now factorize the invariant superpositions of $D$ under {\tt SWAP} transformation between qubits marked therein, so register $D$ can be uncomputed applying step \textit{5.} above. Registers $B$ and $C$ are uncomputed as they were initialized, so we finally have
    \begin{equation}
        \ket{\phi_4} = \ket{000}_D\ket{000}_C\ket{000}_B\left(\ket{q}\ket{ppq}_S\right)
    \end{equation}
\end{enumerate}
Gate costs are discussed on appendix \ref{general-boson-operators} and compiled on table \ref{table:implementation-costs-others}.

With these new operators Eq.~(\ref{eq:one-body-fermions-SWAP}) for bosons reads
\begin{align}
\mathcal{U}^{f}_{10} & = \prod_{j}  \left\{\sum_{\substack{l = 0\\ l\neq j,j-1}}\projector{n}{l} +\hat{\mathcal{S}}^{\dagger}_{j\leftarrow j-1}\cdot\prod_{p} \left[\bXdg{j}{p}\cdot\projector{n-j}{0}\otimes\mathfrak{U}_{10,j}\left(\Delta t,\lambda_p\right)\otimes\projector{j-1}{j-1}\cdot\bX{j}{p}\right]\cdot\hat{\mathcal{S}}_{j\leftarrow j-1}\right\}\nonumber \\
& +\bigO{\Delta t^2}.
\label{eq:one-body-bosons-SWAP}
\end{align}
with the controlled boson-exchange gate that, in this case, only serves one function (bringing the largest momentum to or away from the $j$th register for appropriate ordering) but does not need to operate on an empty register, so
\begin{equation}
     \bX{j}{p} = (\ketbrac{0}{0})_j\otimes\innerid+(\ketbrac{1}{1})_j\otimes\bSWAP{j}{L}{p}.
\label{def:cbX}
\end{equation}
If register $j$ is occupied the boson-{\tt SWAP} tries a permutation between the largest mode (in register $j$ by default) and mode $p$, so it can be annihilated. Mode duplication is not forbidden for bosons, so if register $j$ is empty no {\tt SWAP} but the identity is applied.

\textbf{Gate-level implementation and costs.} The implementation is completely analogous to the fermion case, we just change the fermion operators by their boson counterparts, and elevate the corresponding computing costs to table \ref{table:implementation-costs-others}. They are:
\begin{itemize}
    \item $\bigO{N_p}$ $\bX{j}{p}$, with a total cost of $\bigO{ j^3 N_p\log_2 j + j N_p\log_2 N_p}$ and $\bigO{j^2 N_p}$ oracle-calls.
    \item $\bigO{1}$ $\hat{\mathcal{S}}^{\dagger}_{j\leftarrow j-1}$, with a cost of $\bigO{j^3\log_2 j + j\log_2 N_p}$ {\tt CNOTs} and single-qubit gates and $\bigO{j^2}$ oracle-calls.
    \item $\bigO{N_p}$ $\mathfrak{U}_{10,j}\left(\Delta t,\lambda_p\right)$ with a total cost of $\bigO{N_p\log_2^2N_p}$ {\tt CNOT}s and single-qubit gates.
\end{itemize}
Addition from $j=1$ to $j=n$ gives $\bigO{n^2 N_p\log_2 N_p + N_p \log_2^2 N_p + n^4 N_p\log_2 n}$ {\tt CNOT}s and single-qubit gates and $\bigO{ n^3 N_p}$ order-oracle calls.
\subsection{Splitting operators}
\label{subsec:splitting}
We have now all the elements to work out a splitting operator or vertex interaction. Let us focus, as a master template, on the emission (or absorption) of a boson by a fermion, a process key in QCD gluon emission, nuclear pion emission and more.
\begin{equation}
    \mathcal{U}_{21}(\Delta t, \lambda) = \exp\left[-i\Delta t\sum_{p,k}\lambda_{k-p}\left(a^{\dagger}_{k-p}b^{\dagger}_p b_{k} + h.c.\right)\right].
    \label{def:splitting}
\end{equation}
(The subindices $21$ make the particle-number changing nature of this interaction clear.)
The first step is to express the momenta in the above sums in terms of appropriate indices. Setting the annihilated momenta to $p_m$ and the created ones to $p_r$ and $p_\xi$, the vertex interaction reads
\begin{equation}
     \mathcal{U}_{21}(\Delta t, \lambda) = \exp\left[-i\Delta t \sum_{r,m}\lambda_{\xi}\left(a^{\dagger}_{p_\xi}b^{\dagger}_{p_r} b_{p_m}+h.c.\right)\right]_{\xi=m-r},
     \label{eq:splitting-term}
\end{equation}
where the conditions over the indices should be chosen, as we assume henceforth, so that $p_r$, $p_\xi$ and $p_m$ are all momenta which can be codified in the grid, just as in section \ref{subsec:pexchange}. We may as well then drop the $p$ lettering and keep only the mode indices $r,m,\xi\dots$. 

Boson and fermion states are separately encoded because 
of their different symmetrization.
We thus consider $n_b$ registers of bosons and $n_f$ registers of fermions, so that $n = n_b + n_f$, $a^{\dagger}_\xi = \sum_{j_b}a^{(n_b)\dagger}_{\xi,j_b}\otimes \outeridentity^{\otimes n_f}$ and $b^{\dagger}_\xi = \outeridentity^{\otimes n_b} \otimes\sum_{j_f}b^{(n_f)\dagger}_{\xi,j_f}$, thus
\begin{align}
    a^{(n_b)\dagger}_{\xi,j_b}\otimes b^{(n_f)\dagger}_{r,j_f}b^{(n_f)}_{m,j_f'} &  = a^{(n_b)\dagger}_{\xi,j_b}\otimes\delta_{j_f j_f'} \mathcal{A}_{j_f\leftarrow j_f-1}\cdot \projector{n_f-j_f}{0}\otimes \left(\ketbrac{1}{1}\otimes\set{r}\scrap{m}\right)_{j_f}\otimes\projector{j_f-1}{j_f-1}\cdot \mathcal{A}_{j_f\leftarrow j_f-1}\nonumber \\
    &  = \delta_{j_f, j_f'} \left[\mathcal{S}_{j_b\leftarrow j_b-1}\cdot\projector{n_b-j_b}{0}\otimes \left(\ketbrac{1}{0}\otimes\set{\xi}\right)_{j_b}\otimes\projector{j_b-1}{j_b-1}\right]\nonumber \\
    & \otimes \left[\mathcal{A}_{j_f\leftarrow j_f-1}\cdot \projector{n_f-j_f}{0}\otimes \left(\ketbrac{1}{1}\otimes \set{r}\scrap{m}\right)_{j_f}\otimes\projector{j_f-1}{j_f-1}\cdot \mathcal{A}_{j_f\leftarrow j_f-1}\right],
\label{eq:splitting-operators}
\end{align}
In the following we describe two ways of exponentiation of the preceding equation: the first one is analogous to that already used for other terms in $H$, and particle-exchange symmetry is only partially conserved by the time evolution. Then in the second paragraph below, exchange-symmetry conservation is imposed by construction.

\subsubsection{Exponentiation without exact conservation of fermion antisymmetry \label{subsubsec:splitting-1}}
Several of the techniques developed earlier are used for exponentiation. First, Eq.~(\ref{eq:antisymmetric-action}) can be used to commute the antisymmetrizers since the number of fermions is conserved:
\begin{align}
a^{(n_b)\dagger}_{\xi,j_b}&\otimes b^{(n_f)\dagger}_{r,j_f}b^{(n_f)}_{m,j_f} = \mathcal{S}_{j_b\leftarrow j_b-1}\cdot \projector{n_b-j_b}{0}\otimes\left(\ketbrac{1}{0}\otimes\set{\xi}\right)_{j_b}\otimes\projector{j_b-1}{j_b-1}\nonumber \\
& \otimes\left[\sum^{j_f -1}_{k=0}\projector{n_f+k-j_f}{k}\otimes\left(\ketbrac{1}{1}\otimes\set{r}\scrap{m}\right)_{j_f-k}\otimes\projector{j_f-1-k}{j_f-1-k}\right]\cdot\frac{\mathcal{A}_{j_f\leftarrow j_f-1}}{\sqrt{j_f}}.
\label{}
\end{align}
Note that $\mathcal{A}_{j_f-1\leftarrow j_f-2}$ and $\mathcal{S}_{j_b\leftarrow j_b-1}$ commute as they act on different parts of the memory. To continue, we turn the sums over $j_b$ and $j_f$ in the exponent into products using the commutation between terms with different $j_b$ or $j_f$. The antisymmetrizer can be exponentiated out and simplified as in previous sections:
\begin{align}
\mathcal{U}_{21}(\Delta t, \lambda)  & = \prod_{j_f,j_b=1}\left(\mathbb{I}^{(n)}+\sum^{\infty}_{l=1}\left[-i\Delta t \sum_{m,r}\lambda_\xi \, \mathcal{S}_{j_b\leftarrow j_b-1} \cdot \projector{n_b-j_b}{0}\otimes\left(\ketbrac{1}{0}\otimes\set{\xi}\right)_{j_b}\otimes\projector{j_b-1}{j_b-1}\right.\right.\nonumber \\
&\left.\left.\otimes\left(\sum^{j_f-1}_{k=0}\projector{n_f+k-j_f}{k}\otimes\left(\ketbrac{1}{1}\otimes \set{r}\scrap{m}\right)_{j_f-k}\otimes\projector{j_f-k-1}{j_f-k-1}\right)+h.c.\right]^{l}\right),
\end{align}
we now promote $\mathcal{S}_{j_b\leftarrow j_b-1}$ to a unitary operator by addition of the corresponding inverse operator, and impose an ordering on the boson part of the expression:
\begin{align}
\mathcal{U}_{21}(\Delta t, \lambda)  = & \prod_{j_f,j_b=1}\left(\mathbb{I}^{(n)}+\sum^{\infty}_{l=1}\left[-i\Delta t \sum_{m,r}\lambda_\xi \, \hat{\mathcal{S}}^{\dagger}_{j_b\leftarrow j_b-1} \cdot \left(\projector{n_b-j_b}{0}\otimes\left(\ketbrac{1}{0}\otimes\set{\xi}\right)_{j_b}\otimes\projector{j_b-1}{j_b-1}\right)_{b,ord}\cdot \hat{\mathcal{S}}_{j_b\leftarrow j_b-1}\right.\right.\nonumber \\
&\left.\left.\otimes\left(\sum^{j_f-1}_{k=0}\projector{n_f+k-j_f}{k}\otimes\left(\ketbrac{1}{1}\otimes \set{r}\scrap{m}\right)_{j_f-k}\otimes\projector{j_f-k-1}{j_f-k-1}\right) +h.c.\right]^{l}\right),
\end{align}
since symmetrizers do not transform the fermion part of the memory, we can pass $\hat{\mathcal{S}}_{j_b\leftarrow j_b-1}$ to the right,
\begin{align}
\mathcal{U}_{21}(\Delta t, \lambda)  = & \prod_{j_b j_f} \left(\mathbb{I}^{(n)}+\hat{\mathcal{S}}^{\dagger}_{j_b\leftarrow j_b-1}\cdot\projector{n_b-j_b}{0}\otimes\sum^{\infty}_{l=1}\left[-i\Delta t \sum_{m,r}\lambda_\xi  \sum^{j_f-1}_{k=0}\left(\left(\ketbrac{1}{0}\otimes\set{\xi}\right)_{j_b}\otimes\projector{j_b-1}{j_b-1}\right)_{b,ord}\right.\right. \nonumber \\
&\left.\left.\otimes\,\projector{n_f+k-j_f}{k}\otimes\left(\ketbrac{1}{1}\otimes \set{r}\scrap{m}\right)_{j_f-k}\otimes\projector{j_f-k-1}{j_f-k-1} +h.c.\right]^{l}\cdot \hat{\mathcal{S}}_{j_b\leftarrow j_b-1}\right).
\end{align}
To continue, idempotent terms are factorized from the sum over $l$:
\begin{align}
\mathcal{U}_{21}&(\Delta t, \lambda)  = \prod_{j_b,j_f=1} \left(\mathbb{I}^{(n)}+\hat{\mathcal{S}}^{\dagger}_{j_b\leftarrow j_b-1}\cdot\projector{n_b-j_b}{0}\otimes\left\{\ketbrac{1}{1}^{\otimes(j_b-1)}\otimes\ketbrac{0}{0}^{\otimes(n_f-j_f)}\otimes\ketbrac{1}{1}^{\otimes j_f}\right\}\right.\nonumber\\
&\left.\left\{\sum^{\infty}_{l=1}\left[-i\Delta t \sum_{m,r}\lambda_\xi\sum^{j_f-1}_{k=0}\left(\left(\ketbrac{1}{0}\otimes\set{\xi}\right)_{j_b}\otimes...\otimes(\innerid)_1\right)_{b,ord}\otimes\left(\set{r}\scrap{m}\right)_{j_f-k}+h.c.\right]^l\cdot\hat{\mathcal{S}}_{j\leftarrow j-1}\right\}\right),
\label{U21exact}
\end{align}
where we have omitted the identity operators over the fermion momentum qubits. To resum the exponential we subtract from the identity $\mathbb{I}^{(n)}$ the terms corresponding to $l=0$, and since antisymmetrizers and symmetrizers commute with projectors we arrive at:

\begin{align}
\mathcal{U}_{21}(\Delta t, \lambda)  & = \prod_{j_b,j_f=1} \left(\mathbb{I}^{(n)}-\projector{n_b}{j_b}\otimes \projector{n_f}{j_f}-\projector{n_b}{j_b-1}\otimes \projector{n_f}{j_f}\right.\nonumber\\
&\left.+\hat{\mathcal{S}}^{\dagger}_{j_b\leftarrow j_b-1}\cdot\projector{n_b-j_b}{0}\otimes\left\{\ketbrac{1}{1}^{\otimes(j_b-1)}\otimes\ketbrac{0}{0}^{\otimes(n_f-j_f)}\otimes\ketbrac{1}{1}^{\otimes j_f}\right\}\left\{\mathfrak{U}_{21,(j_b,j_f)}\right\}\right)\cdot\hat{\mathcal{S}}_{j_b\leftarrow j_b-1}\nonumber\\
& = \prod_{j_b,j_f=1} \left(\mathbb{I}^{(n)}-\projector{n_b-j_b}{0}\otimes\mathbb{I}\otimes\projector{j_b-1}{j_b-1}\otimes\projector{n_f}{j_f}\right.\nonumber\\
&\left.+\hat{\mathcal{S}}^{\dagger}_{j_b\leftarrow j_b-1}\cdot\projector{n_b-j_b}{0}\otimes\left\{\ketbrac{1}{1}^{\otimes(j_b-1)}\otimes\ketbrac{0}{0}^{\otimes(n_f-j_f)}\otimes\ketbrac{1}{1}^{\otimes j_f}\right\}\left\{\mathfrak{U}_{21,(j_b,j_f)}\right\}\right)\cdot\hat{\mathcal{S}}_{j_b\leftarrow j_b-1}\nonumber\\
\label{U21approx}
\end{align}
with
\begin{equation}
\mathfrak{U}_{21,(j_b,j_f)}\left(\Delta t,\lambda\right)=\exp\left[-i\Delta t \sum_{m,r}\lambda_\xi\sum^{j_f-1}_{k=0}\left(\left(\ketbrac{1}{0}\otimes\set{\xi}\right)_{j_b}\otimes...\otimes(\innerid)_1\right)_{b,ord}\otimes\left(\set{r}\scrap{m}\right)_{j_f-k}+h.c.\right].
\label{eq:two-fermions-one-boson-1}
\end{equation}
The operator of Eq.~(\ref{eq:two-fermions-one-boson-1}) respects the fermion antisymmetry of input states exactly, but as it acts over the entire fermion memory, we expect its implementation to be expensive. Back on section \ref{subsec:pexchange} the exponentiation of the sum over registers (index $k$) was decomposed exactly in terms of the products of exponentials, since each conmmutator  with $\lambda_\xi$ cancelled a commutator with $\pm\lambda_{-\xi}$. This is not the case here because of the creation/annihilation terms; thus terms with different $k$ do not commute. 

We can then use Trotter's formula, discretizing the time evolution by some $\Delta t$ to obtain an approximate decomposition which implies that exchange symmetry for fermions is only approximated up to $O(\Delta t^2)$. The exchange symmetry for bosons is in contrast exact, as ensured by the application of the step-symmetrizers.

Furthermore, the operator is decomposed in terms of fixed emitted boson-momentum $\xi$ pieces to implement the boson creation and annihilation terms as was done for the tadpole in section \ref{subsubsec:tadpoles}:
\begin{align}
\mathcal{U}_{21}(\Delta t, \lambda)  & = \prod_{j_b,j_f=1} \left(\mathbb{I}^{(n)}-\projector{n_b-j_b}{0}\otimes\mathbb{I}\otimes\projector{j_b-1}{j_b-1}\otimes\projector{n_f}{j_f}+\hat{\mathcal{S}}^{\dagger}_{j_b\leftarrow j_b-1}\cdot\prod_{\xi}\left[\bXdg{j_b}{\xi}\cdot\projector{n_b-j_b}{0}\otimes\right.\right.\nonumber\\
&\left.\left.\left\{\ketbrac{1}{1}^{\otimes(j_b-1)}\otimes\ketbrac{0}{0}^{\otimes(n_f-j_f)}\otimes\ketbrac{1}{1}^{\otimes j_f}\right\}\left\{\prod^{j_f-1}_{k=0}\mathfrak{U}_{21,(j_b,j_f-k)}\right\}\cdot\bX{j_b}{\xi}\right]\cdot\hat{\mathcal{S}}_{j_b\leftarrow j_b-1}\right) + \mathcal{O}(\Delta t^2),
\end{align}
with
\begin{equation}
\mathfrak{U}_{21,(j_b,j_f-k)}\left(p_\xi;\Delta t,\lambda\right)=  \exp\left[-i\Delta t \sum_{m,r}\lambda_\xi\left(\ketbrac{1}{0}\otimes\set{\xi}\right)_{j_b}\otimes\left(\set{r}\scrap{m}\right)_{j_f-k}+h.c.\right].
\end{equation}
Such decomposition of the exponential in a product over the fermion registers indexed by $k$ avoids duplication of momenta \textit{up to second order terms} in the Trotter expansion. That is, the Pauli exclusion principle is recovered for $\Delta t\to 0$.

$\mathcal{U}_{21}(\Delta t, \lambda)$ can be further simplified: operators outside the braces do not depend on $j_f$ and can be factored out of the corresponding product; terms with different $j_f$ are orthogonal, and controlled operations over properly filled memories [as defined below Eq.~(\ref{eq:nReg-commutation})] can be simplified. The final operator reads:
\begin{align}
\mathcal{U}_{21}(\Delta t, \lambda)  & = \prod_{j_b}\left(\mathbb{I}^{(n)}-\projector{n_b-j_b}{0}\otimes\mathbb{I}\otimes\projector{j_b-1}{j_b-1}\otimes\mathbb{I}^{(n_f)}\right.\nonumber \\
& +\left.\hat{\mathcal{S}}^{\dagger}_{j_b\leftarrow j_b-1}\cdot\prod_{\xi}\left[\bXdg{j_b}{\xi}\cdot\hat{\mathfrak{U}}_{21,j_b}(p_\xi;\Delta t, \lambda) \cdot\bX{j_b}{\xi}\right]\cdot\hat{\mathcal{S}}_{j\leftarrow j-1}\right) + \bigO{\Delta t^2},
\label{eq:splitting-1}
\end{align}
where the auxiliary operator $\hat{\mathfrak{U}}_{21,j_b}$ is
\begin{equation}
\hat{\mathfrak{U}}_{21,j_b}(p_\xi;\Delta t, \lambda) = \prod^{n_f}_{j_f=1} \left(\ketbrac{0}{0}\right)_{j_b+1}\otimes \left\{\left(\ketbrac{1}{1}\right)_{j_b-1}\otimes\left[\left(\ketbrac{1}{1}\right)_{j_f}\right\}\left\{\mathfrak{U}_{21,(j_b,j_f)}(p_\xi;\Delta t, \lambda)\right\}+\left.\left(\ketbrac{0}{0}\right)_{j_f}\right\}\left\{\outeridentity\right\}\right],
\label{eq:splitting-2}
\end{equation}
where remaining operators acting on presence qubits are all identities. Note the special combination of square and curly brackets, it denotes that the terms to the left of the first square bracket tensor-multiply both terms of the sum to the right.
  
\begin{figure}[!ht]
\centering
\includegraphics[scale=0.9]{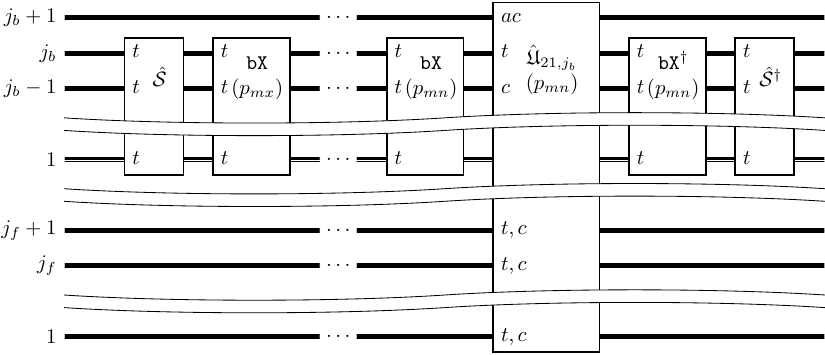}
\caption{Circuit implementation of $\mathcal{U}_{21}(\Delta t, \lambda)$  without exact conservation of fermion antisymmetry, given by Eq.~(\ref{eq:splitting-1}). See Fig.(\ref{circuit:free-evolution}) for an explanation of the $t$, $c$ and $ac$ symbols. \label{circuit:splitting-1}}
\end{figure}

\begin{figure}[!ht]
\centering
\includegraphics[scale=0.9]{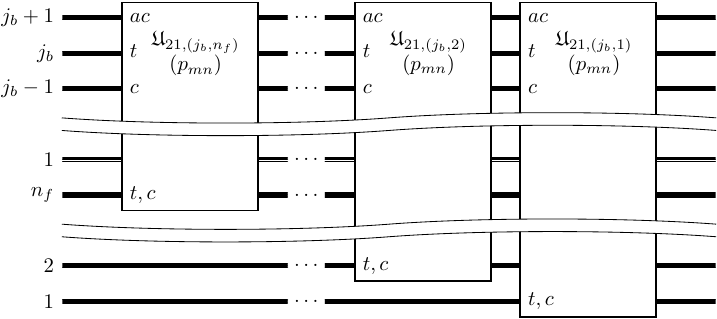}
\caption{Circuit implementation of $\hat{\mathfrak{U}}_{21,j_b}(p_{mn};\Delta t, \lambda)$, subroutine of $\mathcal{U}_{21}(\Delta t, \lambda)$, given by Eq.~(\ref{eq:splitting-2}). See Fig.(\ref{circuit:free-evolution}) for an explanation of the $t$, $c$ and $ac$ symbols.\label{circuit:splitting-2}}
\end{figure}

\textbf{Gate-level implementation and costs.} Fig.~(\ref{circuit:splitting-1}) is the circuit implementation of Eq.~(\ref{eq:splitting-1}), while the subcircuit Eq.~(\ref{eq:splitting-2}) is found in Fig.~(\ref{circuit:splitting-2}). The only unknown costs are those of $\mathfrak{U}_{21,j_b}$, which in turn are composed of $\bigO{n_f^2}$ $\mathfrak{U}_{21,(j_b,j_f)}$ gates, whose implementation is similar to that of the exchange term, see section \ref{subsec:pexchange}.

Each $\mathfrak{U}_{21,(j_b,j_f)}$ gate is a single rotation applied to the registers $j_b$ (on the boson memory section) and $j_f$ (on the fermion one). For a fixed $p_\xi$ we can collect the basis states which are affected by the transformation of the set of registers (only the relevant $j_b$th and $j_f$th registers are explicitly written, in pairs with the $j_b$ boson register either empty or occupied)
\begin{align}
  A_{p_{\xi}}=\left\{ \right. &\ket{\Omega}_{j_{b}}\ket{1\max\left(p_{-\Lambda}+p_{\xi},p_{-\Lambda}\right)}_{j_{f}}\ ,\ \ket{1p_{\xi}}_{j_{b}}\ket{1\max\left(p_{-\Lambda},p_{-\Lambda}-p_{\xi}\right)}_{j_{f}}...\ , 
  \nonumber \\
  & \ket{\Omega}_{j_{b}}\ket{1\max\left(p_{-\Lambda}+p_{\xi},p_{-\Lambda}\right)+p_k}_{j_{f}}\ ,\ \ket{1p_{\xi}}_{j_{b}}\ket{1\max\left(p_{-\Lambda},p_{-\Lambda}-p_{\xi}\right)+p_k}_{j_{f}}\dots 
  \nonumber \\
  & \ket{\Omega}_{j_{b}} \left.\ket{1\text{min}\left(p_{\Lambda}+p_{\xi},p_{\Lambda}\right)}j_{f}\ ,\ \ket{1p_{\xi}}\ket{1\text{min}\left(p_{\Lambda}-p_{\xi},p_{\Lambda}\right)}_{j_{f}}\right\},
\end{align}
where the momentum cutoffs are respected due to  the ``max'' and ``min'' selections. $\mathfrak{U}_{21,(j_b,j_f)}$ is diagonalized with a change of basis $T_{p_\xi,p_k}$ ($p_\xi>0$) to a linear combination of the fermion before the splitting and the fermion+boson pair (sharing the momentum) after it, 
\begin{align}
T_{p_\xi,p_k}^{-1}\ket{\Omega}_{j_{b}}\ket{1p_{-\Lambda}+p_{\xi}+p_{k}}_{j_{f}}=& \frac{1}{\sqrt{2}}\left(\ket{\Omega}_{j_{b}}\ket{1p_{-\Lambda}+p_{\xi}+p_{k}}_{j_{f}}+\ket{1p_{\xi}}_{j_{b}}\ket{1p_{-\Lambda}+p_{k}}_{j_{f}}\right)\nonumber \\
T_{p_\xi,p_k}^{-1}\ket{1p_{\xi}}_{j_{b}}\ket{1p_{-\Lambda}+p_{k}}_{j_{f}}=& \frac{1}{\sqrt{2}}\left(\ket{\Omega}_{j_{b}}\ket{1p_{-\Lambda}+p_{\xi}+p_{k}}_{j_{f}}-\ket{1p_{\xi}}_{j_{b}}\ket{1p_{-\Lambda}+p_{k}}_{j_{f}}\right),
\end{align}
with $p_{k}\in\left[0,p_{\Lambda}-p_{-\Lambda}-p_{\xi}\right]$. For each $p_k$ and $p_\xi$ pair, the $T$ transformation can be implemented using a Gray code, applying a controlled Hadamard and reversing the Gray code, resulting in a cost of $\bigO{\log_2^{2}N_{p}}$. Composition of transformations for different $p_k$ give rise to a gate $T_{p_\xi}$ with a total cost of $\bigO{N_{p}\log_2^{2}N_{p}}$ {\tt CNOTs} and single-qubit gates. $\mathfrak{U}_{21,\left(j_{b},j_{f}\right)}\left(p_{\xi}\right)$ is then diagonal in the new basis, explicitly:
\begin{align}
    T^{-1}_{p_{\xi}}\,\mathfrak{U}_{21,\left(j_{b}, j_{f}\right)} \left(p_{\xi}\right) \, T_{p_{\xi}} &=\exp\left[-i\Delta t\lambda_{\xi}\left(\ket{\Omega}\bra{\Omega}\right)_{j_{b}}\otimes\left(\sum_{p_{r}=\text{max}\left(p_{-\Lambda},p_{-\Lambda}+p_{\xi}\right)}^{\text{min}\left(p_{\Lambda},p_{\Lambda}+p_{\xi}\right)}\set{p_{r}}\scrap{p_{r}}\right)_{j_{f}}\right.\nonumber\\
    & \left.+i\Delta t\lambda_{\xi}\left(\ket{1p_{\xi}}\bra{1p_{\xi}}\right)_{j_{b}}\otimes\left(\sum_{p_{r}=\text{max}\left(p_{-\Lambda},p_{-\Lambda}-p_{\xi}\right)}^{\text{min}\left(p_{\Lambda},p_{\Lambda}-p_{\xi}\right)}\set{p_{r}}\scrap{p_{r}}\right)_{j_{f}}\right], \label{U21transformada}
\end{align}
both terms in the exponential commute and can be separately implemented. 

The $p_r$ sums over fermion $\set{p_{r}}\scrap{p_{r}}$ pairs are projectors, which ensure diagonal transformation affect states for which the change of basis produces codified outputs; if the cutoffs were removed, they would turn to identities by the completeness relation. Thus we can implement Eq.(\ref{U21transformada}) by first testing whether we are in the subspace of allowed states, saving the result in an ancillary qubit, applying the corresponding diagonal rotation controlled on that, and finally returning the ancillary qubit to its initial state, with a total cost of $\bigO{N_{p}\log_2 N_{p}}$ {\tt CNOTs} and single-qubit gates. Thus the costs of $\mathfrak{U}_{21,\left(j_{b},j_{f}\right)}\left(p_{i}\right)$ are dominated by the change of basis, so the operator $\hat{\mathfrak{U}}_{21,\left(j_{b}\right)}\left(p_{i}\right)$ needs $\bigO{n_f^2 N_{p}\log_2^{2}N_{p}}$ {\tt CNOTs} and single-qubit gates.

To conclude, for each $j_b$ on Eq.~(\ref{eq:splitting-1}) there are (see table \ref{table:implementation-costs-others} for individual costs):
\begin{itemize}
    \item $\bigO{N_p}$ $\bX{j}{p}$, with a total cost of $\bigO{ j_b^3 N_p\log_2 j_b + j_b N_p\log_2 N_p}$ and $\bigO{j_b^2 N_p}$ oracle-calls.
    \item $\bigO{1}$ $\hat{\mathcal{S}}^{\dagger}_{j_b\leftarrow j_b-1}$, with a cost of $\bigO{j_b^3\log_2 j_b+j_b\log_2 N_p}$ {\tt CNOTs} and single-qubit gates and $\bigO{j_b^2}$ oracle-calls.
    \item $\bigO{N_p}$ $\hat{\mathfrak{U}}_{21,\left(j_{b}\right)}\left(p_{i}\right)$ with a total cost of $\bigO{n_f^2 N^2_{p}\log_2^{2}N_{p}}$ {\tt CNOT}s and single-qubit gates.
\end{itemize}
Summing over $j_b$, the final cost to implement $\mathcal{U}_{21}$ is $\bigO{n_b n_f^2 N_p^2 \log_2^2 N_p}$ {\tt CNOTs} and single-qubit gates and $\bigO{n_b^3 N_p}$ order-oracle calls.

\subsubsection{Exponentiation with exact conservation of fermion antisymmetry \label{subsubsec:splitting-2}}
We now discuss an alternative route of exponentiation that preserves antisymmetry by construction, although at the expense of an increase in its implementation costs. The key idea is to promote the step antisymmetrizers $\hat{\mathcal{A}}$ to unitary operators instead of commuting them with Eq.~(\ref{eq:antisymmetric-action}), so that \begin{align}
a^{(n_b)\dagger}_{s,j_b}\otimes b^{(n_f)\dagger}_{r,j_f}b^{(n_f)}_{m,j_f-1} & = \hat{\mathcal{S}}^{\dagger}_{j_b\leftarrow j_b-1}\cdot\hat{\mathcal{A}}^{\dagger}_{j_f\leftarrow j_f-1}\cdot \left(\projector{n_b-j_b}{0}\otimes \left(\ketbrac{1}{0}\otimes\set{s}\right)_{j_b}\otimes\projector{j_b-1}{j_b-1}\right)_{b,ord}\nonumber\\
&\otimes\left(\projector{n_f-j_f}{0}\otimes \left(\ketbrac{1}{1}\otimes \set{r}\scrap{m}\right)_{j_f}\otimes\projector{j_f-1}{j_f-1}\right)_{f,ord}\cdot\hat{\mathcal{A}}_{j_f\leftarrow j_f-1}\cdot\hat{\mathcal{S}}_{j_b\leftarrow j_b-1}, \label{splittingops}
\end{align}
adding the Hermitian conjugate and the {\tt cbX} gate
\begin{align}
a^{(n_b)\dagger}_{s,j_b}&\otimes b^{(n_f)\dagger}_{r,j_f}b^{(n_f)}_{m,j_f-1} +h.c.  = \hat{\mathcal{S}}^{\dagger}_{j_b\leftarrow j_b-1}\cdot\hat{\mathcal{A}}^{\dagger}_{j_f\leftarrow j_f-1}\cdot \projector{n_b-j_b}{0}\otimes \left\{\ketbrac{1}{1}^{\otimes(j_b-1)}\otimes\ketbrac{0}{0}^{\otimes(n_f-j_f)}\right\}\nonumber \\
& \left\{\text{\tt cbX}^{\dagger}_{j_b}(s)\cdot\left[\left(\ketbrac{1}{0}\otimes\set{s}\right)_{j_b}\otimes\left(\innerid^{\otimes(n_f-j_f)}\otimes \left(\ketbrac{1}{1}\otimes \set{r}\scrap{m}\right)_{j_f}\otimes\projector{j_f-1}{j_f-1}\right)_{f,ord}+h.c.\right]\cdot\text{\tt cbX}_{j_b}(s)\right\}\nonumber\\
& \cdot\hat{\mathcal{A}}_{j_f\leftarrow j_f-1}\cdot\hat{\mathcal{S}}_{j_b\leftarrow j_b-1},
\end{align}

To avoid  duplication of any fermion momenta and ensure the appropriate ordering of fermion registers upon rotation between states $\ket{\Omega}_{j_b}\ket{1m}_{j_f}$ and $\ket{s}_{j_b}\ket{1r}_{j_f}$ (simultaneously with creation of the new boson, it behooves one to check that the fermion with changed momentum is not yet in the memory) we use the following algorithm:
\begin{enumerate}
\item Apply $\hat{\mathcal{A}}_{j_f\leftarrow j_f-1}$ to unpack from the antisymmetric state  the largest momentum of the fermion memory and move it to the last-used register $j_f$; then apply $\text{\tt fS}_{j_f}(L,m)$ if register $j_b$ is empty and $\text{\tt fS}_{j_f}(L,r)$ if it is occupied.
\item Apply $\text{\tt fS}(r,m)$ if register $j_b$ is occupied or $\text{\tt fS}(m,r)$ if it is empty. In this way,  $\ket{\Omega}_{j_b}\ket{1m}_{j_f}$, originated from step \textit{1.} changes to  $\ket{\Omega}_{j_b}\ket{1r}_{j_f}$ if $r$ is also on memory; similarly, $\ket{1s}_{j_b}\ket{1r}_{j_f}$ changes to $\ket{1s}_{j_b}\ket{1m}_{j_f}$ if $m$ is also on memory.
\item Apply  to registers $j_b$ and $j_f$ the two-register ``splitting'' rotation, that attempts to rotate between $\ket{\Omega}_{j_b}\ket{1m}_{j_f}$ and $\ket{1s}_{j_b}\ket{1r}_{j_f}$; the rotation is thus performed provided $r$ is not already on memory when register $j_b$ is empty; but when it is occupied, if $m$ is not in memory. This avoids the situations in which momenta duplication would occur.
\item Undo the first two {\tt fS}s.
\end{enumerate}
As previously for the fermion and boson {\tt SWAP}s (see \ref{par:fermion-SWAP} and \ref{par:bSWAP-3reg}), the algorithm can be written in terms of fermion-swap-gate controlled multiplications; for the first two steps:
\begin{equation}
\text{\tt fSp}_{(j_b,j_f)}(m,r)=\left(\ketbrac{0}{0}\right)_{j_b}\otimes\left[\text{\tt fS}_{j_f}(m,r)\cdot \text{\tt fS}_{j_f}(L,m)\right]+\left(\ketbrac{1}{1}\right)_{j_b}\otimes\left[\text{\tt fS}_{j_f}(r,m)\cdot \text{\tt fS}_{j_f}(L,r)\right],
\label{def:fSp}
\end{equation}
returning to the splitting operator of Eq.~(\ref{splittingops}), and including its (boson absorption) Hermitian conjugate operator,
\begin{align}
a^{(n_b)\dagger}_{s,j_b}&\otimes b^{(n_f)\dagger}_{r,j_f}b^{(n_f)}_{m,j_f-1} +h.c.  = \hat{\mathcal{S}}^{\dagger}_{j_b\leftarrow j_b-1}\cdot\hat{\mathcal{A}}^{\dagger}_{j_f\leftarrow j_f-1}\cdot \text{\tt cbX}^{\dagger}_{j_b}(s)\cdot\text{\tt fSp}^{\dagger}_{(j_b,j_f)}(m,r)\cdot \nonumber\\
&\projector{n_b-j_b}{0}\otimes\left\{\ketbrac{1}{1}^{\otimes(j_b-1)}\otimes\ketbrac{0}{0}^{\otimes(n_f-j_f)}\right\}\left\{\left[\left(\ketbrac{1}{0}\otimes\set{s}\right)_{j_b}\otimes\left(\ketbrac{1}{1}\otimes \set{r}\scrap{m}\right)_{j_f}+h.c.\right]\right\}\otimes\projector{j_f-1}{j_f-1}\nonumber\\
& \cdot \text{\tt fSp}_{(j_b,j_f)}(m,r)\cdot\text{\tt cbX}_{j_b}(s)\cdot\hat{\mathcal{A}}_{j_f\leftarrow j_f-1}\cdot\hat{\mathcal{S}}_{j_b\leftarrow j_b-1},
\end{align}
where the $\innerid$ operators on fermion and boson registers have been omitted. The exponentiation of this term is now straightforward
\begin{align}
\mathcal{U}_{21}&(\Delta t, \lambda)  =  \prod_{j_b,j_f=1} \left(\right.\mathbb{I}^{(n)}-\projector{n_b-j_b}{0}\otimes\mathbb{I}\otimes\projector{j_b-1}{j_b-1}\otimes\projector{n_f}{j_f}\nonumber \\
& 
\left.+\hat{\mathcal{S}}^{\dagger}_{j_b\leftarrow j_b-1}\cdot\hat{\mathcal{A}}^{\dagger}_{j_f\leftarrow j_f-1}\cdot\prod_{s}\left[\bXdg{j_b}{s}\cdot\hat{\mathfrak{U}}_{21,(j_b,j_f)}\left(s;\lambda,\Delta t\right)\cdot\bX{j_b}{s}\right]\cdot \hat{\mathcal{A}}_{j_f\leftarrow j_f-1}\cdot\hat{\mathcal{S}}_{j_b\leftarrow j_b-1}\right) + \bigO{\Delta t^2}
\label{eq:splitting-3}
\end{align}
where the auxiliary operator $\hat{\mathfrak{U}}_{21,(j_b,j_f)}\left(s\right)$ is
\begin{align}
\hat{\mathfrak{U}}_{21,(j_b,j_f)}&\left(s;\lambda,\Delta t\right) =    \prod_{r}\left(\text{\tt fSp}^{\dagger}_{(j_b,j_f)}(m,r)\right.\nonumber \\
&\left.\cdot\projector{n_b-j_b}{0}\otimes\left\{\ketbrac{1}{1}^{\otimes j_b-1}\otimes\ketbrac{0}{0}^{\otimes j_f+1}\right\}\left\{\mathfrak{U}_{21,(j_b,j_f)}\left(r,s;\lambda,\Delta t\right) \right\}\otimes\projector{j_f-1}{j_f-1}\cdot\text{\tt fSp}_{(j_b,j_f)}(m,r)\right).
\label{eq:splitting-4}
\end{align}

\begin{figure}[!ht]
\centering
\includegraphics[scale=1]{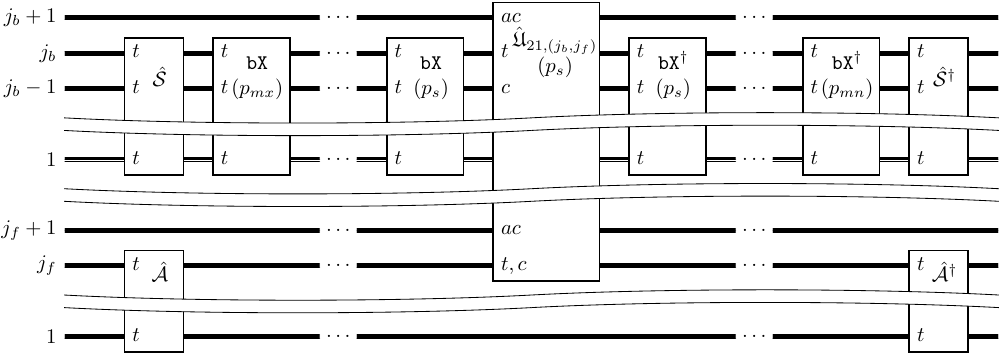}
\caption{Circuit implementation of $\mathcal{U}_{21}(\Delta t, \lambda)$ for fixed $j_b$ and $j_f$ with exact conservation of fermion antisymmetry, given by Eq.~(\ref{eq:splitting-3}). See Fig.(\ref{circuit:free-evolution}) for an explanation of the $t$, $c$ and $ac$ symbols.\label{circ:splitting-3}}
\end{figure}

\begin{figure}[!ht]
\centering
\includegraphics[scale=1]{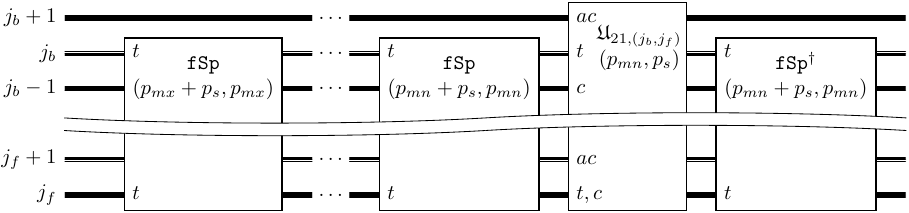}
\caption{Circuit implementation of $\mathfrak{U}_{21,(j_b,j_f)}\left(p_s;\lambda,\Delta t\right)$ given by Eq.~(\ref{circuit:splitting-2}), subroutine of $\mathcal{U}_{21}(\Delta t, \lambda)$. See Fig.(\ref{circuit:free-evolution}) for an explanation of the $t$, $c$ and $ac$ symbols.\label{circ:splitting-4}}
\end{figure}

\textbf{Gate-level implementation and costs.} Eq.~(\ref{eq:splitting-3}) and Eq.~(\ref{eq:splitting-4}) are implemented on circuits Fig.~(\ref{circ:splitting-3}) and Fig.~(\ref{circ:splitting-4}) respectively. The gate count scales as follows, for each $j_b$ and $j_f$ on Eq.~(\ref{eq:splitting-3})
\begin{itemize}
    \item $\bigO{1}$ $\hat{\mathcal{S}}$, $\hat{\mathcal{S}}^{\dagger}$, $\hat{\mathcal{A}}$, $\hat{\mathcal{A}}^{\dagger}$, with a total costs of $\bigO{j_b\log_2 N_p+j_f\log_2 N_p}$  {\tt CNOT}, single-qubit gates and $\bigO{j_b^2+j_f^2}$ order-oracle calls
    \item $\bigO{N_p}$ {\tt cbX} gates, with a total costs of $\bigO{ j_b N_p\log_2 N_p}$ {\tt CNOT} and single-qubit gates + $\bigO{j_b^2 N_p }$ order-oracle calls 
    \item $\bigO{N_p}$ $\mathfrak{U}_{21,(j_b,j_f)}$, each composed of 
    \begin{itemize}
        \item $\bigO{N_p}$ ${\tt fSp}$, in total a cost of $\bigO{j_f N_p  \log_2 N_p}$ {\tt CNOT} and single-qubit gates + $\bigO{j^2_f N_p }$ order-oracle calls.
        \item $\bigO{N_p}$ $\mathfrak{U}_{21,(j_b,j_f)}(p_r,p_s)$, implemented as in subsection \ref{subsubsec:splitting-1}, but with a change of basis both dependent of $p_s$ and $p_k$: $T_{p_s,p_k}$. The cost of each of term is $\bigO{N_p\log_2 N_p}$, giving a total count of $\bigO{N^2_p \log_2 N_p}$.
    \end{itemize}
    Thus the subcircuit costs are $\bigO{N_p^3\log_2 N_p}$ {\tt CNOT} and single-qubit gates  + $\bigO{j_f^2 N^2_p }$ order-oracle calls. 
\end{itemize}
Summing over the $n_f$ values of $j_f$ and $n_b$ values of $j_b$ the dominant costs are $\bigO{ n_f n_b N_p^3 \log_2 N_p}$ {\tt CNOT} and single-qubit gates + $\bigO{ n_f^3 n_b N^2_p}$ order-oracle calls.

This concludes our exemplification of typical terms in a field theoretical Hamiltonian written in the particle basis. Further operators can be encoded as needed following the same principles.

\section{Compilation of results and discussion}
\label{sec:rec&disc}
We are aware that this report makes for a taxing read, given the detailed implementation and discussion, so it is worth providing a clear and concise final overview of what has been achieved. 

In this section we recapitulate the most important operators discussed and their cost 
in memory, computer time surrogated to the number of $R_z$, {\tt CNOT} and Hadamard gates (see Table~\ref{table:implementation-costs-main}), and we offer a few final comments.

\subsection{Memory scalability}

In the first place we observe that the memory requirements are dominated by the momentum discretization; to codify a number of momenta $\displaystyle N_p = \Lambda_{max}-\Lambda_{min} +1 = 2\Lambda + 1$  we need $\left\lceil\log_2\left(N_p\right)\right\rceil$ qubits; with a maximum number of $n$ particles (registers), we have 
\begin{equation}
    \text{\# qubits} = \bigO{n\times \left\lceil\log_2\left(N_p\right)\right\rceil},
\end{equation}
the other quantum numbers of interest (color, spin, flavor, etc) have a fixed and small memory requirement, so they add a subleading $\bigO{n}$ qubits, as do the qubits allotted for presence/absence controls. 
We then proceed to the more complicated counting of the number of gates.

\subsection{Gate costs}
 \label{subsec:costs}
Tables~\ref{table:implementation-costs-main} and~\ref{table:implementation-costs-others}
compile some key operators developed through the manuscript, noting the section where they are defined and their implementation cost in terms of {\tt CNOT}, single-qubit gates and order-oracle calls.
\begin{table}[ht]
\centering
\bgroup
\def\arraystretch{1.5}
\begin{tabular}{|c|ccc|} \hline
\multirow{2}{*}{Operator} & \multicolumn{3}{c|}{Costs}                                                 \\ \cline{2-4} 
                          & \multicolumn{1}{c|}{{\tt CNOT} \& single-qubit}             & \multicolumn{1}{c|}{Order oracle}                  & Section                                   \\ \hline
$\mathcal{U}_{11}$        & \multicolumn{1}{c|}{$\bigO{n N_p}$}                         & \multicolumn{1}{c|}{None}                          & \ref{subsec:free}                         \\
$\mathcal{U}_{22}$        & \multicolumn{1}{c|}{$\bigO{n^2N_p^3\log_2^2 N_p}$}            & \multicolumn{1}{c|}{None}                          & \ref{subsec:pexchange}                    \\
$\mathcal{U}_{10}$        & \multicolumn{1}{c|}{$\bigO{N_p\log_2^2 N_p}$}  & \multicolumn{1}{c|}{$\bigO{n^3N_p}$}               & \ref{subsec:7.1.2}                        \\
$\mathcal{U}_{21}$        & \multicolumn{1}{c|}{$\bigO{n_b n_f^2 N_p^2 \log_2^2 N_p}$}    & \multicolumn{1}{c|}{$\bigO{n_b^3N_p}$}             & \ref{subsubsec:splitting-1}               \\ $\mathcal{U'}_{21}$       & \multicolumn{1}{c|}{$\bigO{n_b n_f  N_p^3 \log_2 N_p}$}       & \multicolumn{1}{c|}{$\bigO{n_b  n_f^3 N^2_p}$}     & \ref{subsubsec:splitting-2}               \\ \hline         
\end{tabular}
\egroup
\caption{
{\tt CNOT} and single-qubit gates costs of implementing one step of the unitary evolution (with associated Hamiltonian terms $\mathcal{H}_{ij}$),  $i$ being the number of outgoing particles and $j$ that of incoming ones (or viceversa). The variable  $n=n_f + n_b$ includes $n_f$, the number of fermion registers and $n_b$, the number of boson registers.  $N_p$ stands for the number of different values of the momentum.\label{table:implementation-costs-main}}
\end{table}

\begin{table}[ht]
\centering
\bgroup
\def\arraystretch{1.5}
\begin{tabular}{|c|ccc|} \hline
\multirow{2}{*}{Operator} & \multicolumn{3}{c|}{Costs}                                                 \\ \cline{2-4} 
                          & \multicolumn{1}{c|}{{\tt CNOT} \& single-qubit}             & \multicolumn{1}{c|}{Order oracle}                  & Section                                   \\ \hline
$\hat{\mathcal{A}}_{j\leftarrow j-1}$        & \multicolumn{1}{c|}{ $\bigO{j^2 + j\log_2 N_p}$}                         & \multicolumn{1}{c|}{$\bigO{j^2}$}                          & \ref{par:antisymmetrizers}                      \\
$\fSWAP{j}{p}{q}$ /  $\fSWAP{j}{p}{L}$       & \multicolumn{1}{c|}{ $\bigO{j\log_2 N_p}$ / $\bigO{j^2+j\log_2 N_p}$}            & \multicolumn{1}{c|}{None / $\bigO{j^2}$}                          & \ref{par:fermion-SWAP}                    \\
$\fX{j}{p}$        & \multicolumn{1}{c|}{$\bigO{j^2+j\log_2 N_p}$}  & \multicolumn{1}{c|}{$\bigO{j^2}$}               & Eq.~(\ref{def:fTd})                        \\
$\fSp{(j_b,j_f)}{m}{r}$        & \multicolumn{1}{c|}{$\bigO{j_f^2+j_f\log_2 N_p}$}    & \multicolumn{1}{c|}{$\bigO{j_f^2}$}             & Eq.~(\ref{def:fSp}) \\ \hline
$\hat{\mathcal{S}}_{j\leftarrow j-1}$       & \multicolumn{1}{c|}{$\bigO{j^3\log_2 j+j\log_2 N_p}$}       & \multicolumn{1}{c|}{$\bigO{j^2}$}     & \ref{par:symmetrizer-3reg} \& \ref{par:symmetrizer-general}              \\
$\bSWAP{j}{p}{q}$ /  $\bSWAP{j}{p}{L}$      & \multicolumn{1}{c|}{$\bigO{j^3\log_2 j+j\log_2 N_p}$}       & \multicolumn{1}{c|}{None / $\bigO{j^2}$}     &\ref{par:bSWAP-3reg} \& \ref{general-boson-operators}             \\
$\bX{j}{p}$       & \multicolumn{1}{c|}{$\bigO{j^3\log_2 j+j\log_2 N_p}$}       & \multicolumn{1}{c|}{$\bigO{j^2}$}     & Eq.~(\ref{def:cbX})               \\ \hline         
\end{tabular}
\egroup
\caption{
{\tt CNOT} and single-qubit gate costs of implementing various operators introduced in section \ref{sec:evolution2};  $j$ is the number of active (occupied) registers and $N_p$ the number of different momenta. The fermion and boson {\tt SWAP}s can be implemented in a cheaper manner when the largest momentum $L$ is not interchanged. \label{table:implementation-costs-others}}
\end{table}

A couple of observations are in order. First, let us note that all the costs listed are at most polynomial in the number of active particles $j$, the size of the momentum discretization implemented $N_p$ and the maximum number of registers $n$. While many algorithms can surely be improved, no exponential-cost hurdles appear. This is plausible in theories where the interactions depend on a finite number of registers (for example, in Chromodynamics the largest number is in the four-gluon vertex; in chiral perturbation 
 theory, the three-body interactions appear at NNLO; there is a large class of few-body problems for which this is the case).
The second thing to note is that, though we need propagators which act on all qubits of the memory, if the number of particles is not saturated ($j<n$) the additional, empty registers, cost little, as the various gates over different combinations of registers can run in parallel.

It is interesting to compare the efficiency of this implementation of a second quantized theory with other traditional approaches, to which we now dedicate a paragraph.

\subsection{Comparison with JW transformation \label{sec:comparison with a JW transformation}}
A word of comparison with a benchmark method is now in order, and when dealing with second quantization, the Jordan-Wigner transformation comes to the fore.
An encoding of a physical problem in a digital quantum computer (by either technique) consists of a mapping between a given finite Hilbert space (a truncation of the full original one as needed) and the quantum computer's memory:
\begin{equation}
    J: H\rightarrow (B^{1/2})^{\otimes n},
\end{equation}
with $B^{1/2}$ the Hilbert space of a qubit and $n$ some number which scales with the dimension of $H$. The states in $H$ should be wisely chosen to represent the different configurations of the system to be simulated. For example, quantum chemistry studies commonly use the STO-nG atomic orbital basis \cite{seeley_12085986_2012}, nuclear shell-model simulations are based on the Slater determinant basis \cite{perez-obiol_nuclear_2023}, whereas for a quark model using the Cornell potential, generalized Laguerre polynomials are enough \cite{gallimore_quantum_2023}, at least for heavy quarks. The common feature of these simulations is that a small number of basis functions or modes are chosen to represent the system, each associated with creation and annihilation operators. Single-particle states are then
\begin{equation}
    \ket{\psi_i} = b^{\dagger}_{i}\ket{0},
\end{equation}
while two-particle states are
\begin{align}
    \ket{\psi_{ij}}=\pm\ket{\psi_{ji}}= b^{\dagger}_{i}b^{\dagger}_{j}\ket{0},
\end{align}
etc. The fermionic or bosonic nature of the states is expressed as the relative sign between $\ket{\psi_{ij}}$ and $\ket{\psi_{ji}}$. Fermions carry a negative sign, which is a way of considering the anticonmutation relations $ \left\{b^{\dagger}_i,b^{\dagger}_j\right\} = \left\{b_i,b_j\right\} = 0,
$
and 
$
    \left\{b_i,b^{\dagger}_j\right\} =\delta_{ij}
$.
Since $\ket{\psi_{ij}}=-\ket{\psi_{ji}}$, states of the form $\ket{\psi_{ii}}$ are excluded, which means that \textit{each mode is either occupied or empty}. A general fermionic state can therefore be written as 
\begin{equation}
    \ket{f_n,...,f_1} = b^{\dagger f_n}_{n}...b^{\dagger f_2}_{2}b^{\dagger f_1}_{1}\ket{0},
\end{equation}
with $f_i\in{0,1}$, $b^{\dagger 0}_{n} = 1$ and $b^{\dagger 1}_{n} = b^{\dagger}_{n}$. For the annihilation operators
\begin{equation}
    b_i|f_n,...,\underset{i}{\underbrace{1}},...,f_1\rangle = (-1)^{\sum^{n}_{j=i+1}f_j}|f_n,...,\underset{i}{\underbrace{0}},...,f_1\rangle,
\end{equation}
the action of $b_i$ is therefore non-local in this basis since it requires a sign that depends on the occupation state of modes with $j>i$.
The Jordan-Wigner mapping identifies the occupation-number basis and that of the direct product of $n$ qubits:
\begin{equation}
    \ket{f_n,...,f_1}\in H \rightarrow J(\ket{f_n,...,f_1}) = \ket{f_n}\otimes ... \otimes \ket{f_1}\in (H^{1/2})^{\otimes n},
\end{equation}
so that the creation and annihilation operators are written as
\begin{align}
    b^{\dagger}_j & = \mathfrak{i}^{\otimes n-j-1}\otimes\sigma^{-j}\otimes Z^{\otimes j-1}\\
    b_j & = \mathfrak{i}^{\otimes n-j-1}\otimes\sigma^{+j}\otimes Z^{\otimes j-1}
\end{align}
which are also non-local. Some improvements are possible in basis such as the one introduced by Bravyi and Kitaev \cite{bravyi_fermionic_2002}, with even indexed qubits codifying occupation numbers and odd indexed ones encoding some partial sums $y_i = \sum^{n}_{j=i+1}f_j$, for more detailse see \cite{seeley_12085986_2012,bravyi_fermionic_2002}. 

The codification therefore uses $N$ qubits to represent $N$ possible single-particle states: the entire Fock space is encoded, from the vacuum to the state obtained from application of all creation operators, i.e, from the $0$-particle state to the $N$-particle state. In contrast, the particle-register codification uses $\mathcal{O}(n\log N)$ qubits, with $n$ the maximum number of particles to be codified and $N$ the number of single-particle states.

It falls off that an implementation based on registers encoding particles is not economical when the number of such particles is high compared to single-particle modes \cite{Barata:2020jtq}. The necessity of ensuring either Bose-Einstein or Fermi-Dirac statistics imposes tight constraints, and renders unphysical most combinations of single-particle states which are encodable, thus increasing the unused fraction of the Hilbert space with the number of particles. For example, using two qubits per particle-register to encode momenta allows for up to four different values; composing these registers to generate up to three-particle states will therefore involve six qubits in total. Of the $2^6$ possible orthogonal states, only the antisymmetric combinations are physical so we have: $\binom{4}{1}=4$ single-particle states,  $\binom{4}{2}=6$ two-particle states and $\binom{4}{3}=4$ three-particle states. A direct Jordan-Wigner transformation or akin may thus be preferable on the grounds of memory alone when the number of momenta is very small, so that the logarithmic advantage of the register encoding is not yet manifest: with the same qubits they are able to represent $\binom{6}{1}=6$ single-particle states, $\binom{6}{2}=15$ two-particle states, $\binom{6}{3}$ = 20 three-particle states, etc.
For larger $N$ but moderate $n$, the JW encoding is inferior to the register encoding.

\subsection{Final comments} \label{sec:conclusions}
We have proposed a transparent encoding of canonical second quantization of quantum theory for digital quantum computers featuring standard gates. 

In traditional field theory much of the emphasis concentrates on the normal modes $a$, $a^\dagger$, and we have striven to provide a straightforward construction thereof in terms of set/scrap $\mathfrak{s}^\dagger$, $\mathfrak{s}$ quantum-number operators as well as memory-control operators $\mathfrak{C}_{ij}$
that allow us to select read-write memory registers as needed.

Still, in quantum computation all operations are unitary, so that it is the exponentiatioon of the creation-destruction algebra which takes the central role, so we have provided detailed representations of all needed operations. 

Much of the work has been devoted to describing an adequate implementation of the (anti)symmetrization of identical quanta and the bookkeeping necessary to carry out the usual operations (free evolution, particle splitting or annihilation, momentum exchange, etc). Note that explicit symmetrization is essential for our implementation; as seen explicitly on Eqs.~(\ref{def:nReg-bosoncreation}) and Eq.(\ref{def:nReg-fermioncreation}), the creation and annihilation operators are written as sums of terms changing only the last register, and for this to work all active modes must ``be'' somehow on the last register. Of course, this has his drawbacks, and we have struggle on section \ref{sec:evolution2} to give an implementation of particle-changing operators while keeping ``local'' transformations, for the Pauli exclusion principle and the creation criterion \ref{criteria:creation} are conditions over the entire memory.

All standard field theory requirements such as commutation relations are satisfied up to boundary terms which arise due to the finite computer memory. In future work we would like to address the necessary (Wilsonian-like) renormalization that arises from varying the memory size. 
This implies that certain symmetries will only be recovered in the limit of an infinite memory (saliently continuous spacetime symmetries, for which the situation is identical to that in established lattice gauge theory).
Internal flavor and color symmetries are exactly implemented in our encoding.

Modern physics circumstances in which having this encoding at hand are many, among them the many-body problem in nuclear physics \cite{Ring}, in hadron physics~\cite{Cotanch:2001mc,Llanes-Estrada:1999nat}, in condensed-matter physics and others~\cite{FetterWalecka}.

\clearpage

\section*{Acknowledgments}
Supported by spanish MICINN under grant number PID2022-137003NB-I00; EU’s 824093 (STRONG2020);  
and Universidad Complutense de Madrid under research group 910309 and the IPARCOS institute.

\appendix

\section{Algorithms}\label{sec:Useful identities}
Generalizations and descriptions of the algorithms mentioned in the main text are collected here for completeness. 

\subsection{``Locate the Largest'' (LL) algorithm}
\label{miscellanea:LL algorithm}

We here address the algorithm which locates the largest momentum $P$ encoded in the memory (in the sense of their previously chosen ordering) and brings it to the front for further operation, taking into account that it may be repeated $m$ times if handling bosons.  

Given a set of $n$ registers in the state $\ket{p_{n},p_{2},...,p_{1}}$ with modes on registers $k_1,k_2,...,k_m$ all equal to $P$ with the largest index, and an auxiliary register with $n$-qubits initially in the 0-ket, the algorithm performs the transformation
\begin{equation}
   \ket{0,...,0}\ket{p_{n},...,p_{1}}\overset{\text{\tt LL}}{\rightarrow}  |0...,\underset{k_m}{\underbrace{1}},...,\underset{k_2}{\underbrace{1}},...,\underset{k_1}{\underbrace{1}},...,0\rangle|p_{n},...,\underset{k_m}{\underbrace{P}},...,\underset{k_2}{\underbrace{P}},...,\underset{k_1}{\underbrace{P}},...,p_{1}\rangle,
\end{equation}
which is obviously reversible. The order relation between states is introduced via an oracle, using the order criteria of \ref{subsubsec:tadpoles}:
\begin{eqnarray}
    O(\ket{x}\ket{p_i}\ket{p_j}) = \begin{cases}
        \ket{x\oplus 1}\ket{p_i}\ket{p_j}&\text{if } i\geq j\\
        \ket{x}\ket{p_i}\ket{p_j}&\text{otherwise\ .}
    \end{cases}
\end{eqnarray}

We now provide a detailed step-by-step description of the algorithm for a three-particle memory with state $\ket{p,q,k}$, and afterwards, but only briefly, discuss its generalization to an arbitrary number of registers.

\begin{itemize}
    \item \textbf{Introduction of auxiliary registers.} Introduce three auxiliary registers: The ``mark'' register, $\mathcal{A}$, with the same number of qubits as the modes in the initial state; the comparison register, $B$, with one qubit less, all in state $0$ and a one qubit-register $C$:
    \begin{equation}
        \ket{0}_C\ket{00}_B\ket{000}_{\mathcal{A}}\ket{p,k,q}
    \end{equation}
    \item \textbf{Third-position test.} Test whether the largest mode is on register $3$:
    \begin{enumerate}
        \item \textbf{Comparison of 3rd and 2nd registers.} Use the oracle with qubit C and the 3rd and 2nd registers of $\ket{p,k,q}$, then, anticontrolling on the state of C, initialize qubit $b_0$ and finally, uncompute C. The result is
        \begin{equation}
        \begin{cases}
            \ket{0}_C\ket{01}_B\ket{000}_{\mathcal{A}}\ket{p,k,q} &\text{if } p\geq k\\
            \ket{0}_C\ket{00}_B\ket{000}_{\mathcal{A}}\ket{p,k,q} &\text{if } p<k
        \end{cases}   
        \end{equation}
        \item \textbf{Comparison of 3nd and 1st registers.} Repeat the process but initialize qubit $b_1$ instead of $b_0$:
        \begin{equation}
        \begin{cases}
            \ket{0}_C\ket{11}_B\ket{000}_{\mathcal{A}}\ket{p,k,q} &\text{if } p\geq k, p \geq q\\
            \ket{0}_C\ket{10}_B\ket{000}_{\mathcal{A}}\ket{p,k,q} &\text{if } p\geq k, p<q\\
            \ket{0}_C\ket{01}_B\ket{000}_{\mathcal{A}}\ket{p,k,q} &\text{if } p<k, p\geq q\\
            \ket{0}_C\ket{00}_B\ket{000}_{\mathcal{A}}\ket{p,k,q} &\text{if } p<k, p<q\\
        \end{cases}   
        \end{equation}
        \item \textbf{End test.} Use the retrieved information to establish whether the momentum on the third register is one of the largest, which is the case only if the state of B is $\ket{11}_B$. Uncompute $B$.
        \begin{equation}
        \begin{cases}
            \ket{0}_C\ket{00}_B\ket{100}_{\mathcal{A}}\ket{p,k,q} &\text{if } p\geq k, p\geq q\\
            \ket{0}_C\ket{00}_B\ket{000}_{\mathcal{A}}\ket{p,k,q} &\text{otherwise} 
        \end{cases}   
        \end{equation}
    \end{enumerate}
    \item \textbf{Second-position test.} Repeat the procedure but now, in the final step, change the second qubit of A instead of the third
            \begin{equation}
        \begin{cases}
            \ket{0}_C\ket{00}_B\ket{100}_{\mathcal{A}}\ket{p,k,q} &\text{if } p>k, p\geq q\\
            \ket{0}_C\ket{00}_B\ket{010}_{\mathcal{A}}\ket{p,k,q} &\text{if } k>p, k\geq q\\
            \ket{0}_C\ket{00}_B\ket{110}_{\mathcal{A}}\ket{p,k,q} &\text{if } p=k \geq q \\
        \ket{0}_C\ket{00}_B\ket{000}_{\mathcal{A}}\ket{p,k,q} &\text{if     
} q>k,q>p
        \end{cases}   
        \end{equation}
\item \textbf{Third-position test.} Finally, repeat for the third qubit of $A$
            \begin{equation}
        \begin{cases}
            \ket{0}_C\ket{00}_B\ket{100}_{\mathcal{A}}\ket{p,k,q} &\text{if } p>k, p>q\\
            \ket{0}_C\ket{00}_B\ket{010}_{\mathcal{A}}\ket{p,k,q} &\text{if } k>p, k>q\\
        \ket{0}_C\ket{00}_B\ket{001}_{\mathcal{A}}\ket{p,k,q} &\text{if     
} q>k,q>p\\
            \ket{0}_C\ket{00}_B\ket{110}_{\mathcal{A}}\ket{p,k,q} &\text{if } p=k > q\\
            \ket{0}_C\ket{00}_B\ket{101}_{\mathcal{A}}\ket{p,k,q} &\text{if } p=q > k\\            
            \ket{0}_C\ket{00}_B\ket{011}_{\mathcal{A}}\ket{p,k,q} &\text{if } k=q > p\\
            \ket{0}_C\ket{00}_B\ket{111}_{\mathcal{A}}\ket{p,k,q} &\text{if } k=q = p\\            
        \end{cases}   
        \end{equation}
\end{itemize}
The generalization of the algorithm is straightforward, for $n$ initial modes,  register A has $n$ qubits, register B $n-1$ qubits and register C still one, and $n$ position tests are necessary, each one requiring:
\begin{itemize}
    \item $\bigO{n}$ comparisons $\rightarrow\bigO{n}$ oracle calls and $\bigO{n}$ {\tt CNOT}s.
    \item An end test: $\bigO{n}$ Toffoli gates for a general control $\rightarrow\bigO{n}$ {\tt CNOT}s and single-qubit gates.
    \item Total: $\bigO{n}$ oracle calls + $\bigO{n}$ {\tt CNOT}s + $\bigO{n}$ single-qubit gates.
\end{itemize}
Thus, the algorithm requires at most $\bigO{n^2}$ oracle calls {\tt CNOT}s and single-qubit gates.

\subsection{Generalization of symmetrizers and boson-{\tt SWAP}s. \label{general-boson-operators}}
This subsection, introducing two useful auxiliary operators, is divided in three parts.
We start with a subroutine that is used by both of those operators.

\paragraph{Symmetric-Digit-Decomposition ``{\tt SDD}'' algorithm.} \ \ \ This is useful in the following two algorithms: for step \textit{5} in the implementation of the generalized symmetrizer presented shortly, and for step \textit{4} of the boson-{\tt SWAP}. Suppose a register $B$ of $n$ qubits in the state $\ket{m}_B$ and call $m_1$,$m_2$,...,$m_k$ the binary digits with only one qubit in state $\ket{1}$, so that $m=m_1+m_2+...+m_k$ is the decomposition of $m$ in powers of 2 (e.g. if $m=101$, we would have $m_1=001$ and $m_2=100$), then we seek an algorithm that efficiently performs the transformation
\begin{equation}
    \ket{0,...,0}_C\ket{m}_B\rightarrow\frac{1}{\sqrt{k}}\left(\ket{m_1}_C+\ket{m_2}_C+...+\ket{m_k}_C\right)\ket{m}_B\ .
\end{equation}
 If $n$ is a small number, we can simply control on each of the possibilities of $m$ and perform the corresponding transformation, as in \ref{subsubsec:boson-tadpole}, but the possible values of $m$ grow as $2^n$ with the number of digits $n$. Instead we propose the following:
\begin{enumerate}
\item Add two auxiliary registers $D$ and $E$ with $\log_2 n$ qubits
\begin{equation}
\ket{0}_C\ket{m}_B\rightarrow \underbrace{\ket{0}_E\ket{0}_D}\ket{0}_C\ket{m}_B
\end{equation}
\item Initialize $D$ with the number of qubits of B in state $1$; i.e., for each qubit of $B$ on state $1$, add one to $D$. The computing costs up to this point become: $n$ unit-adders over $\log_2 n$ qubits which cost $\bigO{n\log_2^2 n}$ {\tt CNOT} and single-qubit gates.
\begin{equation}
\ket{0}_E\ket{0}_D\ket{0}_C\ket{m}_B\rightarrow\ket{0}_E\ket{k}_D\ket{0}_C\ket{m}_B
\end{equation}
\item Each qubit $b_i$ of $B$ starting from the right may perform a controlled transformation as follows.
\begin{enumerate}
\item[$3.1$] Controlling on $b_i = 1$, on $E=\ket{l}_E>0$, on $D = \ket{k}_D$ and on qubits $c_j$ with $j<i$ (count right to left) apply the following single-qubit rotation on qubit $i$ of register $C$:
\begin{equation}
\ket{0,...,0}_C\rightarrow \sqrt{\frac{k-l}{k-l+1}}\ket{0,...,0}_C + \frac{1}{\sqrt{k-l+1}}\ket{m_i}.
\end{equation}
Costs: Registers $E$ and $D$ have  $\bigO{2\log_2 n}$ qubits in common, which generates $\bigO{n^2}$ states. For each of these we apply a controlled (over up most $\bigO{3\log_2 n}$ qubits) single-qubit rotation, which give a total cost of $\bigO{n^2\log_2 n}$ {\tt CNOT}s and single-qubit gates.
\item[$3.2$] Controlled on $b_i=1$, increase $E$ by one. Costs: $\bigO{\log_2^2 n}$ {\tt CNOT}s and single-qubit gates.
\end{enumerate}
In total, step 3 costs $\bigO{n^3\log_2 n + n\log_2^2n}$ {\tt CNOT} and single-qubit gates and generates the state
\begin{equation}
\ket{0}_E\ket{k}_D\ket{0}_C\ket{m}_B\rightarrow \ket{k}_E\ket{k}_D\frac{1}{\sqrt{k}}\left(\ket{0}_C+\ket{m_2}_C+...+\ket{m_k}_C\right)\ket{m}_B
\end{equation}
\item Using register $B$, do the transformation $\ket{0,...,,0}_C\leftrightarrow\ket{m_1}_C$ using the following procedure
\begin{enumerate}
\item[$4.1$] Controlled on $b_1=1$ and $c_2=...=c_n=0$, apply an {\tt X} gate to $c_1$
\item[$4.2$] Controlled on $b_2=1$, $b_1=0$ and $c_1=c_3=...=c_n=0$, apply an {\tt X} gate to $c_2$
\item[$4.n$] Controlled on $b_n=1$, $b_1=...=b_{n-1}=0$ and $c_1=...=c_{n-1}=0$, apply an {\tt X} gate to $c_n$
\end{enumerate}
Step $4.n$ is the most expensive, requiring a controlled {\tt X} gate  over $2n$ qubits, which can be decomposed in $\bigO{n^2}$ {\tt CNOT}s and single-qubit gates. Since there are $n$ steps, this gives a total cost of $\bigO{n^3}$ {\tt CNOT}s and single-qubit gates and produces the desired state
\begin{equation}
\ket{k}_E\ket{k}_D\frac{1}{\sqrt{k}}\left(\ket{0}_C+\ket{m_2}_C+...+\ket{m_k}_C\right)\ket{m}_B\rightarrow\ket{k}_E\ket{k}_D\frac{1}{\sqrt{k}}\left(\ket{m_1}_C+\ket{m_2}_C+...+\ket{m_k}_C\right)\ket{m}_B
\end{equation}
\item Return registers $E$ and $D$ to their initial states. Costs: $\bigO{\log_2 n}$ {\tt CNOT} and single-qubit gates to uncompute $E$ from $D$ and $\bigO{n \log_2^2 n}$ {\tt CNOT}s and single-qubit gates to uncompute $D$ from $B$. 
\end{enumerate}
The algorithm in total requires $\bigO{n^3\log_2 n}$ {\tt CNOT}s and single-qubit gates. $\blacksquare$

\paragraph{Generalization of the symmetrizer $\hat{\mathcal{S}}^{\dagger}_{j\leftarrow j-1}$. \label{par:symmetrizer-general}}\ \ \  Consider a quantum memory $A$ with $j$ occupied registers, each composed of $n_p=\log_2 N_p$ qubits, symmetric under the interchange of any two registers from $1$ to $j-1$. Consider also that the largest mode sits on register $j$ and that it is repeated $k$ times. The algorithm proceeds as follows:
\begin{enumerate}
\item Add two auxiliary registers $C$ and $B$ of $j$ qubits. Initialize register $C$ to $\ket{1,0,...,0}_C$ and register $B$ to $\ket{0,...,0}_B$. 
\item Apply the {\tt LL} algorithm to the first $j$ registers of $A$ and store the locations of the largest modes on register $B$. This entangles $B$ and $A$. Costs: $\bigO{j^2}$ oracle calls, {\tt CNOT}s and single-qubit gates.
\item Symmetrize register $C$. Costs: $\bigO{\log_2 j}$ single-qubit gates.
\item Permute the $j$th-mode register of $A$ and the $j$th qubit of $B$ with that pair of mode-register of $A$ and qubit of $B$ marked by the qubits on $C$. This applies the different permutations that defines the symmetrizer Eq.~(\ref{def:step-symmetrizer}), so the quantum memory $A$ ends up being symmetric, but entangled now also with $C$. To analyze its costs note that register $C$ has $j$ qubits, each one used as control for a two mode-register {\tt SWAP}, which costs  $\bigO{n_p}$ {\tt Toffoli} gates, in total decomposed in $\bigO{n_p}$ {\tt CNOT}s and single-qubit gates. This gives a total cost of $\bigO{j n_p}$ {\tt CNOT}s and single-qubit gates. The costs of swapping two qubits on $B$ are negligible compared with the cost of swapping two mode-registers.
\item We now want to disentangle the auxiliary registers. We first uncompute $C$. For this note that step \textit{2.} leaves $B$ on a superposition of the form (we omit the corresponding registers of $A$ for conciseness)
\begin{equation}
|1,(0,...,0,\underset{k-1}{\underbrace{1}},...,1)_S\rangle_B
\end{equation}
where the superposition is symmetric under the interchange of the qubits inside the parenthesis. The permutations of step 4. just applies the symmetrizer to this state (and the mode-registers), mixing the first $1$ in all the remaining positions (and symmetrizing the $j$th register mode). If there is already a $1$ in that position, the permutation leaves the corresponding state on the superposition of $B$ invariant. Consider for example $|1,0,...,0,\underset{k-1}{\underbrace{1}},...,1\rangle_B$, then, grouping all the states of $C$ that corresponds to a permutation which leaves that state invariant we have
\begin{small}
\begin{equation}
\frac{1}{\sqrt{k}}\left(\ket{1,0,...,0}_C+|0,..,\underset{k-1}{\underbrace{1}},...,0\rangle _C+ ...+|...,0,1\rangle_C\right)|1,0,...,0,\underset{k-1}{\underbrace{1}},...,1\rangle_B,
\end{equation}
\end{small}
if instead there is a $0$ in the position of $B$ to permute, the permutation interchanges that for the $1$ of the $j$th qubit. Since $B$ is symmetric under any interchange of its first $j-1$ qubits, we can arrive to the same final state from a states of $B$ that differ in a permutation, for example, consider 
\begin{equation}
|0,..,\underset{k}{\underbrace{1}},...,0\rangle _C|1,...,0,\underset{k-1}{\underbrace{1}},...,1\rangle_B\rightarrow |0,..,\underset{k}{\underbrace{1}},...,0\rangle _C|0,...,0,\underset{k}{\underbrace{1}},...,1\rangle_B
\end{equation} 
the same final state can be reached with another state of $C$ from an state of $B$ with some of their final 1s permuted to the $k$th position, for example:
\begin{equation}
|0,..,\underset{m}{\underbrace{1}},...,0\rangle _C|1,...,0,\underset{k}{\underbrace{1}},1,...,\underset{m}{\underbrace{0}},...,1\rangle_B\rightarrow |0,..,\underset{m}{\underbrace{1}},,...,0\rangle _C|0,...,0,\underset{k}{\underbrace{1}},...,1\rangle_B,
\end{equation} 
we can group all this possibilities as before:
\begin{small}
\begin{equation}
\frac{1}{\sqrt{k}}\left(|0,..,\underset{k}{\underbrace{1}},...,0\rangle _C+ |0,..,\underset{k-1}{\underbrace{1}},...,0\rangle_C+|0,..,\underset{k-2}{\underbrace{1}},...,0\rangle_C + ...+|...,0,1\rangle_C\right)|0,0,...,0,\underset{k}{\underbrace{1}},...,1\rangle_B,
\end{equation}
\end{small}
we see that for all possible interchanges the states of $C$ and $B$ can be grouped together as the registers $C$ and $B$ of the {\tt SDD} algorithm at the beginning of this section, thus we can apply it in reverse to arrive to 
\begin{equation}
\ket{0}_C|(1,0,...,0,\underset{k-1}{\underbrace{1}},...,1)_S\rangle_B,
\end{equation}
Register $B$ can now be disentangled using the {\tt LL} algorithm in reverse. Costs: $\bigO{j^3\log_2 j}$ {\tt CNOT}s and single-qubit gates to disentangle $C$ and $\bigO{j^2}$ order-oracle calls, {\tt CNOT}s and single-qubit gates to disentangle $B$. 
\end{enumerate} 
The total costs of the algorithm are: $\bigO{j^3\log_2 j+j\log_2 N_p}$ {\tt CNOT} and single-qubit gates + $\bigO{j^2}$ order-oracle calls. $\blacksquare$

\paragraph{Generalization of boson-{\tt SWAP} $\bSWAP{j}{p}{q}$ to allow for repeated momenta.\label{par:boson-swap}}\ \ \  Consider a quantum memory $A$ with $j$ occupied registers, each composed of $n_p = \log_2 N_p$ qubits, symmetric under the interchange of any two registers $1$ to $j-1$. To be specific, mode $p$ sits at register $j$ and it is repeated $k$ times, while mode $q$ is repeated $l$ times (this situation does not arise for fermions):
\begin{equation}
\mathcal{S}_{j-1\leftarrow 1}\left[\ket{p}_j...\ket{p}_{l+k-1}...\ket{p}_{l+1}\ket{q}_l...\ket{q}_1\right]_A,
\end{equation}
we proceed as follows:
\begin{enumerate}
\item Add three auxiliary registers with $j-1$ qubits named $B$, $C$ and $D$. Initialize $B$ with the positions of mode $q$ and $C$ with the positions of mode $p$, if $p$ is the largest mode, the {\tt LL} algorithm is applied instead. Both registers become entangled with $A$ and among themselves. Without explicitly writing register $A$, $B$ and $C$ are combined in the state
\begin{equation}
\mathcal{S}_{j-1\leftarrow 1}\left[\ket{00}_{j-1}...\ket{00}_{l+k}\ket{10}_{l+k-1}...\ket{10}_{l+1}\ket{01}_l...\ket{01}_1\right],
\end{equation} 
where the rightmost qubit  corresponds to register $B$ and the left one to register $C$. So far, the starting costs are: $\bigO{n_p}$ {\tt CNOT}s and single-qubit gates for a controlled operation over $n_p$ qubits. This is repeated $j$ times, so  $\bigO{j n_p}$ {\tt CNOT}s and single-qubit gates, and $\bigO{j^2}$ order-oracle calls, {\tt CNOT} and single-qubit gates should the {\tt LL} algorithm be applied. 
\item Register $B$ and the {\tt SDD} algorithm from the first paragraph of this section are used to initialize register $D$:
\begin{align}
\mathcal{S}_{j-1\leftarrow 1}&\frac{1}{\sqrt{l}}\left(|0,...,\underset{l}{\underbrace{1}},...,0\rangle_D+...+\ket{0,...,1,0}_D+\ket{0,..,1}_D\right)\nonumber\\
&\times\left[\ket{00}_{j-1}...\ket{00}_{l+k}\ket{10}_{l+k-1}...\ket{10}_{l+1}\ket{01}_l...\ket{01}_1\right]\ .
\end{align}
The accrued costs are: $\bigO{j^3\log_2 j}$ {\tt CNOT}s and single-qubit gates.
\item Register $D$ now marks the positions of the modes to be interchanged; if $d_i=1$ (so $b_i=1$), mode $q$ sits on register $i$. Thus registers $i$ and $j$ should be interchanged. The qubits $b_i$ and $c_i$ are also interchanged, since now mode $p$, instead of $q$, occupies register $i$
\begin{align}
\mathcal{S}_{j-1\leftarrow 1}&\frac{1}{\sqrt{l}}\left\{\ket{0,..,1}_D\left[\ket{00}_{j-1}...\ket{00}_{l+k}\ket{10}_{l+k-1}...\ket{10}_{l+1}\ket{01}_l...\ket{01}_2\ket{10}_1\right]\right.\nonumber\\
&+\ket{0,..1,0}_D\left[\ket{00}_{j-1}...\ket{00}_{l+k}\ket{10}_{l+k-1}...\ket{10}_{l+1}\ket{01}_l...\ket{10}_2\ket{01}_1\right]+...+\nonumber\\
&\left.+|0,...,\underset{l}{\underbrace{1}},...,0\rangle_D\left[\ket{00}_{j-1}...\ket{00}_{l+k}\ket{10}_{l+k-1}...\ket{10}_{l+1}\ket{10}_l...\ket{01}_2\ket{01}_1\right]\right\},
\end{align}
this can be rearranged. Register $D$ now points to a qubit which is $1$ on register $C$, and since the state is symmetric with respect to the interchange of any two qubits grom $1$ to $j-1$, we can obtain a given state above, say the first, from a set of interchanges, e.g. from the swap of registers $1$ and $l+m$, $0<m<k$, register $D$ also swaps from $|0,...,1\rangle_D$ to $|0,...,\underset{l+m}{\underbrace{1}},..,0\rangle_D$, so that this step produces
\begin{align}
&|0,...,\underset{l+m}{\underbrace{1}},...,0\rangle_D\left[\ket{00}_{j-1}...\ket{00}_{l+k}\ket{10}_{l+k-1}...\ket{01}_{l+m}...\ket{10}_{l+1}\ket{01}_l...\ket{01}_2\ket{10}_1\right]\nonumber\\
&\rightarrow |0,...,\underset{l+m}{\underbrace{1}},...,0\rangle_D\left[\ket{00}_{j-1}...\ket{00}_{l+k}\ket{10}_{l+k-1}...\ket{10}_{l+m}...\ket{10}_{l+1}\ket{01}_l...\ket{01}_2\ket{10}_1\right],
\end{align}
the same as the state of the first line above. Grouping all the interchanges that produce the same states we have
\begin{align}
\mathcal{S}_{j-1\leftarrow 1}&\frac{1}{\sqrt{k}}\left(|0,...,\underset{l+k-1}{\underbrace{1}},...,0\rangle_D+...+|0,...,\underset{l+1}{\underbrace{1}},...,0\rangle_D+|0,...,1\rangle_D\right)\nonumber\\
&\times\left[\ket{00}_{j-1}...\ket{00}_{l+k}\ket{10}_{l+k-1}...\ket{10}_{l+m}...\ket{10}_{l+1}\ket{01}_l...\ket{01}_2\ket{10}_1\right],
\end{align}
which preludes the next step. Costs: Register $D$ has $j-1$ qubits, each one is used as control for a two mode-register {\tt SWAP}, which costs  $\bigO{n_p}$ {\tt Toffoli} gates, each of which can be decomposed in $\bigO{n_p}$ {\tt CNOT}s and single-qubit gates. This gives a total cost of $\bigO{j n_p}$ {\tt CNOT}s and single-qubit gates. The costs of swapping two qubits of $B$ and $C$ are negligible compared with the cost of swapping two mode-registers
\item Disentangle register $D$ with the {\tt SDD} algorithm using register $C$. 
Costs: $\bigO{j^3\log_2 j}$ {\tt CNOT}s and single-qubit gates
\item Disentangle registers $C$ and $B$ applying again step \textit{1.} Costs: $\bigO{jn_p}$ {\tt CNOT}s and single-qubit gates, $\bigO{j^2}$ order-oracle calls, {\tt CNOT} and single-qubit gates if the {\tt LL} algorithm is applied. 
\end{enumerate}
The total cost of the algorithm is $\bigO{j^3\log_2 j+j\log_2 N_p}$  {\tt CNOT} and single-qubit gates + $\bigO{j^2}$ order-oracle calls,  {\tt CNOT} and single-qubit gates, to advance the sytem forward in time by one step. $\blacksquare$

\newpage
\section{Proof of anti-commutation relations}\label{proof:commutation}

This appendix is devoted to prove Eq.~(\ref{eq:nReg-commutation}) and Eq.~(\ref{eq:nReg-anticommutation}). The demonstration of both equations is similar, we give the main steps for the anticommutation relations.

\textbf{Proof.} The demonstration is based on induction over $n$ (the number of active particles). Take the creation and annihilation operators over $n$ registers, $b^{(n)}_{q},b^{(n)\dagger}_{q}$, and assume that the anticommutation relation holds,
\begin{equation}
\left\{b^{(n)}_{q},b^{(n)\dagger}_{p}\right\}  = \delta_{q p}\left(\ketbrac{0}{0}\otimes\innerid\right)_{n}\otimes\mathbb{I}^{(n-1)}  + \,\mathcal{A}_{n\leftarrow n-1}\cdot \left(\ketbrac{1}{1}\otimes \mathfrak{s}^{\dagger}_{q}\mathfrak{s}_{p}\right)_n\otimes \projector{n-1}{n-1}\cdot \mathcal{A}_{n\leftarrow n-1}\ .    
\end{equation}

This is now used to prove the $(n+1)^{\rm th}$ case (for bosons there is a  negative sign in the second term and symmetrizers instead of antisymmetrizers). Operators over $n+1$ particles can be written in terms of those for $n$ particles; first, by Eq.(\ref{def:nReg-fermioncreation})
\begin{align}
b^{(n+1)\dagger}_{q} = & \sum^{n+1}_{i=1} \mathbb{I}^{\otimes (n+1-i)}\otimes\stepasym{i}\cdot \projector{n+1-i}{0}\otimes \left(\ketbrac{1}{0}\otimes \mathfrak{s}^{\dagger}_{q}\right)_{i}
\otimes \,\projector{i-1}{i-1}\nonumber \\
  =  & + \sum^{n}_{i = 1}\mathbb{I}\otimes\left(\mathbb{I}^{\otimes (n-i)}\otimes\mathcal{A}_{i\leftarrow i-1}\right)\cdot\projector{1}{0}\otimes\left(\projector{n-i}{0}\otimes\left(\ketbrac{1}{0}\otimes\set{q}\right)_i\otimes\projector{i-1}{i-1}\right)\nonumber \\
  \, & + \mathcal{A}_{n+1\leftarrow n}\cdot \left(\ketbrac{1}{0}\otimes \mathfrak{s}^{\dagger}_{q}\right)_{n+1}\otimes\projector{n}{n}\nonumber \\
  \equiv & \, \left[\left(\mathbb{I}\otimes\right)\cdot\left(\projector{1}{0}\otimes\right)\right]\left[b^{(n)\dagger}_{q}\right] + \mathcal{A}_{n+1\leftarrow n}\cdot \left[\left(\ketbrac{1}{0}\otimes \set{q}\right)_{n+1}\otimes\right]\left[\projector{n}{n}\right],
\label{eq:dag-com-dep}
\end{align}

similarly 
\begin{equation}
b^{(n+1)}_{p} = \left[\left(\mathbb{I}\otimes\right)\cdot\left(\projector{1}{0}\otimes\right)\right]\left[b^{(n)}_{p}\right] +  \left[\left(\ketbrac{0}{1}\otimes \scrap{p}\right)_{n+1}\otimes\right]\left[\projector{n}{n}\right]\cdot \mathcal{A}_{n+1\leftarrow n},
\label{eq:com-dep}
\end{equation}
the corresponding expressions for bosons are completely analogous. We now calculate $\left\{b^{(n+1)}_{p},b^{(n+1)\dagger}_{q}\right\}$:
\begin{align}
b^{(n+1)\dagger}_{q}b^{(n+1)}_{p} & = \mathcal{A}_{n+1\leftarrow n}\cdot  \left( \left[\left(\ketbrac{1}{0}\otimes \mathfrak{s}^{\dagger}_{q}\right)_{n+1}\otimes\right]\left[\projector{n}{n}\right]\right)\left(\left[\left(\ketbrac{0}{1}\otimes \mathfrak{s}_{p}\right)_{n+1}\otimes\right] \left[\projector{n}{n}\right]\right)\cdot \mathcal{A}_{n+1\leftarrow n}\nonumber \\
& + \mathcal{A}_{n+1\leftarrow n}\cdot \left( \left[\left(\ketbrac{1}{0}\otimes \mathfrak{s}^{\dagger}_{q}\right)_{n+1}\otimes\right]\left[\projector{n}{n}\right]\right)\left(\left[\left(\outeridentity\otimes\right)\cdot\left(\projector{1}{0}\otimes\right)\right] \left[b^{(n)}_{p}\right]\right) \nonumber \\
& + \left(\left[\left(\outeridentity\otimes\right)\cdot\left(\projector{1}{0}\otimes\right)\right] \left[b^{(n)\dagger}_{q}\right]\right)\left(\left[\left(\ketbrac{0}{1}\otimes \mathfrak{s}_{p}\right)_{n+1}\otimes\right]\left[\projector{n}{n}\right]\right)\cdot \mathcal{A}_{n+1\leftarrow n}\nonumber \\
& + \left(\left[\left(\outeridentity\otimes\right)\cdot\left(\projector{1}{0}\otimes\right)\right] \left[b^{(n)\dagger}_{q}\right]\right)\left(\left[\left(\outeridentity\otimes\right)\cdot\left(\projector{1}{0}\otimes\right)\right] \left[b^{(n)}_{p}\right]\right)\nonumber \\
& = \mathcal{A}_{n+1\leftarrow n}\cdot \left[\left(\ketbrac{1}{1}\otimes \set{q}\scrap{p}\right)_{n+1}\otimes\right]\left[\projector{n}{n}\right]\cdot \mathcal{A}_{n+1\leftarrow n} \nonumber \\
& + \left[\left(\outeridentity\otimes\right)\cdot\left(\projector{1}{0}\otimes\right)\right]\left[b^{(n)\dagger}_{q}b^{(n)}_{p}\right]
\end{align}
where the second and third terms vanish because, on the one hand, $\projector{n}{n}b^{(n)}_{q}$ represents the action of $\projector{n}{n}$ over an state that does not have $n$-particles because of the annihilation operator, which gives 0. On the other hand, $b^{(n)\dagger}_{p}\projector{n}{n}$ represents the action of $b^{(n)\dagger}_{p}$ over an state with exactly $n$-particles, which is again 0. The other product reads
\begin{align}
b^{(n+1)}_{p}b^{(n+1)\dagger}_{q} & = \left[\left(\ketbrac{0}{1}\otimes \mathfrak{s}_{p}\right)_{n+1}\otimes\right] \left[\projector{n}{n}\right]   \cdot  \mathcal{A}^{2}_{n+1\leftarrow n} \cdot  \left[\left(\ketbrac{1}{0}\otimes \mathfrak{s}^{\dagger}_{q}\right)\otimes\right]\left[\projector{n}{n}\right]\nonumber \\
& + \left[\left(\outeridentity\otimes\right)\cdot\left(\projector{1}{0}\otimes\right)\right]\left[ b^{(n)}_{p}\right] \cdot \mathcal{A}_{n+1\leftarrow n}\cdot  \left[\left(\ketbrac{1}{0}\otimes \mathfrak{s}^{\dagger}_{q}\right)\otimes\right]\left[\projector{n}{n}\right] \nonumber \\
& +  \left[\left(\ketbrac{0}{1}\otimes \mathfrak{s}_{p}\right)_{n+1}\otimes\right] \left[\projector{n}{n}\right] \cdot \mathcal{A}_{n+1\leftarrow n}\cdot\left[\left(\outeridentity\otimes\right)\cdot\left(\projector{1}{0}\otimes\right)\right] \left[b^{(n)\dagger}_{q}\right]\nonumber \\
& + \left(\left[\left(\outeridentity\otimes\right)\cdot\left(\projector{1}{0}\otimes\right)\right]\left[ b^{(n)}_{p}\right] \right)\left(\left[\left(\outeridentity\otimes\right)\cdot\left(\projector{1}{0}\otimes\right)\right] \left[b^{(n)\dagger}_{q}\right]\right) \nonumber \\
& = \sqrt{n+1}\left[\left(\ketbrac{0}{1}\otimes \mathfrak{s}_{p}\right)_{n+1}\otimes\right] \left[\projector{n}{n}\right]  \cdot  \mathcal{A}_{n+1\leftarrow n} \cdot   \left[\left(\ketbrac{1}{0}\otimes \mathfrak{s}^{\dagger}_{q}\right)_{n+1}\otimes\right]\left[\projector{n}{n}\right] \nonumber \\
&+ \left[\left(\outeridentity\otimes\right)\cdot \left(\projector{1}{0}\otimes\right)\right] \left[b^{(n)}_{p}b^{(n)\dagger}_{q}\right].
\end{align}
in this case the products of the second and third lines also give zero:  on one hand,  in the second line we see that the left parenthesis will be 0 if the $n+1$ register is occupied, while the operators in the parenthesis to the right give an occupied $n+1$ register, and their multiplication gives 0. On the other hand, the left parentheses in the third line will be 0 if the $n+1$ register is empty, but looking at the parenthesis to the right we see that this is always the case, so multiplication of both parenthesis gives again 0; finally, $\mathcal{A}^{2}_{n+1\leftarrow n} = \sqrt{n+1}\mathcal{A}_{n+1\leftarrow n}$.
The anticommutator therefore reads
\begin{align}
\left\{b^{(n+1)}_{p}, b^{(n+1)\dagger}_{q}\right\} & = \left[\left(\outeridentity\otimes\right)\cdot \left(\projector{1}{0}\otimes\right)\right] \left[\left\{b^{(n)}_{p}b^{(n)\dagger}_{q}\right\}\right]  \nonumber \\
& + \sqrt{n+1}\, \left[\left(\ketbrac{0}{1}\otimes \mathfrak{s}_{p}\right)_{n+1}\otimes\right] \left[\projector{n}{n}\right]  \cdot  \mathcal{A}_{n+1\leftarrow n} \cdot   \left[\left(\ketbrac{1}{0}\otimes \mathfrak{s}^{\dagger}_{q}\right)_{n+1}\otimes\right]\left[\projector{n}{n}\right] \nonumber \\
& + \mathcal{A}_{n+1\leftarrow n}\cdot \left[\left(\ketbrac{1}{1}\otimes \set{q}\scrap{p}\right)_{n+1}\otimes\right]\left[\projector{n}{n}\right]\cdot \mathcal{A}_{n+1\leftarrow n}
\end{align}
Using the induction hyphotesis on the first line,
\begin{align}
\left\{b^{(n+1)}_{p}, b^{(n+1)\dagger}_{q}\right\} & = \delta_{pq}\outeridentity^{\otimes n+1} \cdot \projector{2}{0}\otimes\mathbb{I}^{\otimes(n-1)}\nonumber \\
& + \outeridentity\otimes \mathcal{A}_{n\leftarrow n-1}\cdot \projector{1}{0}\otimes\left(\ketbrac{1}{1}\otimes \mathfrak{s}^{\dagger}_{q}\mathfrak{s}_{p}\right)_n\otimes\projector{n-1}{n-1}\cdot \outeridentity\otimes \mathcal{A}_{n\leftarrow n-1}\nonumber \\
& + \sqrt{n+1}\,\left[\left(\ketbrac{0}{1}\otimes \mathfrak{s}_{p}\right)_{n+1}\otimes\right] \left[\projector{n}{n}\right]  \cdot  \mathcal{A}_{n+1\leftarrow n} \cdot   \left[\left(\ketbrac{1}{0}\otimes \mathfrak{s}^{\dagger}_{q}\right)_{n+1}\otimes\right]\left[\projector{n}{n}\right]\nonumber  \\
& +\mathcal{A}_{n+1\leftarrow n}\cdot \left(\ketbrac{1}{1}\otimes \mathfrak{s}^{\dagger}_{q}\mathfrak{s}_{p}\right)_{n+1}\otimes\projector{n}{n}\cdot \mathcal{A}_{n+1\leftarrow n},
\label{proof:anticommutation-induction}
\end{align}
where the last line is the boundary term in Eq.~(\ref{eq:nReg-anticommutation}). For the commutator the result is similar, but with negative signs in the second and fourth lines. The first line can then be rewritten as
\begin{equation}
    \delta_{p q}\outeridentity^{\otimes n+1} \cdot \projector{2}{0}\otimes\mathbb{I}^{\otimes(n-1)} = \delta_{q_1q_2}\projector{2}{0}\otimes\sum^{n-1}_{j= 0}\projector{n-1}{k} = \delta_{q_1q_2}\left(\ketbrac{0}{0}\otimes\innerid\right)_{n+1}\otimes\sum^{n-1}_{k= 0}\projector{n}{k}\ .
\end{equation}
To complete the induction step we need the $k = n$ term of the sum, i.e., $\delta_{q_1q_2}\left(\ketbrac{0}{0}\otimes \innerid\right)_{n+1}\otimes\projector{n}{n}$. Applying it to a general $n$-particles state (the only ones for which the result is not null) gives
\begin{equation}
\delta_{q_{1}q_{2}}\left(\ketbrac{0}{0}\otimes \innerid\right)_{n+1}\otimes\projector{n}{n} \ket{\Omega}_{n+1}\left(\ket{1p_{n}}_n...\ket{1p_1}_1\right)_{\mathcal{A}} =  \delta_{q_{1}q_{2}}\ket{\Omega}_{n+1}\left(\ket{1p_{n}}_n...\ket{1p_1}_1\right)_{\mathcal{A}}.
\end{equation}
We now prove that the action of the second and third lines of Eq.~(\ref{proof:anticommutation-induction}) is the same upon arbitrary states, (we concentrate in the case with $n$ filled registers, for the rest the result is null). Starting from the second line (identity operators multiplying antisymmetrizers are omitted)
\begin{align}
\mathcal{A}_{n\leftarrow n-1} & \cdot \projector{1}{0}\otimes\left(\ketbrac{1}{1}\otimes \mathfrak{s}^{\dagger}_{q}\mathfrak{s}_{p}\right)_n\otimes\projector{n-1}{n-1}\cdot \mathcal{A}_{n\leftarrow n-1}\ket{\Omega}_{n+1}\left(\ket{1p_{n}}_n...\ket{1p_1}_1\right)_{\mathcal{A}}\nonumber \\
& = \sqrt{n}\, \mathcal{A}_{n\leftarrow n-1}\cdot \left(\ketbrac{0}{0}\otimes \innerid\right)_{n+1}\otimes\left(\ketbrac{1}{1}\otimes \mathfrak{s}^{\dagger}_{q}\mathfrak{s}_{p}\right)_n\otimes\projector{n-1}{n-1}\ket{\Omega}_{n+1}\left(\ket{1p_{n}}_n...\ket{1p_1}_1\right)_{\mathcal{A}} \nonumber \\
& =  \mathcal{A}_{n\leftarrow n-1}\sum^{n-1}_{k=0} (-1)^{k} \delta_{p p_{n-k}}\ket{\Omega}_{n+1}\ket{1q}_{n}\left(\ket{p_{n}}_{n-1}...\ket{p_{n-k+1}}_{n-k}\ket{p_{n-k-1}}_{n-k-1}...\ket{1p_1}_1\right)_{\mathcal{A}}\nonumber \\
& = \sum^{n-1}_{k=0} (-1)^{k} \delta_{q_2 p_{n-k}}\ket{\Omega}_{n+1}\left(\ket{1q_1}_{n}\ket{p_{n}}_{n-1}...\ket{p_{n-k+1}}_{n-k}\ket{p_{n-k-1}}_{n-k-1}...\ket{1p_1}_1\right)_{\mathcal{A}},
\end{align}
we continue with the other term of the right-hand side
\begin{align}
\sqrt{n+1}\, & \left[\left(\ketbrac{0}{1}\otimes \mathfrak{s}_{p}\right)_{n+1}\otimes\right] \left[\projector{n}{n}\right]  \cdot  \mathcal{A}_{n+1\leftarrow n} \cdot \left[\left(\ketbrac{1}{0}\otimes \mathfrak{s}^{\dagger}_{q}\right)_{n+1}\otimes\right]\left[\projector{n}{n}\right]\ket{\Omega}_{n+1}\left(\ket{1p_{n}}...\ket{1p_1}\right)_{\mathcal{A}} \nonumber \\
& =\sqrt{n+1} \left[\left(\ketbrac{0}{1}\otimes \mathfrak{s}_{p}\right)_{n+1}\otimes\right] \left[\projector{n}{n}\right]  \cdot  \mathcal{A}_{n+1\leftarrow n} \ket{1q}_{n+1}\left(\ket{1p_{n}}_n...\ket{1p_1}_1\right)_{\mathcal{A}} \nonumber \\
& = \sqrt{n+1} \, \left[\left(\ketbrac{0}{1}\otimes \mathfrak{s}_{p}\right)_{n+1}\otimes\right] \left[\projector{n}{n}\right] \left(\ket{1q}_{n+1}\ket{1p_{n}}_n...\ket{1p_1}_1\right)_{\mathcal{A}}\nonumber \\
& = \delta_{p q}\ket{\Omega}_{n+1}\left(\ket{p_{n}}_{n}...\ket{1p_1}_1\right)_{\mathcal{A}} \nonumber\\
& + \sum^{n}_{k=1} (-1)^{k} \delta_{p p_{n+1-k}}\ket{\Omega}_{n+1}\left(\ket{1q}_{n}\ket{p_{n}}_{n-1}...\ket{p_{n-k+2}}_{n+1-k}\ket{p_{n-k}}_{n-k}...\ket{1p_1}_1\right)_{\mathcal{A}},
\end{align}
changing $k\rightarrow k+1$ in the last line
\begin{align}
\sqrt{n+1}\, & \left(\ketbrac{0}{1}\otimes \mathfrak{s}_{p}\right)_{n+1}\otimes \projector{n}{n} \cdot  \mathcal{A}_{n+1\leftarrow n} \cdot  \left(\ketbrac{1}{0}\otimes \mathfrak{s}^{\dagger}_{q}\right)_{n+1}\otimes\projector{n}{n}\ket{\Omega}_{n+1}\left(\ket{1p_{n}}_n...\ket{1p_1}_1\right)_{\mathcal{A}} \nonumber \\
& = \delta_{p q}\ket{\Omega}_{n+1}\left(\ket{p_{n}}_{n}...\ket{1p_1}_1\right)_{\mathcal{A}} \nonumber\\
& - \sum^{n-1}_{k=0} (-1)^{k} \delta_{p p_{n-k}}\ket{\Omega}_{n+1}\left(\ket{1q}_{n}\ket{p_{n}}_{n-1}...\ket{p_{n-k+1}}_{n-k}\ket{p_{n-k-1}}_{n-k-1}...\ket{1p_1}_1\right)_{\mathcal{A}},
\end{align}
adding both terms:
\begin{align}
&\left(\mathcal{A}_{n\leftarrow n-1} \cdot \projector{1}{0}\otimes\left(\ketbrac{1}{1}\otimes \mathfrak{s}^{\dagger}_{q}\mathfrak{s}_{p}\right)_n\otimes\projector{n-1}{n-1}\cdot \mathcal{A}_{n\leftarrow n-1}\right.\nonumber \\
& + \left.\sqrt{n+1}\, \left[\left(\ketbrac{0}{1}\otimes \mathfrak{s}_{p}\right)_{n+1}\otimes\right] \left[\projector{n}{n}\right]  \cdot  \mathcal{A}_{n+1\leftarrow n} \cdot \left[\left(\ketbrac{1}{0}\otimes \mathfrak{s}^{\dagger}_{q}\right)_{n+1}\otimes\right]\left[\projector{n}{n}\right]\ket{\Omega}_{n+1}\right)\left(\ket{1p_{n}}_n...\ket{1p_1}_1\right)_{\mathcal{A}} \nonumber \\
& =  \delta_{p q}\ket{\Omega}_{n+1}\left(\ket{p_{n}}_{n}...\ket{1p_1}_1\right)_{\mathcal{A}}.
\end{align}
We thus have
\begin{align}
\left\{b^{(n+1)}_{p}, b^{(n+1)\dagger}_{q}\right\} & = \delta_{q p}\left(\ketbrac{0}{0}\otimes\innerid\right)_{n+1}\otimes\sum^{n-1}_{k= 0}\projector{n}{k} + \delta_{q_{1}q_{2}}\left(\ketbrac{0}{0}\otimes \innerid\right)_{n+1}\otimes\projector{n}{n}\nonumber \\
& +\mathcal{A}_{n+1\leftarrow n}\cdot \left(\ketbrac{1}{1}\otimes \mathfrak{s}^{\dagger}_{q}\mathfrak{s}_{p}\right)_{n+1}\otimes\projector{n}{n}\cdot \mathcal{A}_{n+1\leftarrow n}\nonumber \\
& = \delta_{q p}\left(\ketbrac{0}{0}\otimes\innerid\right)_{n+1}\otimes\sum^{n}_{k= 0}\projector{n}{k} + \mathcal{A}_{n+1\leftarrow n}\cdot \left(\ketbrac{1}{1}\otimes \mathfrak{s}^{\dagger}_{q}\mathfrak{s}_{p}\right)_{n+1}\otimes\projector{n}{n} \cdot \mathcal{A}_{n+1\leftarrow n}\nonumber \\
& = \delta_{q p}\left(\ketbrac{0}{0}\otimes\innerid\right)_{n+1}\otimes\mathbb{I}^{(n)} + \mathcal{A}_{n+1\leftarrow n}\cdot \left(\ketbrac{1}{1}\otimes \mathfrak{s}^{\dagger}_{q}\mathfrak{s}_{p}\right)_{n+1}\otimes\projector{n}{n} \cdot \mathcal{A}_{n+1\leftarrow n},
\end{align}
which concludes the induction step. $\blacksquare$
\newpage
\section{Glossary of symbols} \label{app:glossary}
Given the extensive number of mathematical symbols employed through the article, we believe the reader might find useful to have this reference at hand when looking for a particular one. They are given in Latin alphabetic order, independently of the font in which they are encoded.
\subsection{General}
\begin{table}[!ht]
\centering
\footnotesize
\begin{tabular}{l|l|c}
Symbol & Meaning & Defining equation \\ \hline 
$\mathfrak{C}_{ij}$ & Control operator $\ket{i}\bra{j}$ wih $i,j\in\left\{0,1\right\}$ standing for absence/presence. & Eq.~(\ref{def:control-operators})
\\
$\Delta$ & Momentum-grid spacing. & Under Eq.~(\ref{Momentum-discretization})\\
$\Delta t$ & Time interval. & \\
$E_q$ & Energy of a particle with momentum $q$. & \\
$\innerid$ & Identity over one-particle  qubits encoding momentum. & Eq.~(\ref{2bodyauxiliary})\\
$\outeridentity$  &  Identity over entire one-particle registers. & Eq.~(\ref{2bodycreator}) \\
$\mathbb{I}^{(n)}$ & Identity over $n$ ordered registers. & \\ 
$\mathfrak{h}_{22,l.k}$ & Auxiliary potential exchanging momentum over registers $l$, $k$. & Eq.~(\ref{h22lk}) \\
$\eta_\sigma$ & Parity of permutation $\sigma$; 0 if $\sigma$ is even and 1 if $\sigma$ is odd. & \\
$j,k,l,m$ & Indices over the register number. \\
$\lambda_\xi$ & Coupling intensity of terms exchanging momentum $p_{\xi}$.& \\
$\Lambda_{\rm max},\ \Lambda_{\rm min}$ & Extreme values of the integer indexing  the momentum variable. & Eq.~(\ref{Momentum-discretization})
\\
$\hat{N}$ & Number operator. & Eq.~(\ref{Number-operator}).\\
$N_p$ & Number of encoded momenta. & Eq.~(\ref{Momentum-discretization}) \\
$n_p$ & Number of qubits to encode momenta. & Eq.~(\ref{sec:1body}) \\
$n_b/n_f$ & Number of boson/fermion registers & Eq.~(\ref{subsec:splitting}) \\
$n$ & Total number of registers, boson + fermion & Eq.~(\ref{subsec:splitting}) \\
$|\Omega \rangle $ & Empty registers, in the state $\ket{0}_{P/A}\otimes\ket{0...0}_{momentum}$. & Eq.~(\ref{eq:creation-1})\\
$|0\rangle_{P/A} $ & Presence/absence qubit, 0 indicating absence. & Eq.~(\ref{eq:creation-1})\\
$|0...0\rangle_{\rm momentum}$ & State to which empty momentum registers are assigned. & Eq.~(\ref{eq:creation-1})\\
$|p_i\rangle$, $|q_i\rangle$ & States representing $p_i$ and $q_i$ momenta, respectively. & Eq.~(\ref{eq:creation-3})\\
$|p_{min}\rangle$, $|p_{mn}\rangle$ & Lowest codified momenta & Eq.~(\ref{Momentum-discretization})\\
$|p_{max}\rangle$, $|p_{mx}\rangle$ & Largest codified momenta & Eq.~(\ref{Momentum-discretization})\\
$\overline{\ket{p_i}}_l$ & Denotes that momentum on register $l$  is fixed to $p_i$ & Eq.~(\ref{def:momentum-fixing})\\
$\mathcal{P}_{ij}$ & Permutation operator of registers $i$ and $j$. & Eq.~(\ref{def:12-permutator})\\
$\projector{n}{j}$ & Projector over exactly $j$-occupied registers for a total of $n$ registers & Eq.~(\ref{def:projectors})\\
$\mathfrak{s}_p$, $\mathfrak{s}_p^\dagger$ & Scrap and set operators reassigning momentum qubits &Eq.~(\ref{eq:creation-3})\\
& from $\ket{p}$ to $\ket{0...0}$ and viceversa. & \\
$\mathcal{U}^{f/b}_{ij}(t)$ & Evolution operator for fermion/boson quanta, & Eq.~(\ref{eq:free-evolution},\ref{4bodypotential},\\
& it creates/annihilates $i/j$ particles and viceversa. & \& \ref{def:tadpole},\ref{def:splitting})\\
$\mathfrak{U}_{ij,l}(t)$ & Part of the operator $\mathcal{U}^{f/b}_{ij}(t)$ that performs memory changes & Eq.~(\ref{def:free-evolution-onregister},\ref{eq:four-body-I-fermions-exp},\\
& $l$ designates registers, it can be a number or a tuple.  & \& \ref{eq:one-body-fermions-exp},\ref{eq:two-fermions-one-boson-1},\ref{eq:splitting-2})\\
$\xi$ & Momentum discretization. & \\ 
$\otimes$ & Tensor product splitting an operator within a register & \\
& into components over its subspaces. & \\
$\cdot$ (dot) & Successive application of operators over the entire memory. & \\
\hline
\end{tabular}
\end{table}

\newpage

\subsection{Boson symbols}
\begin{table}[!ht]
\centering
\begin{tabular}{l|l|c}
Symbol & Meaning & Definition \\ \hline 
$a^{(n)}_{p,j}$, $a^{(n)\dagger}_{p,j}$ & Destructor and creator operators for a boson of momentum $p$ & Eq.~(\ref{def:nReg-bosoncreator-j}) \& Eq.~(\ref{def:nReg-bosonannihilation-j}) \\ 
& at the highest occupied register, $j$. & \\
$a^{(n)}_{p}$, $a^{(n)\dagger}_{p}$  & Boson destructor/creator operators with momentum $p$, at any register. & Eq.~(\ref{def:nReg-bosoncreation}) \&  Eq.~(\ref{def:nReg-bosonannihilation}) \\ &   (The sum over $j$ of  $a^{(n)}_{p,j},a^{(n)\dagger}_{p,j}$: they represent the field theoretical $a$, $a^\dagger$.) & \\
$\bSWAP{j}{p}{q}$ & Interchanges modes $p$ and $q$ in a symmetric memory with $j$  & \ref{par:bSWAP-3reg} \& \ref{general-boson-operators}\\
& occupied registers, the first mode ($p$) should sit on register $j$ &\\
$\bX{j}{p}$ & Auxiliary exchange operator for creation/annihilation of bosons  & Eq.~(\ref{def:cbX})\\
$\mathcal{S}_{n\leftarrow m}$ & Non-unitary step symmetrizer; it outputs an $n$-particle symmetric  & Eq.~(\ref{def:step-symmetrizer})\\
&state from $\textit{any}$ $m$-particle symmetric one &\\
$\hat{\mathcal{S}}^{\dagger}_{n\leftarrow m}$ & Unitary step symmetrizer; it outputs an $n$-particle symmetric  & \ref{par:symmetrizer-3reg} \& \ref{par:symmetrizer-general}\\
&state from a $\textit{semi-ordered}$ $m$-particle symmetric one &\\
\hline
\end{tabular}
\end{table}

\subsection{Fermion symbols}
\begin{table}[!ht]
\centering
\begin{tabular}{l|l|c}
Symbol & Meaning & Definition \\ \hline 
$\mathcal{A}_{n\leftarrow m}$ & Non-unitary step antisymmetrizer; it outputs an $n$-particle antisymmetric  & Eq.~(\ref{def:step-asym})\\
&state from $\textit{any}$ $m$-particle antisymmetric one &\\
$\hat{\mathcal{A}}^{\dagger}_{n\leftarrow m}$ & Unitary step antisymmetrizer; outputs an $n$-particle antisymmetric& \ref{par:antisymmetrizers} \\
&state from an $\textit{ordered}$ $m$-particle antisymmetric one &  \\
$b^{(n)}_{p,j},b^{(n)\dagger}_{p,j}$ & 
Destructor and creator operators for a fermion of momentum $p$ & Eq.~(\ref{def:nReg-fermioncreation-j}), Eq.~(\ref{def:nreg-fermionannihilation-j})\\   & at the highest occupied register, $j$. \\
$b^{(n)}_{p}$, $b^{(n)\dagger}_{p}$  & Destructor and creator operators for a fermion of momentum $p$ & Eq.~(\ref{def:nReg-fermioncreation}), Eq.~(\ref{def:nReg-fermionannihilation})\\ & at any register.  (The sum over $j$ of  $b^{(n)}_{p,j},b^{(n)\dagger}_{p,j}$.) & \\
$\fSWAP{j}{p}{q}$ & Interchanges modes $p$ and $q$ in an antisymmetric memory with $j$  & \ref{par:fermion-SWAP}\\
& occupied registers, the first mode ($p$) should sit on register $j$ &\\
$\fSWAP{(j_b,j_f)}{p}{q}$ & Controlled fermion {\tt SWAP} depending on   & Eq.~(\ref{def:fSp})\\
& whether split boson is in register $j$ &   \\
$\fX{j}{p}$ & Auxiliary exchange operator for creation/annihilation of fermions  & Eq.~(\ref{def:fTd})\\
\hline
\end{tabular}
\end{table}

\clearpage
\bibliography{BibliografiaPaper2024}

\begin{thebibliography}{46}%
\makeatletter
\providecommand \@ifxundefined [1]{%
 \@ifx{#1\undefined}
}%
\providecommand \@ifnum [1]{%
 \ifnum #1\expandafter \@firstoftwo
 \else \expandafter \@secondoftwo
 \fi
}%
\providecommand \@ifx [1]{%
 \ifx #1\expandafter \@firstoftwo
 \else \expandafter \@secondoftwo
 \fi
}%
\providecommand \natexlab [1]{#1}%
\providecommand \enquote  [1]{``#1''}%
\providecommand \bibnamefont  [1]{#1}%
\providecommand \bibfnamefont [1]{#1}%
\providecommand \citenamefont [1]{#1}%
\providecommand \href@noop [0]{\@secondoftwo}%
\providecommand \href [0]{\begingroup \@sanitize@url \@href}%
\providecommand \@href[1]{\@@startlink{#1}\@@href}%
\providecommand \@@href[1]{\endgroup#1\@@endlink}%
\providecommand \@sanitize@url [0]{\catcode `\\12\catcode `\$12\catcode
  `\&12\catcode `\#12\catcode `\^12\catcode `\_12\catcode `\%12\relax}%
\providecommand \@@startlink[1]{}%
\providecommand \@@endlink[0]{}%
\providecommand \url  [0]{\begingroup\@sanitize@url \@url }%
\providecommand \@url [1]{\endgroup\@href {#1}{\urlprefix }}%
\providecommand \urlprefix  [0]{URL }%
\providecommand \Eprint [0]{\href }%
\providecommand \doibase [0]{https://doi.org/}%
\providecommand \selectlanguage [0]{\@gobble}%
\providecommand \bibinfo  [0]{\@secondoftwo}%
\providecommand \bibfield  [0]{\@secondoftwo}%
\providecommand \translation [1]{[#1]}%
\providecommand \BibitemOpen [0]{}%
\providecommand \bibitemStop [0]{}%
\providecommand \bibitemNoStop [0]{.\EOS\space}%
\providecommand \EOS [0]{\spacefactor3000\relax}%
\providecommand \BibitemShut  [1]{\csname bibitem#1\endcsname}%
\let\auto@bib@innerbib\@empty
\bibitem [{\citenamefont {Beck}\ \emph {et~al.}(2023)\citenamefont {Beck},
  \citenamefont {Carlson}, \citenamefont {Davoudi} \emph
  {et~al.}}]{Beck:2023xhh}%
  \BibitemOpen
  \bibfield  {author} {\bibinfo {author} {\bibfnamefont {D.}~\bibnamefont
  {Beck}}, \bibinfo {author} {\bibfnamefont {J.}~\bibnamefont {Carlson}},
  \bibinfo {author} {\bibfnamefont {Z.}~\bibnamefont {Davoudi}}, \emph
  {et~al.},\ }\href {http://arxiv.org/abs/2303.00113} {\bibinfo {title}
  {Quantum {Information} {Science} and {Technology} for {Nuclear} {Physics}.
  {Input} into {U}.{S}. {Long}-{Range} {Planning}}} (\bibinfo {year} {2023}),\
  \bibinfo {note} {arXiv:2303.00113 [nucl-ex, physics:nucl-th,
  physics:quant-ph]}\BibitemShut {NoStop}%
\bibitem [{\citenamefont {Lanyon}\ \emph {et~al.}(2010)\citenamefont {Lanyon},
  \citenamefont {Whitfield}, \citenamefont {Gillett} \emph
  {et~al.}}]{Lanyon:2009tha}%
  \BibitemOpen
  \bibfield  {author} {\bibinfo {author} {\bibfnamefont {B.~P.}\ \bibnamefont
  {Lanyon}}, \bibinfo {author} {\bibfnamefont {J.~D.}\ \bibnamefont
  {Whitfield}}, \bibinfo {author} {\bibfnamefont {G.~G.}\ \bibnamefont
  {Gillett}}, \emph {et~al.},\ }\href {https://doi.org/10.1038/nchem.483}
  {\bibfield  {journal} {\bibinfo  {journal} {Nature Chem}\ }\textbf {\bibinfo
  {volume} {2}},\ \bibinfo {pages} {106} (\bibinfo {year} {2010})}\BibitemShut
  {NoStop}%
\bibitem [{\citenamefont {Wang}\ \emph {et~al.}(2024)\citenamefont {Wang},
  \citenamefont {Du}, \citenamefont {Zuo},\ and\ \citenamefont
  {Vary}}]{Wang:2024scd}%
  \BibitemOpen
  \bibfield  {author} {\bibinfo {author} {\bibfnamefont {P.}~\bibnamefont
  {Wang}}, \bibinfo {author} {\bibfnamefont {W.}~\bibnamefont {Du}}, \bibinfo
  {author} {\bibfnamefont {W.}~\bibnamefont {Zuo}},\ and\ \bibinfo {author}
  {\bibfnamefont {J.~P.}\ \bibnamefont {Vary}},\ }\href
  {http://arxiv.org/abs/2401.17138} {\bibinfo {title} {Nuclear scattering via
  quantum computing}} (\bibinfo {year} {2024}),\ \bibinfo {note}
  {arXiv:2401.17138 [nucl-th, physics:quant-ph]}\BibitemShut {NoStop}%
\bibitem [{\citenamefont {Pérez-Obiol}\ \emph {et~al.}(2023)\citenamefont
  {Pérez-Obiol}, \citenamefont {Romero}, \citenamefont {Menéndez} \emph
  {et~al.}}]{perez-obiol_nuclear_2023}%
  \BibitemOpen
  \bibfield  {author} {\bibinfo {author} {\bibfnamefont {A.}~\bibnamefont
  {Pérez-Obiol}}, \bibinfo {author} {\bibfnamefont {A.~M.}\ \bibnamefont
  {Romero}}, \bibinfo {author} {\bibfnamefont {J.}~\bibnamefont {Menéndez}},
  \emph {et~al.},\ }\href {https://doi.org/10.1038/s41598-023-39263-7}
  {\bibfield  {journal} {\bibinfo  {journal} {Scientific Reports}\ }\textbf
  {\bibinfo {volume} {13}},\ \bibinfo {pages} {12291} (\bibinfo {year}
  {2023})}\BibitemShut {NoStop}%
\bibitem [{\citenamefont {Drissi}\ \emph {et~al.}(2024)\citenamefont {Drissi},
  \citenamefont {Keeble}, \citenamefont {Sarmiento},\ and\ \citenamefont
  {Rios}}]{Drissi:2024cnn}%
  \BibitemOpen
  \bibfield  {author} {\bibinfo {author} {\bibfnamefont {M.}~\bibnamefont
  {Drissi}}, \bibinfo {author} {\bibfnamefont {J.~W.~T.}\ \bibnamefont
  {Keeble}}, \bibinfo {author} {\bibfnamefont {J.~R.}\ \bibnamefont
  {Sarmiento}},\ and\ \bibinfo {author} {\bibfnamefont {A.}~\bibnamefont
  {Rios}},\ }\href {http://arxiv.org/abs/2401.17550} {\bibinfo {title}
  {Second-order optimisation strategies for neural network quantum states}}
  (\bibinfo {year} {2024}),\ \bibinfo {note} {arXiv:2401.17550 [nucl-th,
  physics:quant-ph]}\BibitemShut {NoStop}%
\bibitem [{\citenamefont {Abrams}\ and\ \citenamefont
  {Lloyd}(1997)}]{abrams_009711_1997}%
  \BibitemOpen
  \bibfield  {author} {\bibinfo {author} {\bibfnamefont {D.~S.}\ \bibnamefont
  {Abrams}}\ and\ \bibinfo {author} {\bibfnamefont {S.}~\bibnamefont {Lloyd}},\
  }\href {https://doi.org/10.1103/PhysRevLett.79.2586} {\bibfield  {journal}
  {\bibinfo  {journal} {Physical Review Letters}\ }\textbf {\bibinfo {volume}
  {79}},\ \bibinfo {pages} {2586} (\bibinfo {year} {1997})}\BibitemShut
  {NoStop}%
\bibitem [{\citenamefont {Ortiz}\ \emph {et~al.}(2001)\citenamefont {Ortiz},
  \citenamefont {Gubernatis}, \citenamefont {Knill},\ and\ \citenamefont
  {Laflamme}}]{ortiz_quantum_2001}%
  \BibitemOpen
  \bibfield  {author} {\bibinfo {author} {\bibfnamefont {G.}~\bibnamefont
  {Ortiz}}, \bibinfo {author} {\bibfnamefont {J.~E.}\ \bibnamefont
  {Gubernatis}}, \bibinfo {author} {\bibfnamefont {E.}~\bibnamefont {Knill}},\
  and\ \bibinfo {author} {\bibfnamefont {R.}~\bibnamefont {Laflamme}},\ }\href
  {https://doi.org/10.1103/PhysRevA.64.022319} {\bibfield  {journal} {\bibinfo
  {journal} {Phys. Rev. A}\ }\textbf {\bibinfo {volume} {64}},\ \bibinfo
  {pages} {022319} (\bibinfo {year} {2001})}\BibitemShut {NoStop}%
\bibitem [{\citenamefont {Bravyi}\ and\ \citenamefont
  {Kitaev}(2002)}]{bravyi_fermionic_2002}%
  \BibitemOpen
  \bibfield  {author} {\bibinfo {author} {\bibfnamefont {S.~B.}\ \bibnamefont
  {Bravyi}}\ and\ \bibinfo {author} {\bibfnamefont {A.~Y.}\ \bibnamefont
  {Kitaev}},\ }\href {https://doi.org/10.1006/aphy.2002.6254} {\bibfield
  {journal} {\bibinfo  {journal} {Annals of Physics}\ }\textbf {\bibinfo
  {volume} {298}},\ \bibinfo {pages} {210} (\bibinfo {year}
  {2002})}\BibitemShut {NoStop}%
\bibitem [{\citenamefont {Whitfield}\ \emph {et~al.}(2011)\citenamefont
  {Whitfield}, \citenamefont {Biamonte},\ and\ \citenamefont
  {Aspuru-Guzik}}]{Whitfield:2011mp}%
  \BibitemOpen
  \bibfield  {author} {\bibinfo {author} {\bibfnamefont {J.~D.}\ \bibnamefont
  {Whitfield}}, \bibinfo {author} {\bibfnamefont {J.}~\bibnamefont
  {Biamonte}},\ and\ \bibinfo {author} {\bibfnamefont {A.}~\bibnamefont
  {Aspuru-Guzik}},\ }\href {https://doi.org/10.1080/00268976.2011.552441}
  {\bibfield  {journal} {\bibinfo  {journal} {Molecular Physics}\ }\textbf
  {\bibinfo {volume} {109}},\ \bibinfo {pages} {735} (\bibinfo {year}
  {2011})}\BibitemShut {NoStop}%
\bibitem [{\citenamefont {Toloui}\ and\ \citenamefont
  {Love}(2013)}]{toloui_quantum_2013}%
  \BibitemOpen
  \bibfield  {author} {\bibinfo {author} {\bibfnamefont {B.}~\bibnamefont
  {Toloui}}\ and\ \bibinfo {author} {\bibfnamefont {P.~J.}\ \bibnamefont
  {Love}},\ }\href {http://arxiv.org/abs/1312.2579} {\bibinfo {title} {Quantum
  {Algorithms} for {Quantum} {Chemistry} based on the sparsity of the
  {CI}-matrix}} (\bibinfo {year} {2013}),\ \bibinfo {note} {arXiv:1312.2579
  [quant-ph]}\BibitemShut {NoStop}%
\bibitem [{\citenamefont {Di~Matteo}\ \emph {et~al.}(2021)\citenamefont
  {Di~Matteo}, \citenamefont {McCoy}, \citenamefont {Gysbers} \emph
  {et~al.}}]{di_matteo_improving_2021}%
  \BibitemOpen
  \bibfield  {author} {\bibinfo {author} {\bibfnamefont {O.}~\bibnamefont
  {Di~Matteo}}, \bibinfo {author} {\bibfnamefont {A.}~\bibnamefont {McCoy}},
  \bibinfo {author} {\bibfnamefont {P.}~\bibnamefont {Gysbers}}, \emph
  {et~al.},\ }\href {https://doi.org/10.1103/PhysRevA.103.042405} {\bibfield
  {journal} {\bibinfo  {journal} {Phys. Rev. A}\ }\textbf {\bibinfo {volume}
  {103}},\ \bibinfo {pages} {042405} (\bibinfo {year} {2021})}\BibitemShut
  {NoStop}%
\bibitem [{\citenamefont {Kirby}\ \emph {et~al.}(2022)\citenamefont {Kirby},
  \citenamefont {Fuller}, \citenamefont {Hadfield},\ and\ \citenamefont
  {Mezzacapo}}]{kirby_second-quantized_2022}%
  \BibitemOpen
  \bibfield  {author} {\bibinfo {author} {\bibfnamefont {W.}~\bibnamefont
  {Kirby}}, \bibinfo {author} {\bibfnamefont {B.}~\bibnamefont {Fuller}},
  \bibinfo {author} {\bibfnamefont {C.}~\bibnamefont {Hadfield}},\ and\
  \bibinfo {author} {\bibfnamefont {A.}~\bibnamefont {Mezzacapo}},\ }\href
  {https://doi.org/10.1103/PRXQuantum.3.020351} {\bibfield  {journal} {\bibinfo
   {journal} {PRX Quantum}\ }\textbf {\bibinfo {volume} {3}},\ \bibinfo {pages}
  {020351} (\bibinfo {year} {2022})}\BibitemShut {NoStop}%
\bibitem [{\citenamefont {Shee}\ \emph {et~al.}(2022)\citenamefont {Shee},
  \citenamefont {Tsai}, \citenamefont {Hong} \emph
  {et~al.}}]{shee_qubit-efficient_2022}%
  \BibitemOpen
  \bibfield  {author} {\bibinfo {author} {\bibfnamefont {Y.}~\bibnamefont
  {Shee}}, \bibinfo {author} {\bibfnamefont {P.-K.}\ \bibnamefont {Tsai}},
  \bibinfo {author} {\bibfnamefont {C.-L.}\ \bibnamefont {Hong}}, \emph
  {et~al.},\ }\href {https://doi.org/10.1103/PhysRevResearch.4.023154}
  {\bibfield  {journal} {\bibinfo  {journal} {Phys. Rev. Research}\ }\textbf
  {\bibinfo {volume} {4}},\ \bibinfo {pages} {023154} (\bibinfo {year}
  {2022})}\BibitemShut {NoStop}%
\bibitem [{\citenamefont {Huang}\ \emph {et~al.}(2023)\citenamefont {Huang},
  \citenamefont {Sheng}, \citenamefont {Govoni},\ and\ \citenamefont
  {Galli}}]{huang_quantum_2023}%
  \BibitemOpen
  \bibfield  {author} {\bibinfo {author} {\bibfnamefont {B.}~\bibnamefont
  {Huang}}, \bibinfo {author} {\bibfnamefont {N.}~\bibnamefont {Sheng}},
  \bibinfo {author} {\bibfnamefont {M.}~\bibnamefont {Govoni}},\ and\ \bibinfo
  {author} {\bibfnamefont {G.}~\bibnamefont {Galli}},\ }\href
  {https://doi.org/10.1021/acs.jctc.2c01119} {\bibfield  {journal} {\bibinfo
  {journal} {J. Chem. Theory Comput.}\ }\textbf {\bibinfo {volume} {19}},\
  \bibinfo {pages} {1487} (\bibinfo {year} {2023})}\BibitemShut {NoStop}%
\bibitem [{\citenamefont {McArdle}\ \emph {et~al.}(2020)\citenamefont
  {McArdle}, \citenamefont {Endo}, \citenamefont {Aspuru-Guzik}, \citenamefont
  {Benjamin},\ and\ \citenamefont {Yuan}}]{mcardle_quantum_2020}%
  \BibitemOpen
  \bibfield  {author} {\bibinfo {author} {\bibfnamefont {S.}~\bibnamefont
  {McArdle}}, \bibinfo {author} {\bibfnamefont {S.}~\bibnamefont {Endo}},
  \bibinfo {author} {\bibfnamefont {A.}~\bibnamefont {Aspuru-Guzik}}, \bibinfo
  {author} {\bibfnamefont {S.~C.}\ \bibnamefont {Benjamin}},\ and\ \bibinfo
  {author} {\bibfnamefont {X.}~\bibnamefont {Yuan}},\ }\href
  {https://doi.org/10.1103/RevModPhys.92.015003} {\bibfield  {journal}
  {\bibinfo  {journal} {Reviews of Modern Physics}\ }\textbf {\bibinfo {volume}
  {92}},\ \bibinfo {pages} {015003} (\bibinfo {year} {2020})}\BibitemShut
  {NoStop}%
\bibitem [{\citenamefont {Babbush}\ \emph {et~al.}(2016)\citenamefont
  {Babbush}, \citenamefont {Berry}, \citenamefont {Kivlichan} \emph
  {et~al.}}]{babbush_exponentially_2016}%
  \BibitemOpen
  \bibfield  {author} {\bibinfo {author} {\bibfnamefont {R.}~\bibnamefont
  {Babbush}}, \bibinfo {author} {\bibfnamefont {D.~W.}\ \bibnamefont {Berry}},
  \bibinfo {author} {\bibfnamefont {I.~D.}\ \bibnamefont {Kivlichan}}, \emph
  {et~al.},\ }\href {https://doi.org/10.1088/1367-2630/18/3/033032} {\bibfield
  {journal} {\bibinfo  {journal} {New J. Phys.}\ }\textbf {\bibinfo {volume}
  {18}},\ \bibinfo {pages} {033032} (\bibinfo {year} {2016})}\BibitemShut
  {NoStop}%
\bibitem [{\citenamefont {Bravyi}\ \emph {et~al.}(2017)\citenamefont {Bravyi},
  \citenamefont {Gambetta}, \citenamefont {Mezzacapo},\ and\ \citenamefont
  {Temme}}]{bravyi_tapering_2017}%
  \BibitemOpen
  \bibfield  {author} {\bibinfo {author} {\bibfnamefont {S.}~\bibnamefont
  {Bravyi}}, \bibinfo {author} {\bibfnamefont {J.~M.}\ \bibnamefont
  {Gambetta}}, \bibinfo {author} {\bibfnamefont {A.}~\bibnamefont
  {Mezzacapo}},\ and\ \bibinfo {author} {\bibfnamefont {K.}~\bibnamefont
  {Temme}},\ }\href {http://arxiv.org/abs/1701.08213} {\bibinfo {title}
  {Tapering off qubits to simulate fermionic {Hamiltonians}}} (\bibinfo {year}
  {2017}),\ \bibinfo {note} {arXiv:1701.08213 [quant-ph]}\BibitemShut {NoStop}%
\bibitem [{\citenamefont {Berry}\ \emph {et~al.}(2018)\citenamefont {Berry},
  \citenamefont {Kieferová}, \citenamefont {Scherer} \emph
  {et~al.}}]{berry_improved_2018}%
  \BibitemOpen
  \bibfield  {author} {\bibinfo {author} {\bibfnamefont {D.~W.}\ \bibnamefont
  {Berry}}, \bibinfo {author} {\bibfnamefont {M.}~\bibnamefont {Kieferová}},
  \bibinfo {author} {\bibfnamefont {A.}~\bibnamefont {Scherer}}, \emph
  {et~al.},\ }\href {https://doi.org/10.1038/s41534-018-0071-5} {\bibfield
  {journal} {\bibinfo  {journal} {npj Quantum Inf}\ }\textbf {\bibinfo {volume}
  {4}},\ \bibinfo {pages} {22} (\bibinfo {year} {2018})}\BibitemShut {NoStop}%
\bibitem [{\citenamefont {Babbush}\ \emph {et~al.}(2019)\citenamefont
  {Babbush}, \citenamefont {Berry}, \citenamefont {McClean},\ and\
  \citenamefont {Neven}}]{babbush_quantum_2019}%
  \BibitemOpen
  \bibfield  {author} {\bibinfo {author} {\bibfnamefont {R.}~\bibnamefont
  {Babbush}}, \bibinfo {author} {\bibfnamefont {D.~W.}\ \bibnamefont {Berry}},
  \bibinfo {author} {\bibfnamefont {J.~R.}\ \bibnamefont {McClean}},\ and\
  \bibinfo {author} {\bibfnamefont {H.}~\bibnamefont {Neven}},\ }\href
  {https://doi.org/10.1038/s41534-019-0199-y} {\bibfield  {journal} {\bibinfo
  {journal} {npj Quantum Inf}\ }\textbf {\bibinfo {volume} {5}},\ \bibinfo
  {pages} {92} (\bibinfo {year} {2019})}\BibitemShut {NoStop}%
\bibitem [{\citenamefont {Su}\ \emph {et~al.}(2021)\citenamefont {Su},
  \citenamefont {Berry}, \citenamefont {Wiebe}, \citenamefont {Rubin},\ and\
  \citenamefont {Babbush}}]{su_fault-tolerant_2021}%
  \BibitemOpen
  \bibfield  {author} {\bibinfo {author} {\bibfnamefont {Y.}~\bibnamefont
  {Su}}, \bibinfo {author} {\bibfnamefont {D.~W.}\ \bibnamefont {Berry}},
  \bibinfo {author} {\bibfnamefont {N.}~\bibnamefont {Wiebe}}, \bibinfo
  {author} {\bibfnamefont {N.}~\bibnamefont {Rubin}},\ and\ \bibinfo {author}
  {\bibfnamefont {R.}~\bibnamefont {Babbush}},\ }\href
  {https://doi.org/10.1103/PRXQuantum.2.040332} {\bibfield  {journal} {\bibinfo
   {journal} {PRX Quantum}\ }\textbf {\bibinfo {volume} {2}},\ \bibinfo {pages}
  {040332} (\bibinfo {year} {2021})}\BibitemShut {NoStop}%
\bibitem [{\citenamefont {Kogut}\ and\ \citenamefont
  {Susskind}(1975)}]{Kogut:1974ag}%
  \BibitemOpen
  \bibfield  {author} {\bibinfo {author} {\bibfnamefont {J.}~\bibnamefont
  {Kogut}}\ and\ \bibinfo {author} {\bibfnamefont {L.}~\bibnamefont
  {Susskind}},\ }\href {https://doi.org/10.1103/PhysRevD.11.395} {\bibfield
  {journal} {\bibinfo  {journal} {Phys. Rev. D}\ }\textbf {\bibinfo {volume}
  {11}},\ \bibinfo {pages} {395} (\bibinfo {year} {1975})}\BibitemShut
  {NoStop}%
\bibitem [{\citenamefont {Jordan}\ \emph {et~al.}(2012)\citenamefont {Jordan},
  \citenamefont {Lee},\ and\ \citenamefont {Preskill}}]{jordan_quantum_2012}%
  \BibitemOpen
  \bibfield  {author} {\bibinfo {author} {\bibfnamefont {S.~P.}\ \bibnamefont
  {Jordan}}, \bibinfo {author} {\bibfnamefont {K.~S.~M.}\ \bibnamefont {Lee}},\
  and\ \bibinfo {author} {\bibfnamefont {J.}~\bibnamefont {Preskill}},\ }\href
  {https://doi.org/10.1126/science.1217069} {\bibfield  {journal} {\bibinfo
  {journal} {Science}\ }\textbf {\bibinfo {volume} {336}},\ \bibinfo {pages}
  {1130} (\bibinfo {year} {2012})}\BibitemShut {NoStop}%
\bibitem [{\citenamefont {Zohar}\ \emph {et~al.}(2013)\citenamefont {Zohar},
  \citenamefont {Cirac},\ and\ \citenamefont {Reznik}}]{Zohar:2013zla}%
  \BibitemOpen
  \bibfield  {author} {\bibinfo {author} {\bibfnamefont {E.}~\bibnamefont
  {Zohar}}, \bibinfo {author} {\bibfnamefont {J.~I.}\ \bibnamefont {Cirac}},\
  and\ \bibinfo {author} {\bibfnamefont {B.}~\bibnamefont {Reznik}},\ }\href
  {https://doi.org/10.1103/PhysRevA.88.023617} {\bibfield  {journal} {\bibinfo
  {journal} {Phys. Rev. A}\ }\textbf {\bibinfo {volume} {88}},\ \bibinfo
  {pages} {023617} (\bibinfo {year} {2013})}\BibitemShut {NoStop}%
\bibitem [{\citenamefont {Macridin}\ \emph {et~al.}(2018)\citenamefont
  {Macridin}, \citenamefont {Spentzouris}, \citenamefont {Amundson},\ and\
  \citenamefont {Harnik}}]{macridin_digital_2018}%
  \BibitemOpen
  \bibfield  {author} {\bibinfo {author} {\bibfnamefont {A.}~\bibnamefont
  {Macridin}}, \bibinfo {author} {\bibfnamefont {P.}~\bibnamefont
  {Spentzouris}}, \bibinfo {author} {\bibfnamefont {J.}~\bibnamefont
  {Amundson}},\ and\ \bibinfo {author} {\bibfnamefont {R.}~\bibnamefont
  {Harnik}},\ }\href {https://doi.org/10.1103/PhysRevA.98.042312} {\bibfield
  {journal} {\bibinfo  {journal} {Phys. Rev. A}\ }\textbf {\bibinfo {volume}
  {98}},\ \bibinfo {pages} {042312} (\bibinfo {year} {2018})}\BibitemShut
  {NoStop}%
\bibitem [{\citenamefont {Farrell}\ \emph
  {et~al.}(2023{\natexlab{a}})\citenamefont {Farrell}, \citenamefont
  {Chernyshev}, \citenamefont {Powell} \emph {et~al.}}]{Farrell:2022wyt}%
  \BibitemOpen
  \bibfield  {author} {\bibinfo {author} {\bibfnamefont {R.~C.}\ \bibnamefont
  {Farrell}}, \bibinfo {author} {\bibfnamefont {I.~A.}\ \bibnamefont
  {Chernyshev}}, \bibinfo {author} {\bibfnamefont {S.~J.~M.}\ \bibnamefont
  {Powell}}, \emph {et~al.},\ }\href
  {https://doi.org/10.1103/PhysRevD.107.054512} {\bibfield  {journal} {\bibinfo
   {journal} {Phys. Rev. D}\ }\textbf {\bibinfo {volume} {107}},\ \bibinfo
  {pages} {054512} (\bibinfo {year} {2023}{\natexlab{a}})}\BibitemShut
  {NoStop}%
\bibitem [{\citenamefont {Stetina}\ \emph {et~al.}(2022)\citenamefont
  {Stetina}, \citenamefont {Ciavarella}, \citenamefont {Li},\ and\
  \citenamefont {Wiebe}}]{Stetina:2022simulatingeffective}%
  \BibitemOpen
  \bibfield  {author} {\bibinfo {author} {\bibfnamefont {T.~F.}\ \bibnamefont
  {Stetina}}, \bibinfo {author} {\bibfnamefont {A.}~\bibnamefont {Ciavarella}},
  \bibinfo {author} {\bibfnamefont {X.}~\bibnamefont {Li}},\ and\ \bibinfo
  {author} {\bibfnamefont {N.}~\bibnamefont {Wiebe}},\ }\href
  {https://doi.org/10.22331/q-2022-01-18-622} {\bibfield  {journal} {\bibinfo
  {journal} {Quantum}\ }\textbf {\bibinfo {volume} {6}},\ \bibinfo {pages}
  {622} (\bibinfo {year} {2022})}\BibitemShut {NoStop}%
\bibitem [{\citenamefont {Macridin}\ \emph {et~al.}(2022)\citenamefont
  {Macridin}, \citenamefont {Li}, \citenamefont {Mrenna},\ and\ \citenamefont
  {Spentzouris}}]{macridin_bosonic_2022}%
  \BibitemOpen
  \bibfield  {author} {\bibinfo {author} {\bibfnamefont {A.}~\bibnamefont
  {Macridin}}, \bibinfo {author} {\bibfnamefont {A.~C.~Y.}\ \bibnamefont {Li}},
  \bibinfo {author} {\bibfnamefont {S.}~\bibnamefont {Mrenna}},\ and\ \bibinfo
  {author} {\bibfnamefont {P.}~\bibnamefont {Spentzouris}},\ }\href
  {https://doi.org/10.1103/PhysRevA.105.052405} {\bibfield  {journal} {\bibinfo
   {journal} {Physical Review A}\ }\textbf {\bibinfo {volume} {105}},\ \bibinfo
  {pages} {052405} (\bibinfo {year} {2022})}\BibitemShut {NoStop}%
\bibitem [{\citenamefont {Farrell}\ \emph
  {et~al.}(2023{\natexlab{b}})\citenamefont {Farrell}, \citenamefont
  {Chernyshev}, \citenamefont {Powell}, \citenamefont {Zemlevskiy},
  \citenamefont {Illa},\ and\ \citenamefont
  {Savage}}]{farrell_preparations_2023}%
  \BibitemOpen
  \bibfield  {author} {\bibinfo {author} {\bibfnamefont {R.~C.}\ \bibnamefont
  {Farrell}}, \bibinfo {author} {\bibfnamefont {I.~A.}\ \bibnamefont
  {Chernyshev}}, \bibinfo {author} {\bibfnamefont {S.~J.}\ \bibnamefont
  {Powell}}, \bibinfo {author} {\bibfnamefont {N.~A.}\ \bibnamefont
  {Zemlevskiy}}, \bibinfo {author} {\bibfnamefont {M.}~\bibnamefont {Illa}},\
  and\ \bibinfo {author} {\bibfnamefont {M.~J.}\ \bibnamefont {Savage}},\
  }\href {https://doi.org/10.1103/PhysRevD.107.054513} {\bibfield  {journal}
  {\bibinfo  {journal} {Physical Review D}\ }\textbf {\bibinfo {volume}
  {107}},\ \bibinfo {pages} {054513} (\bibinfo {year}
  {2023}{\natexlab{b}})}\BibitemShut {NoStop}%
\bibitem [{\citenamefont {Farrell}\ \emph {et~al.}(2024)\citenamefont
  {Farrell}, \citenamefont {Illa}, \citenamefont {Ciavarella},\ and\
  \citenamefont {Savage}}]{farrell_scalable_2024}%
  \BibitemOpen
  \bibfield  {author} {\bibinfo {author} {\bibfnamefont {R.~C.}\ \bibnamefont
  {Farrell}}, \bibinfo {author} {\bibfnamefont {M.}~\bibnamefont {Illa}},
  \bibinfo {author} {\bibfnamefont {A.~N.}\ \bibnamefont {Ciavarella}},\ and\
  \bibinfo {author} {\bibfnamefont {M.~J.}\ \bibnamefont {Savage}},\ }\href
  {https://doi.org/10.1103/PRXQuantum.5.020315} {\bibfield  {journal} {\bibinfo
   {journal} {PRX Quantum}\ }\textbf {\bibinfo {volume} {5}},\ \bibinfo {pages}
  {020315} (\bibinfo {year} {2024})}\BibitemShut {NoStop}%
\bibitem [{\citenamefont {Weinberg}(2014)}]{weinberg_quantum_2014}%
  \BibitemOpen
  \bibfield  {author} {\bibinfo {author} {\bibfnamefont {S.}~\bibnamefont
  {Weinberg}},\ }\href@noop {} {\emph {\bibinfo {title} {The quantum theory of
  fields. 1: {Foundations}}}}\ (\bibinfo  {publisher} {Cambridge Univ. Press},\
  \bibinfo {address} {Cambridge},\ \bibinfo {year} {2014})\BibitemShut
  {NoStop}%
\bibitem [{\citenamefont {Barata}\ \emph {et~al.}(2022)\citenamefont {Barata},
  \citenamefont {Du}, \citenamefont {Li}, \citenamefont {Qian},\ and\
  \citenamefont {Salgado}}]{barata_220806750v1_2022}%
  \BibitemOpen
  \bibfield  {author} {\bibinfo {author} {\bibfnamefont {J.}~\bibnamefont
  {Barata}}, \bibinfo {author} {\bibfnamefont {X.}~\bibnamefont {Du}}, \bibinfo
  {author} {\bibfnamefont {M.}~\bibnamefont {Li}}, \bibinfo {author}
  {\bibfnamefont {W.}~\bibnamefont {Qian}},\ and\ \bibinfo {author}
  {\bibfnamefont {C.~A.}\ \bibnamefont {Salgado}},\ }\href
  {https://doi.org/10.1103/PhysRevD.106.074013} {\bibfield  {journal} {\bibinfo
   {journal} {Physical Review D}\ }\textbf {\bibinfo {volume} {106}},\ \bibinfo
  {pages} {074013} (\bibinfo {year} {2022})},\ \bibinfo {note}
  {arXiv:2208.06750 [hep-ph, physics:quant-ph]}\BibitemShut {NoStop}%
\bibitem [{\citenamefont {Barata}\ \emph {et~al.}(2023)\citenamefont {Barata},
  \citenamefont {Du}, \citenamefont {Li}, \citenamefont {Qian},\ and\
  \citenamefont {Salgado}}]{barata_230701792v1_2023}%
  \BibitemOpen
  \bibfield  {author} {\bibinfo {author} {\bibfnamefont {J.}~\bibnamefont
  {Barata}}, \bibinfo {author} {\bibfnamefont {X.}~\bibnamefont {Du}}, \bibinfo
  {author} {\bibfnamefont {M.}~\bibnamefont {Li}}, \bibinfo {author}
  {\bibfnamefont {W.}~\bibnamefont {Qian}},\ and\ \bibinfo {author}
  {\bibfnamefont {C.~A.}\ \bibnamefont {Salgado}},\ }\href
  {https://doi.org/10.1103/PhysRevD.108.056023} {\bibfield  {journal} {\bibinfo
   {journal} {Physical Review D}\ }\textbf {\bibinfo {volume} {108}},\ \bibinfo
  {pages} {056023} (\bibinfo {year} {2023})},\ \bibinfo {note}
  {arXiv:2307.01792 [hep-ph, physics:nucl-th, physics:quant-ph]}\BibitemShut
  {NoStop}%
\bibitem [{\citenamefont {Głazek}\ \emph {et~al.}(2017)\citenamefont
  {Głazek}, \citenamefont {Gómez-Rocha}, \citenamefont {More},\ and\
  \citenamefont {Serafin}}]{glazek_renormalized_2017}%
  \BibitemOpen
  \bibfield  {author} {\bibinfo {author} {\bibfnamefont {S.~D.}\ \bibnamefont
  {Głazek}}, \bibinfo {author} {\bibfnamefont {M.}~\bibnamefont
  {Gómez-Rocha}}, \bibinfo {author} {\bibfnamefont {J.}~\bibnamefont {More}},\
  and\ \bibinfo {author} {\bibfnamefont {K.}~\bibnamefont {Serafin}},\ }\href
  {https://doi.org/10.1016/j.physletb.2017.08.018} {\bibfield  {journal}
  {\bibinfo  {journal} {Physics Letters B}\ }\textbf {\bibinfo {volume}
  {773}},\ \bibinfo {pages} {172} (\bibinfo {year} {2017})}\BibitemShut
  {NoStop}%
\bibitem [{\citenamefont {Barata}\ \emph {et~al.}(2021)\citenamefont {Barata},
  \citenamefont {Mueller}, \citenamefont {Tarasov},\ and\ \citenamefont
  {Venugopalan}}]{Barata:2020jtq}%
  \BibitemOpen
  \bibfield  {author} {\bibinfo {author} {\bibfnamefont {J.}~\bibnamefont
  {Barata}}, \bibinfo {author} {\bibfnamefont {N.}~\bibnamefont {Mueller}},
  \bibinfo {author} {\bibfnamefont {A.}~\bibnamefont {Tarasov}},\ and\ \bibinfo
  {author} {\bibfnamefont {R.}~\bibnamefont {Venugopalan}},\ }\href
  {https://doi.org/10.1103/PhysRevA.103.042410} {\bibfield  {journal} {\bibinfo
   {journal} {Phys. Rev. A}\ }\textbf {\bibinfo {volume} {103}},\ \bibinfo
  {pages} {042410} (\bibinfo {year} {2021})}\BibitemShut {NoStop}%
\bibitem [{\citenamefont {Kirby}\ \emph {et~al.}(2021)\citenamefont {Kirby},
  \citenamefont {Hadi}, \citenamefont {Kreshchuk},\ and\ \citenamefont
  {Love}}]{Kirby:2021PRA}%
  \BibitemOpen
  \bibfield  {author} {\bibinfo {author} {\bibfnamefont {W.~M.}\ \bibnamefont
  {Kirby}}, \bibinfo {author} {\bibfnamefont {S.}~\bibnamefont {Hadi}},
  \bibinfo {author} {\bibfnamefont {M.}~\bibnamefont {Kreshchuk}},\ and\
  \bibinfo {author} {\bibfnamefont {P.~J.}\ \bibnamefont {Love}},\ }\href
  {https://doi.org/10.1103/PhysRevA.104.042607} {\bibfield  {journal} {\bibinfo
   {journal} {Phys. Rev. A}\ }\textbf {\bibinfo {volume} {104}},\ \bibinfo
  {pages} {042607} (\bibinfo {year} {2021})}\BibitemShut {NoStop}%
\bibitem [{\citenamefont {Qian}\ \emph {et~al.}(2022)\citenamefont {Qian},
  \citenamefont {Basili}, \citenamefont {Pal}, \citenamefont {Luecke},\ and\
  \citenamefont {Vary}}]{Qian:2021jxp}%
  \BibitemOpen
  \bibfield  {author} {\bibinfo {author} {\bibfnamefont {W.}~\bibnamefont
  {Qian}}, \bibinfo {author} {\bibfnamefont {R.}~\bibnamefont {Basili}},
  \bibinfo {author} {\bibfnamefont {S.}~\bibnamefont {Pal}}, \bibinfo {author}
  {\bibfnamefont {G.}~\bibnamefont {Luecke}},\ and\ \bibinfo {author}
  {\bibfnamefont {J.~P.}\ \bibnamefont {Vary}},\ }\href
  {https://doi.org/10.1103/PhysRevResearch.4.043193} {\bibfield  {journal}
  {\bibinfo  {journal} {Phys. Rev. Research}\ }\textbf {\bibinfo {volume}
  {4}},\ \bibinfo {pages} {043193} (\bibinfo {year} {2022})}\BibitemShut
  {NoStop}%
\bibitem [{\citenamefont {Yao}(2022)}]{Yao:2022eqm}%
  \BibitemOpen
  \bibfield  {author} {\bibinfo {author} {\bibfnamefont {X.}~\bibnamefont
  {Yao}},\ }\href {http://arxiv.org/abs/2205.07902} {\bibinfo {title} {Quantum
  {Simulation} of {Light}-{Front} {QCD} for {Jet} {Quenching} in {Nuclear}
  {Environments}}} (\bibinfo {year} {2022}),\ \bibinfo {note} {arXiv:2205.07902
  [hep-lat, physics:hep-ph, physics:nucl-th, physics:quant-ph]}\BibitemShut
  {NoStop}%
\bibitem [{\citenamefont {Gómez~Rocha}\ \emph {et~al.}(2010)\citenamefont
  {Gómez~Rocha}, \citenamefont {Llanes-Estrada}, \citenamefont {Schütte},\
  and\ \citenamefont {Villalba-Chávez}}]{Rocha:2010nab}%
  \BibitemOpen
  \bibfield  {author} {\bibinfo {author} {\bibfnamefont {M.}~\bibnamefont
  {Gómez~Rocha}}, \bibinfo {author} {\bibfnamefont {F.~J.}\ \bibnamefont
  {Llanes-Estrada}}, \bibinfo {author} {\bibfnamefont {D.}~\bibnamefont
  {Schütte}},\ and\ \bibinfo {author} {\bibfnamefont {S.}~\bibnamefont
  {Villalba-Chávez}},\ }\href {https://doi.org/10.1140/epja/i2010-10949-3}
  {\bibfield  {journal} {\bibinfo  {journal} {Eur. Phys. J. A}\ }\textbf
  {\bibinfo {volume} {44}},\ \bibinfo {pages} {411} (\bibinfo {year}
  {2010})}\BibitemShut {NoStop}%
\bibitem [{\citenamefont {Chawdhry}\ and\ \citenamefont
  {Pellen}(2023)}]{Chawdhry:2023jks}%
  \BibitemOpen
  \bibfield  {author} {\bibinfo {author} {\bibfnamefont {H.}~\bibnamefont
  {Chawdhry}}\ and\ \bibinfo {author} {\bibfnamefont {M.}~\bibnamefont
  {Pellen}},\ }\href {https://doi.org/10.21468/SciPostPhys.15.5.205} {\bibfield
   {journal} {\bibinfo  {journal} {SciPost Phys.}\ }\textbf {\bibinfo {volume}
  {15}},\ \bibinfo {pages} {205} (\bibinfo {year} {2023})}\BibitemShut
  {NoStop}%
\bibitem [{\citenamefont {Nielsen}\ and\ \citenamefont
  {Chuang}(2012)}]{Nielsen:2012yss}%
  \BibitemOpen
  \bibfield  {author} {\bibinfo {author} {\bibfnamefont {M.~A.}\ \bibnamefont
  {Nielsen}}\ and\ \bibinfo {author} {\bibfnamefont {I.~L.}\ \bibnamefont
  {Chuang}},\ }\href {https://doi.org/10.1017/CBO9780511976667} {\emph
  {\bibinfo {title} {Quantum {Computation} and {Quantum} {Information}: 10th
  {Anniversary} {Edition}}}},\ \bibinfo {edition} {1st}\ ed.\ (\bibinfo
  {publisher} {Cambridge University Press},\ \bibinfo {year}
  {2012})\BibitemShut {NoStop}%
\bibitem [{\citenamefont {Seeley}\ \emph {et~al.}(2012)\citenamefont {Seeley},
  \citenamefont {Richard},\ and\ \citenamefont {Love}}]{seeley_12085986_2012}%
  \BibitemOpen
  \bibfield  {author} {\bibinfo {author} {\bibfnamefont {J.~T.}\ \bibnamefont
  {Seeley}}, \bibinfo {author} {\bibfnamefont {M.~J.}\ \bibnamefont
  {Richard}},\ and\ \bibinfo {author} {\bibfnamefont {P.~J.}\ \bibnamefont
  {Love}},\ }\href {https://doi.org/10.1063/1.4768229} {\bibfield  {journal}
  {\bibinfo  {journal} {The Journal of Chemical Physics}\ }\textbf {\bibinfo
  {volume} {137}},\ \bibinfo {pages} {224109} (\bibinfo {year}
  {2012})}\BibitemShut {NoStop}%
\bibitem [{\citenamefont {Gallimore}\ and\ \citenamefont
  {Liao}(2023)}]{gallimore_quantum_2023}%
  \BibitemOpen
  \bibfield  {author} {\bibinfo {author} {\bibfnamefont {D.}~\bibnamefont
  {Gallimore}}\ and\ \bibinfo {author} {\bibfnamefont {J.}~\bibnamefont
  {Liao}},\ }\href {https://doi.org/10.1103/PhysRevD.107.074012} {\bibfield
  {journal} {\bibinfo  {journal} {Physical Review D}\ }\textbf {\bibinfo
  {volume} {107}},\ \bibinfo {pages} {074012} (\bibinfo {year}
  {2023})}\BibitemShut {NoStop}%
\bibitem [{\citenamefont {Ring}\ and\ \citenamefont {Schuck}(2004)}]{Ring}%
  \BibitemOpen
  \bibfield  {author} {\bibinfo {author} {\bibfnamefont {P.}~\bibnamefont
  {Ring}}\ and\ \bibinfo {author} {\bibfnamefont {P.}~\bibnamefont {Schuck}},\
  }\href@noop {} {\emph {\bibinfo {title} {The nuclear many body problem}}},\
  \bibinfo {edition} {1st}\ ed.,\ Texts and monographs in physics\ (\bibinfo
  {publisher} {Springer},\ \bibinfo {address} {Berlin Heidelberg},\ \bibinfo
  {year} {2004})\BibitemShut {NoStop}%
\bibitem [{\citenamefont {Cotanch}\ and\ \citenamefont
  {Llanes-Estrada}(2001)}]{Cotanch:2001mc}%
  \BibitemOpen
  \bibfield  {author} {\bibinfo {author} {\bibfnamefont {S.}~\bibnamefont
  {Cotanch}}\ and\ \bibinfo {author} {\bibfnamefont {F.}~\bibnamefont
  {Llanes-Estrada}},\ }\href {https://doi.org/10.1016/S0375-9474(01)00886-7}
  {\bibfield  {journal} {\bibinfo  {journal} {Nuclear Physics A}\ }\textbf
  {\bibinfo {volume} {689}},\ \bibinfo {pages} {481} (\bibinfo {year}
  {2001})}\BibitemShut {NoStop}%
\bibitem [{\citenamefont {Llanes-Estrada}\ and\ \citenamefont
  {Cotanch}(2000)}]{Llanes-Estrada:1999nat}%
  \BibitemOpen
  \bibfield  {author} {\bibinfo {author} {\bibfnamefont {F.~J.}\ \bibnamefont
  {Llanes-Estrada}}\ and\ \bibinfo {author} {\bibfnamefont {S.~R.}\
  \bibnamefont {Cotanch}},\ }\href
  {https://doi.org/10.1103/PhysRevLett.84.1102} {\bibfield  {journal} {\bibinfo
   {journal} {Phys. Rev. Lett.}\ }\textbf {\bibinfo {volume} {84}},\ \bibinfo
  {pages} {1102} (\bibinfo {year} {2000})}\BibitemShut {NoStop}%
\bibitem [{\citenamefont {Fetter}\ and\ \citenamefont
  {Walecka}(2003)}]{FetterWalecka}%
  \BibitemOpen
  \bibfield  {author} {\bibinfo {author} {\bibfnamefont {A.~L.}\ \bibnamefont
  {Fetter}}\ and\ \bibinfo {author} {\bibfnamefont {J.~D.}\ \bibnamefont
  {Walecka}},\ }\href@noop {} {\emph {\bibinfo {title} {Quantum theory of
  many-particle systems}}},\ \bibinfo {edition} {corr. repr}\ ed.\ (\bibinfo
  {publisher} {Dover Publ},\ \bibinfo {address} {Mineola, N.Y},\ \bibinfo
  {year} {2003})\BibitemShut {NoStop}%
\end{thebibliography}%
\bibliographystyle{apsrev4-2}
\end{document}